%% file: main.tex
\documentclass[preprint,12pt,preprintnumbers]{elsarticle}
\usepackage[utf8]{inputenc}
\usepackage[T1]{fontenc}
\usepackage{graphicx}
\usepackage{amsmath,amssymb}
\usepackage{amsfonts}
\usepackage{bm}
\usepackage{braket}
\usepackage{siunitx}
\usepackage{tikz}
\usepackage{microtype}

\journal{Physics Reports}

\makeatletter
\let\oldps@pprintTitle\ps@pprintTitle
\def\ps@pprintTitle{%
  \oldps@pprintTitle
  \def\@oddhead{%
    \hfill\small FERMILAB-PUB-26-0274-T\hspace{1cm}}%
  \let\@evenhead\@oddhead
}
\makeatother

\newcommand{\fig}[1]{Fig.~\ref{fig:#1}}

\newcommand{\eq}[1]{Eq.~\eqref{eq:#1}}
\newcommand{\Sec}[1]{Sec.~\ref{sec:#1}}

\usepackage[colorlinks=true,backref=false, linktocpage=true,
citecolor=blue,urlcolor=blue,linkcolor=blue,pdfpagemode=UseOutlines]{hyperref}
\hypersetup{%
  bookmarksnumbered=true,
  pdftitle = {},
  pdfsubject = {},
  pdfauthor = {},
  pdfkeywords = {}
}

\newcommand{\D}{\mathrm{d}}

\newcommand{\be}{\begin{equation}}
\newcommand{\ee}{\end{equation}}
\newcommand{\bea}{\begin{eqnarray}}
\newcommand{\eea}{\end{eqnarray}}

\usepackage{amsmath}
\usepackage{amssymb}
\usepackage{mathrsfs}
\usepackage{multirow}
\usepackage{tabularx}
\usepackage{placeins}
\usepackage{graphicx}
\usepackage{subcaption}

\begin{document}
\begin{frontmatter}
\title{Particle Physics Driven by Quantum Technology – Quantum Simulations and Quantum Sensing}
\author[a,b,n,o]{Itay M. Bloch}
\affiliation[a]{Theoretical Physics Group, Lawrence Berkeley National Laboratory, Berkeley, CA 94720, U.S.A.}
\affiliation[b]{Berkeley Center for Theoretical Physics, University of California, Berkeley, CA 94720, U.S.A.}
\affiliation[n]{Theoretical Physics Department, CERN, 1 Esplanade des Particules, CH-1211 Geneva 23, Switzerland}
\affiliation[o]{Physics Department, Technion – Israel Institute of Technology, Haifa 3200003, Israel}

\ead{ItayBlochM@Gmail.com}
\author[c,d,e,f,g]{Marcela Carena}
\affiliation[c]{Perimeter Institute for Theoretical Physics, 31 Caroline St. N., Waterloo, Ontario N2L 2Y5, Canada}
\affiliation[d]{Fermi National Accelerator Laboratory, Batavia,  Illinois, 60510, USA}
\affiliation[e]{Enrico Fermi Institute, University of Chicago, Chicago, Illinois, 60637, USA}
\affiliation[f]{Kavli Institute for Cosmological Physics, University of Chicago, Chicago, Illinois, 60637, USA}
\affiliation[g]{Department of Physics, University of Chicago, Chicago, Illinois, 60637, USA}
\ead{mcarena@perimeterinstitute.ca}
\author[h,i]{Yifan Chen}
\affiliation[h]{State Key Laboratory of Dark Matter Physics, Tsung-Dao Lee Institute, Shanghai Jiao Tong University, Shanghai 200240, China} 
\affiliation[i]{Key Laboratory for Particle Astrophysics and Cosmology (MOE) \& Shanghai Key Laboratory for Particle Physics and Cosmology, Shanghai Jiao Tong University, Shanghai 200240, China}
\ead{chen.yifan@sjtu.edu.cn}
\author[j]{Xinran Li}
\affiliation[j]{School of Physics, 
	Peking University, Beijing 100871, China}
\ead{xrl@pku.edu.cn}
\author[k]{Ying-Ying Li}
\affiliation[k]{Institute of High Energy Physics, Chinese Academy of Sciences, Beijing 100049, China}
\ead{liyingying@ihep.ac.cn}
\author[j,l,m]{Jing Shu}
\affiliation[l]{Center for High Energy Physics, Peking University, Beijing 100871, China}
\affiliation[m]{Beijing Laser Acceleration Innovation Center, Huairou, Beijing, 101400, China}
\ead{jshu@pku.edu.cn}
\date{\today}
\begin{abstract}

We review recent advances in particle physics enabled and motivated by the rapid progress of quantum technologies. The continued development of quantum computing toward large-scale, fault-tolerant systems has the potential to address dynamical problems that remain extremely challenging for classical computational approaches, opening new avenues for studying nonperturbative dynamical processes.
Tabletop detectors and quantum-enhanced technologies have reached unprecedented sensitivities, enabling novel searches for ultralight particles, gravitational waves, and other physics beyond the Standard Model in parameter regimes previously inaccessible to conventional instrumentation. Taken together, these developments highlight a rapidly evolving landscape in which quantum technologies are beginning to reshape fundamental physics. We aim to synthesize recent progress and illuminate the opportunities they present for advancing particle physics in the coming years.
\end{abstract}

\end{frontmatter}

\tableofcontents

\section{Introduction}
Particle physics is the scientific discipline devoted to understanding the fundamental building blocks of matter and the forces that govern their interactions. Over more than a century of exploration, it has achieved remarkable successes, ranging from the discovery of the electron in 1897 \cite{Thomson:1897cm} to the discovery of the Higgs boson in 2012 \cite{ATLAS:2012yve, CMS:2012qbp} by the ATLAS and CMS collaborations at the Large Hadron Collider (LHC), as well as the establishment of the quantum field theory (QFT) framework underlying the Standard Model (SM) throughout the latter half of the twentieth century. The SM provides a unified description of all known fundamental particles and their interactions. 

Despite these triumphs, profound mysteries remain. The mechanism of hadronization—the process by which quarks and gluons form bound states—has challenged theorists for more than fifty years \cite{Gross:2022hyw}; the origin of the universe’s matter–antimatter asymmetry has puzzled scientists since the 1940s, when it became clear that the universe was hot during the early stages of its history \cite{Gamow:1946eb, Alpher:1948ve}; the nature of dark matter has remained elusive since its astrophysical evidence was first identified in 1933 by Fritz Zwicky through studies of galaxy clusters~\cite{Zwicky:1933gu}. Addressing these mysteries requires a concerted and collaborative effort, combining precision measurements of SM processes, continued advancement of theoretical and computational tools capable of modeling and predicting complex particle phenomena in challenging regimes and exploration of physics beyond the Standard Model (BSM) across all experimental frontiers.

Meanwhile, the ability to precisely control and manipulate quantum entanglement and superposition has driven transformative progress across quantum information science, turning foundational quantum phenomena into practical tools. 
These advances have opened the prospect of using quantum computing for calculations that are classically intractable \cite{QC-40} and quantum sensing for detecting extremely weak signals with unprecedented sensitivity \cite{Montenegro:2024agq}, thereby forging a powerful connection to particle physics.

\begin{figure}
    \centering
\includegraphics[width=\linewidth]{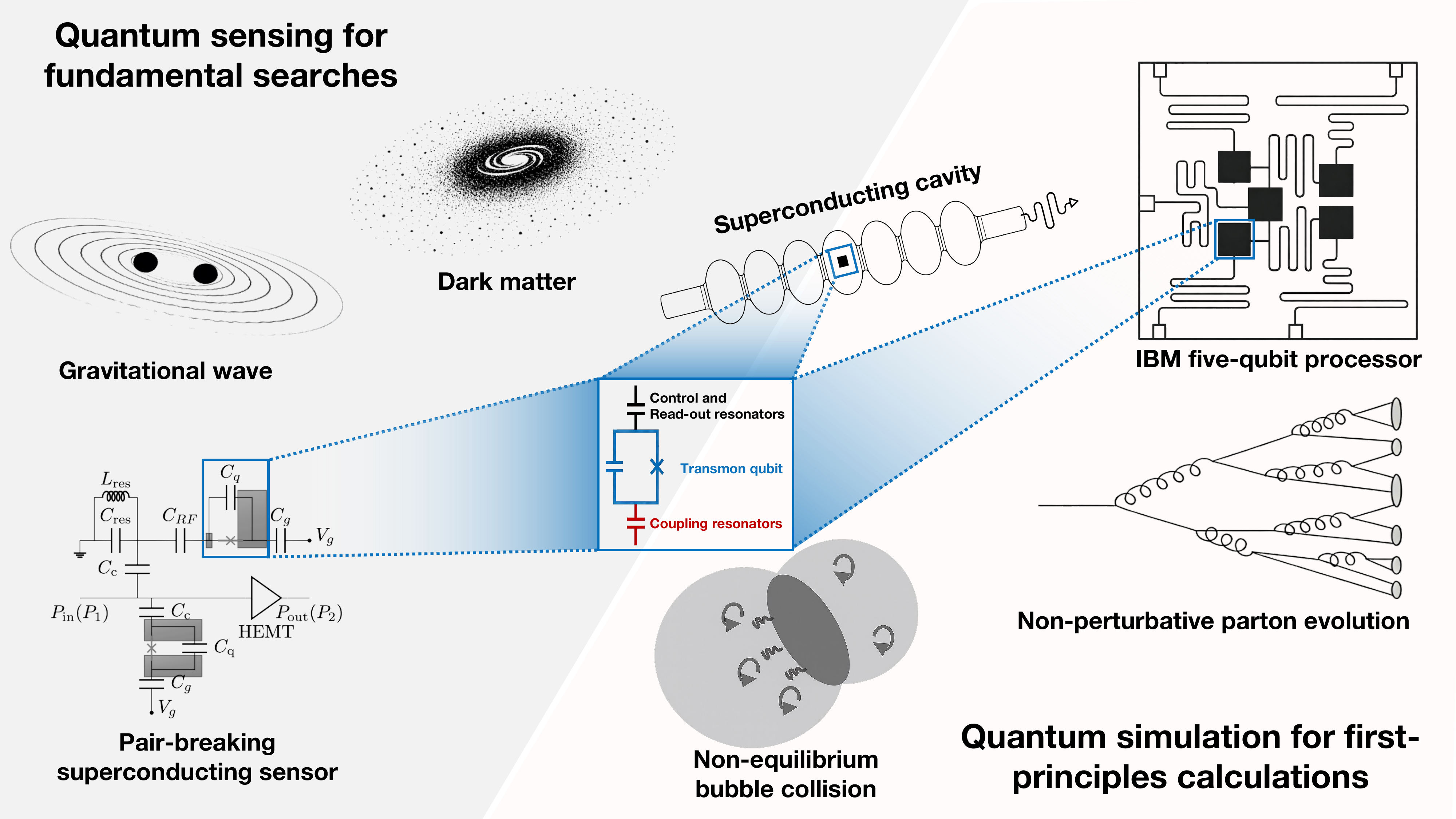}
    \caption{Quantum sensing and quantum simulation with superconducting platforms. 
Left: quantum sensing with superconducting cavities and pair-breaking detectors, probing weak signals from dark matter and gravitational waves.
Right: quantum simulations using transmon-based superconducting processors, enabling first-principles studies of non-perturbative quantum field dynamics. 
Images include a superconducting quantum chip (IBM) and conceptual cavity-based sensor designs (adapted from~\cite{Ramanathan:2024hsf}).}
    \label{fig:overview}
\end{figure}
Transmon-based superconducting devices provide a concrete example of this connection, as illustrated in \fig{overview}. On the computing side, transmon-based devices developed by IBM—now reaching $\mathcal{O}(10^3)$ physical qubits \cite{AbuGhanem:2024atm}, whereas a five-qubit processor is shown in \fig{overview}—as well as efforts \cite{Roy:2024uro} aimed at enabling qudit-based computing systems using superconducting cavities, have the potential to enable first-principles quantum simulations of non-perturbative dynamics in quantum field theories and out-of-equilibrium physics in the early universe. These capabilities open new avenues for studying quantum phenomena ranging from parton evolution to bubble collisions associated with early-universe phase transitions, which remain inaccessible to classical computation.

At the same time, the same quantum-control and readout techniques underpin quantum sensors based on superconducting cavities and pair-breaking superconducting detectors. These sensors enable ultra-sensitive measurements of faint signals, opening new pathways for probing BSM physics through subtle effects potentially generated by light BSM particles such as axion dark matter, while also advancing detection frontiers for gravitational waves. As these capabilities mature, quantum technologies are emerging as a uniquely powerful platform for addressing some of the deepest open questions in particle physics. 

In this report, we thus focus on two central aspects of quantum technologies—quantum simulation and quantum sensing—providing a pedagogical overview of their underlying principles, together with selected examples of their applications in particle physics. We aim to clarify their current capabilities and limitations, and to outline future directions toward fully realizing their potential for advancing our understanding of the fundamental laws of nature. 

We begin in \Sec{simulation} by discussing the interplay between quantum simulation and particle physics. We then turn to quantum sensing strategies for probing particle-physics phenomena, with emphasis on cavity- and circuit-based platforms in \Sec{cavity}, quantum magnetometry in \Sec{magnetometry}, and superconducting quantum devices in \Sec{superconducting}. We conclude in \Sec{summary} with a summary.

We note that the area of quantum sensing is much broader than those covered in this review. For example, atomic clocks~\cite{Filzinger:2023zrs,Banerjee:2023bjc,Madge:2024aot,Kennedy:2020bac,Sherrill:2023zah,Kobayashi:2022vsf,Brzeminski:2026rox}, interferometers~\cite{Crescini:2023zyl,Gottel:2024cfj,Vermeulen:2021epa,Aiello:2021wlp,Murgui:2026mlr} and many other quantum technologies~\cite{Delaunay:2026ymq,Arvanitaki:2021wjk,Chigusa:2025rqs,Ye:2024qdg,VanTilburg:2015oza,Alarcon:2022ero,Safronova:2017xyt,Aharony:2019iad,Antypas:2022asj,Gan:2025nlu} have been used or proposed in a wide range of new-physics searches. For broader perspectives and complementary coverage, we refer the reader to existing reviews on quantum sensing for fundamental physics~\cite{Clerk:2008tlb,Graham:2015ouw,Safronova:2017xyt,Irastorza:2018dyq,Agrawal:2021dbo,Semertzidis:2021rxs,Irastorza:2021tdu,Chadha-Day:2021szb,Alonso:2022oot,Buchmueller:2023nll,Bass:2023hoi,Arrowsmith-Kron:2023hcr,Cong:2024qly,Berlin:2024pzi,Ye:2024qdg,Malo:2024tye,Jiang:2024boi,Rybka:2024zoe,Baryakhtar:2025jwh,Aybas:2026rwu}.

\input{sections/quantum-simulation}

\input{sections/quantum-cavity}

\input{sections/quantum-magnetometer}

\input{sections/quantum-superconducting_updated}
\input{sections/conclusions}

\bibliographystyle{elsarticle-num-et-al-5}
\bibliography{main_published_clean}
\end{document}

%% file: sections/quantum-simulation.tex
  \section{Quantum simulation and particle physics}
\label{sec:simulation} 
In particle physics, investigating the nonperturbative dynamics of quantum field theory (QFT) is of fundamental importance. Such studies encompass a broad range of challenging problems, including real-time dynamics, finite-density systems, and out-of-equilibrium phenomena. They are essential for elucidating the mechanisms of hadronization, exploring the phase structure of strongly interacting matter at finite density, testing the SM with high precision, and probing potential new physics. They also provide a pathway toward understanding longstanding puzzles such as the origin of matter in the universe. However, QFT describes quantum many-body systems whose dynamics involve an enormous number of degrees of freedom, making generic first-principles simulations intractable with classical computational resources. 

Euclidean lattice field theory, a first-principles nonperturbative formulation of quantum field theory on a discretized spacetime lattice and evaluated using Monte Carlo importance sampling, has achieved remarkable success in computing static and equilibrium properties of quantum field theories, particularly in QCD. However, its application to nonperturbative dynamical processes remains severely limited by the sign problem, which leads to an exponential degradation of the signal-to-noise ratio and renders efficient classical computation in these regimes intractable.

The idea of simulating quantum systems with quantum devices was first proposed by Richard Feynman \cite{Feynman:1981tf}. It was later shown by Seth Lloyd \cite{Lloyd:1996aai} that the unitary time evolution generated by local Hamiltonians can be efficiently simulated on a quantum computer, establishing the theoretical foundation for quantum simulation. This provides a fundamentally different approach to the real-time dynamics of QFT, which can be formulated in terms of a Hamiltonian and its unitary time evolution. By encoding quantum states in controllable quantum hardware, quantum computers can, in principle, simulate local quantum systems with computational resources that can scale polynomially with system size and evolution time. This capability provides a potential route toward addressing the exponential computational bottlenecks encountered in classical first-principles approaches.

Recent advances in quantum algorithms and quantum hardware have brought this vision closer to practical realization, stimulating efforts to apply quantum computing to first-principles studies of particle physics. In particular, the development of quantum algorithms tailored to simulating dynamics of QFT, together with the establishment of benchmarking protocols to assess quantum hardware capabilities and to build physical intuition for particle-physics processes, constitutes a central direction of current and future research. We further refer the reader to roadmap articles and general reviews of the field \cite{Bauer:2022hpo, DiMeglio:2023nsa, Fang:2024ple}.

\subsection{Basics of Quantum Simulation}
Quantum-simulation platforms can operate in analogue, digital, or hybrid modes, each offering distinct advantages. Analog simulators implement continuous time evolution via engineered Hamiltonians and can realize certain interactions, such as quantum link models for Abelian theories \cite{Chandrasekharan:1997,Wiese:2013,Banerjee:2012}. However, extending this approach to non-Abelian gauge theories with fermions in $3+1d$ poses significant challenges for simulating full SM dynamics. Hybrid digital–analog schemes combine native Hamiltonian evolution with programmable control and may provide near-term advantages on NISQ hardware. Nevertheless, fully digital, gate-based simulation—enabled by the universality of one- and two-qubit operations—offers the most systematic and scalable framework for addressing the complexity of SM physics. We therefore focus on digital quantum simulation.

\subsubsection{Digital quantum computation}
Quantum computing can be built upon two-level quantum bits or qubits. A qubit can encode information in a superposition of two basis states, written as
\begin{equation}
    \ket{\psi} = e^{i\gamma}\big(\cos\frac{\theta}{2}\ket{0}+ e^{i\phi} \sin\frac{\theta}{2} \ket{1}\big) \equiv \alpha \ket{0} + \beta \ket{1}
\end{equation}
where $\ket{0}$ and $\ket{1}$ form the computational basis. $\theta, \phi, \gamma$ are real numbers, and as there is no observable effects related to the factor $e^{i\gamma}$, we often neglect it. One natural question concerns how much information a qubit can store. Although a qubit state is specified by continuous parameters, such as the complex amplitudes $\alpha$ and $\beta$, a projective measurement in the computational basis collapses the state to either $\ket{0}$ or $\ket{1}$, which limits the accessible classical information from an individual qubit.

Quantum computing manipulates qubits with unitary quantum gates to realize the desired evolution of the quantum system. Of these, some of the most important are the Pauli matrices:
\begin{eqnarray}
    X\equiv\begin{pmatrix}
        0&1\\
        1&0
\end{pmatrix},\,\,
Y\equiv\begin{pmatrix}
        0&-i\\
        i&0
    \end{pmatrix},\,\,
    Z\equiv\begin{pmatrix}
        1&0\\
        0&-1
    \end{pmatrix}.
\end{eqnarray}
The Pauli matrices give rise to three useful classes of unitary matrices when exponentiated: the rotation operators about the $x$, $y$, and $z$ axes, defined by the equations:
\begin{eqnarray}
    &&R_x(\theta) \equiv e^{-i\frac{\theta}{2} X} =\cos\frac{\theta}{2} - i \sin\frac{\theta}{2} X = \begin{pmatrix}
        \cos\frac{\theta}{2} & -i \sin\frac{\theta}{2}\\
        -i\sin\frac{\theta}{2} &\cos\frac{\theta}{2}
    \end{pmatrix},\nonumber\\
    &&R_y(\theta) \equiv e^{-i \frac{\theta}{2} Y} =\cos\frac{\theta}{2} - i \sin\frac{\theta}{2} Y= \begin{pmatrix}
        \cos\frac{\theta}{2} & -\sin\frac{\theta}{2}\\
        \sin\frac{\theta}{2} &\cos\frac{\theta}{2}
    \end{pmatrix},\nonumber\\
    &&R_z(\theta) \equiv e^{-i \frac{\theta}{2} Z} =\cos\frac{\theta}{2} - i \sin\frac{\theta}{2} Z = \begin{pmatrix}
        e^{-i \theta/2} & 0 \\
        0 &e^{i \theta/2}
    \end{pmatrix}.
\end{eqnarray}
One-qubit gates represented by $2\times 2$ unitary matrices $U$ can generally be decomposed as 
\begin{equation}\scalebox{0.87}{$U = e^{i\alpha} \begin{pmatrix}e^{-i\beta/2}&0\\
    0& e^{i \beta/2}\end{pmatrix}\begin{pmatrix}\cos\frac{\gamma}{2}&-\sin\frac{\gamma}{2}\\
    \sin\frac{\gamma}{2}& \cos\frac{\gamma}{2}\end{pmatrix}
    \begin{pmatrix}e^{-i\delta/2}&0\\
    0& e^{i \delta/2}\end{pmatrix} \equiv e^{i\alpha} R_z(\beta) R_y(\gamma) R_z(\delta)$}
\end{equation}

Three other single-qubit gates that will play a large part in the following discussions for universal quantum computation are the Hadamard gate denoted as $H$, phase gate denoted as $S$ and $\pi/8$ gate denoted as $T$:
\begin{eqnarray}
    H= \frac{1}{\sqrt{2}}\begin{pmatrix}
        1&1\\
        1&-1
    \end{pmatrix},\,\,
    S = \begin{pmatrix}
        1&0\\
        0&i
    \end{pmatrix},\,\,
    T =e^{i\pi/8}\begin{pmatrix}
        e^{-i \pi/8}&0\\
        0&e^{i\pi/8}
    \end{pmatrix}.
\end{eqnarray}
with the relation: $H = e^{i\pi/2} R_z(\pi)\, R_y\Big(\frac{\pi}{2}\Big)$ and $S = T^2 = e^{i\pi/4}R_z(\frac{\pi}{2})$.

\begin{figure}
    \centering
\includegraphics[width=0.9\linewidth]{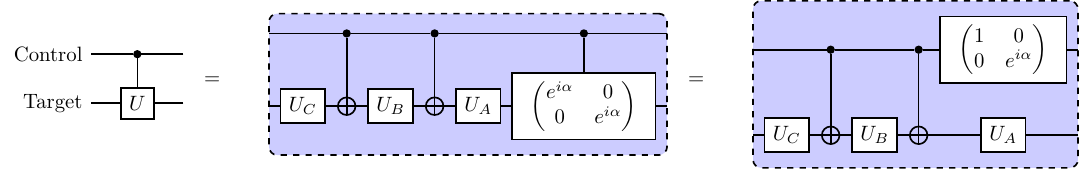}
    \caption{The construction of Controlled-$U$ gate using single-qubit gates and CNOT gates, where $U = U_AU_BU_C$ \cite{nielsen_chuang_2010}.}
    \label{fig:contrl-U}
\end{figure}

Two-qubit gates are operations that act on two qubits. Common ones include the Controlled-$X$ gate (CNOT), SWAP gate, Controlled-$Z$ (CZ) gate, and others. The Controlled-$U$ gate has one controlled qubit and a target qubit. If the control qubit is set then $U$ is applied to the target qubit, otherwise the target qubits are left alone; that is $\ket{c}\ket{t}\to\ket{c}U^c\ket{t}$. One important corollary to construct controlled unitary operations is that: for a unitary gate on a single qubit $U$, there exist unitary operators $U_A, U_B, U_C$ on a single qubit such that $U_A U_B U_C=I$ and $U= e^{i\alpha} U_AXU_BXU_C$. Using this corollary, a controlled-$U$ gate for an arbitrary single-qubit unitary $U$ can be implemented using a circuit composed solely of single-qubit gates and CNOT gates, as shown in \fig{contrl-U}.

With a similar decomposition applicable to multi-controlled-$U$ gates, any unitary operation on $n$ qubits can be implemented using a circuit containing only single-qubit and CNOT gates. The circuit depth is $\mathcal{O}(4^n)$, which grows exponentially with $n$; see the proof in \cite{nielsen_chuang_2010}. Nevertheless, it is sufficient to only approximate the given unitary operation. A set of gates is said to be universal quantum  gates if any unitary operation can be approximated to arbitrary accuracy by a quantum circuit involving only those gates. 
One universal gate set, denoted by $\mathfrak{U}_{\rm set}$, consists of the CNOT, Hadamard, and $T$ gates. Since the $S$ gate admits a more natural fault-tolerant implementation, including it directly in the universal set—rather than constructing it from $T$ gates—makes fault-tolerant quantum simulations more efficient.

Measurement provides the essential link between the abstract mathematical formalism of quantum states and experimentally accessible physical observables. In quantum mechanics, physical observables are represented by Hermitian operators, and extracting information from a quantum system corresponds to measuring these operators. The possible measurement outcomes are given by the eigenvalues of the corresponding Hermitian operator, with probabilities determined by the Born rule.

Because each measurement yields a single stochastic outcome, expectation values of observables must be estimated from repeated experimental trials, often referred to as shots. The statistical uncertainty of such estimates decreases inversely with the square root of the number of shots. Consequently, for quantum simulations to remain computationally efficient and scalable, the number of measurements required to achieve a desired precision must scale at most polynomially with the system size and the inverse target precision. Ensuring such polynomial scaling is essential for maintaining controlled statistical errors while preserving practical feasibility.

Quantum computers are highly susceptible to errors due to the delicate nature of quantum states, which are prone to decoherence and noise. Unlike classical systems, quantum errors are more challenging to detect and correct, particularly because quantum information cannot be directly copied (due to the no-cloning theorem), and direct measurement generally disturbs the encoded superposition.

The quantum threshold theorem, first established by Peter Shor and subsequently developed by Aharonov and Ben-Or, as well as Knill, Laflamme, and Zurek, provides a rigorous theoretical foundation for fault tolerance in quantum computing \cite{Shor:1996if,Aharonov:1996an,Knill:1996bj}. It states that there exists a critical error rate for quantum operations (such as quantum gates) below which quantum error correction can protect quantum information and enable large-scale quantum computations to be performed reliably. The threshold theorem is foundational for the practical realization of large-scale fault-tolerant quantum computing \cite{Preskill:1997zz}. Without a robust error correction scheme, quantum computers would not be able to perform useful computations for many applications, particularly complex algorithms such as Shor's factoring algorithm \cite{Shor:1997we}.

The threshold is not a single fixed value, but rather depends on the specific quantum error correction code, the architecture, and the noise model considered. However, theoretical and numerical studies have shown that, for many commonly studied schemes, the error threshold for fault-tolerant quantum computation typically lies in the range of approximately 
$10^{-2}$ to $10^{-3}$ (i.e., an error rate of 1\% to 0.1\% per gate) \cite{errorthreshold}. This implies that physical quantum systems must achieve error rates below this range to make scalable fault-tolerant quantum computing feasible. In fault-tolerant quantum computation, a universal set of logical gates must be implemented in a manner that suppresses error propagation, ensuring that reliable computation can be maintained even in the presence of noise.

Recent technical advances have brought us closer to realizing practical large-scale quantum simulators: engineered quantum many-particle systems that can controllably simulate complex quantum phenomena. Platforms being pursued include superconducting circuits, neutral atoms in optical lattices or tweezer arrays, trapped ions, superconducting radiofrequency cavities, and photonic networks (see \cite{Altman:2019vbv} for a review).

The distinctive features of these architectures—ranging from the all-to-all connectivity of trapped ions to the high-dimensional quantum systems of cavity platforms—together with advances in quantum error correction and mitigation techniques, may ultimately enable large-scale scientific applications in particle physics. Such progress could significantly advance our understanding of the dynamical properties of matter that remain intractable for classical computers.

\subsubsection{An illustrative example of quantum speedup}

Before exploring potential applications in particle physics, we provide an example based on the Discrete Fourier Transform (DFT), which underlies quantum algorithms that achieve exponential speedup over the best known classical algorithms. The classical Fast Fourier Transform (FFT) \cite{Cooley:1965zz} requires $\mathcal{O}(N \log N)$ operations, which for $N = 2^n$ data points corresponds to $\mathcal{O}(n 2^n)$ operations when expressed in terms of $n$.

The DFT takes a vector of $N$ complex numbers, $x_0, \dots,x_j,\dots, x_{N-1}$, and outputs a vector of complex numbers $y_0, \dots, y_k, \dots, y_{N-1}$, defined by
\begin{equation}
    y_k = \frac{1}{\sqrt{N}}\sum_{j=0}^{N-1} x_j e^{2\pi i j k/N}.
    \label{eq:FT}
\end{equation}

The DFT on the orthonormal basis $\ket{0}, \dots,\ket{j},\dots, \ket{N-1}$ is a linear operator acting on a basis state $\ket{j}$ as
\begin{equation}
    U_{\rm FFT} \ket{j} = \frac{1}{\sqrt{N}} \sum_{k=0}^{N-1} e^{2\pi i j k/N}\ket{k},
\end{equation}
and on an arbitrary state as
\begin{equation}
    U_{\rm FFT} \left(\sum_{j=0}^{N-1} x_j\ket{j} \right)= \sum_{k=0}^{N-1} y_k \ket{k},
\end{equation}
where the amplitudes $y_k$ are given by the DFT in Eq.~(\ref{eq:FT}).

For quantum algorithms to implement DFT, we represent the basis state $\ket{j}$ as the computational basis state $\ket{j} = \ket{j_1 j_2 \dots j_n}$ for $n$ qubits, with the identification
\begin{equation}
    j = j_1 2^{n-1} + \dots + j_n 2^0 .
\end{equation}

In this basis, the DFT can be implemented efficiently using quantum Fourier transform,
\begin{equation}
    U_{\rm QA} \ket{j}
    = \frac{1}{2^{n/2}}
    \bigotimes_{l=1}^{n}
    \left(
    \ket{0}
    + e^{2\pi i\, 0.j_l j_{l+1}\dots j_n}
    \ket{1}
    \right),
    \label{eq:QFT}
\end{equation}
where the binary fraction is defined as
\[
0.j_l j_{l+1}\dots j_n
=
\frac{j_l}{2}
+
\frac{j_{l+1}}{4}
+
\dots
+
\frac{j_n}{2^{n-l+1}} .
\]

With this decomposition, one can construct quantum circuits to implement the quantum Fourier transform using Hadamard gates and controlled phase-rotation gates. The controlled phase-rotation gate acting on qubit $k$ is given by
\[
R_k =
\begin{pmatrix}
1 & 0 \\
0 & e^{2\pi i / 2^k}
\end{pmatrix}.
\]
The full quantum Fourier transform circuit will require $\mathcal{O}(n^2)$ elementary gates, involving $\mathcal{O}(n)$ Hadamard gate and $\mathcal{O}(n^2)$ controlled phase-rotation gates.

As shown in \fig{contrl-U}, a controlled phase gate can be decomposed into two CNOT gates and single-qubit rotations. The latter can be approximated to accuracy $\epsilon$ using a number of gates from $\mathfrak{U}_{\rm set}$ that scales polylogarithmically with $1/\epsilon$, as guaranteed by the Solovay--Kitaev theorem \cite{Dawson:2005mz}. More efficient synthesis methods for single-qubit $Z$-rotations over the $\mathfrak{U}_{\rm set}$ set achieve the optimal scaling $\mathcal{O}(\log(1/\epsilon))$ \cite{Kliuchnikov:2012cm, Ross:2014zfa}.

Thus, the quantum Fourier transform can be implemented with $\mathcal{O}(n^2)$ operations, which is exponentially smaller than the $\mathcal{O}(n 2^n)$ cost of the classical FFT. It should be emphasized, however, that the quantum Fourier transform outputs a quantum state whose amplitudes encode the Fourier coefficients rather than a classical list of coefficients. Consequently, the exponential reduction in circuit complexity does not translate into an exponential speedup for extracting the complete Fourier spectrum as classical data.
In algorithmic contexts where quantum Fourier transform is used as a subroutine, this exponential separation in circuit complexity forms the basis of quantum speedup.

The quantum Fourier transform is a central subroutine in quantum algorithms for order finding and integer factorization, most notably Shor's algorithm \cite{shor1994algorithm}. It also plays an important role in quantum algorithms for simulating scalar field theories, as discussed in a subsequent section, and has recently been employed in gauge-theory simulations to transform between conjugate field variables, see \textit{e.g.} \cite{Buser:2020cvn, Li:2026ppp}. These applications illustrate how quantum Fourier transform can contribute to substantial quantum speedups for specific computational tasks that are intractable using currently known classical methods.

\subsection{Quantum algorithms for simulating dynamics in QFT}
Simulating QFT dynamics using quantum computers requires translating the continuum Lagrangian formalism into the discrete, operational language of qubits and quantum gates. One way to realize this translation is through the equal-time Hamiltonian formalism. 
This translation involves two essential layers of discretization. The first step is spatial discretization, where continuous space is replaced by a discrete lattice. For bosonic fields, a second layer of discretization is required. This involves truncating the infinite-dimensional local Hilbert space at each lattice site to a finite set of discrete values, which can then be encoded into a register of qubits. Within this digitized framework, quantum algorithms for computing dynamical observables generally follow three essential stages:
\begin{enumerate}
    \item \textbf{State preparation:} Construct the initial state $\ket{\psi(0)}$, for example a bound-state proton, encoded in terms of qubit-represented partonic degrees of freedom.  
    \item \textbf{Time evolution:} Simulate the system's dynamics by applying the unitary operator $\mathcal{U}(t) = e^{-iHt}$, where $H$ is the Hamiltonian governing the interactions among the qubits in the digitized representation, to obtain the evolved state $\ket{\psi(t)}$.  
    \item \textbf{Measurement:} Evaluate physical observables by performing measurements on the evolved state $\ket{\psi(t)}$.  
\end{enumerate}
The overarching objective of such simulations is to determine the system’s state at later times given an initial configuration, enabling the study of dynamics in QFT within a controlled computational framework. To illustrate the implementation of these steps within a quantum algorithm, we will consider a lattice $\Omega$ with the lattice site index $\bold{x} = (x_1, \cdots, x_d)$ for $0 \leq x_i \leq N-1$ labeling the sites of lattice in three spatial dimensions, while lattice of $d$ spatial dimensions are indexed with $d$ variables. 

\subsubsection{Hamiltonians and digitizations}
\label{sec:Ham-digitization}
\textbf{Scalar fields --- }
Quantum algorithms for simulating the $\phi^4$ theory were first investigated by Jordan, Lee, and Preskill in 2011 \cite{Jordan_2012}, and later examined in \cite{Klco_2019} to assess the effectiveness of different digitization methods, with a particular focus on noisy intermediate-scale quantum (NISQ) devices.
The lattice Hamiltonian for $\phi^4$ theory on lattice a $d$-dimensional $\Omega$ is defined as
\begin{equation}
    H_\phi = \sum_{\bold{x} \in \Omega} a^d \bigg[\frac{1}{2}\hat{\pi}(\bold{x})^2 - \frac{1}{2}\hat{\phi}(\bold{x})\nabla^2_a \hat{\phi}(\bold{x}) + \frac{1}{2}m^2_0\hat{\phi}(\bold{x})^2 +\frac{\lambda_0}{4!}\hat{\phi}(\bold{x})^4\bigg]
    \label{eq:scalarham}
\end{equation}
where $\nabla^2_a \hat{\phi}(\bold{x}) = \sum^d_{i=1}(\hat{\phi}(\bold{x} + a \vec{i}) + \hat{\phi}(\bold{x} - a \vec{i}) - 2 \hat{\phi}(\bold{x}))/a^2$ with $\vec{i}$ the unit vector in the $i^{\rm th}$ direction, denoting a discretized derivative. $\hat{\pi}(\bold{x})$ is the canonically conjugate momentum, satisfying the commutation relation 
\begin{equation}
    [\hat{\phi}(\bold{x}), \hat{\pi}(\bold{x})] = i a^{-d} \delta_{\bold{x, y}} \mathbb{I}.
\end{equation}
The field value $\phi(\bold{x})$, corresponding to the eigenvalue of the operator $\hat{\phi}(\bold{x})$, is an unbounded continuous variable.
One way to digitize the $\phi(\bold{x})$ field is to truncate the field value to a maximum magnitude $\phi_{\rm max}$ and discrete it in a separation of $\delta_\phi$. This digitization leads to $n_\phi = 2\phi_{\rm max/}\delta_\phi + 1$ field values taking values of $\phi_k = -\phi_{\rm max} + k \delta\phi$ for $k = 0, ..., n_\phi -1$.  $\phi_k$ can be mapped to the integer $\ket{k}$ state in the binary representation of the $n_q$ qubits on each lattice site $\bold{x}$ with $n_q = \log n_\phi$, $\hat{\phi}$ operator can be decomposed as $\hat{\phi} = \frac{\phi_{\rm max}}{n_\phi - 1} \sum^{n_q -1}_{i=0} 2^j \sigma^z_j$, with the relation $\hat{\phi} \ket{k} = \phi_k \ket{k}$.
Given the commutation relation in \eq{scalarham}, the eigenbasis of $a^d \hat{\pi}(\bold{x})$ is the Fourier transformation of the eigenbasis of $\hat{\phi}(\bold{x})$, which can be approximated by a finite-dimensional quantum Fourier transform after field digitization, with a total number of $\mathcal{O}(n^2_q)$ quantum gates. Other frequently discussed bases for field digitization include the Fock-state basis and the harmonic oscillator basis, with the latter consisting of a finite set of harmonic oscillator eigenstates. These bases often involve nonlocal interactions, which can require circuit depth that grows exponentially with the number of qubits $n_q$ to implement the unitary time-evolution operator \cite{Klco_2019, Ingoldby:2025bdb}. Recently, new methods have been developed to offer exponential improvements in circuit depth \cite{Cao:2026nyj} for the Fock-state basis. New ideas are still required to address the exponential scaling in large systems.

\textbf{Gauge fields --- }
The lattice Hamiltonian describing the dynamics of gauge fields was formulated by Kogut and Susskind in the 1970s~\cite{PhysRevD.11.395}. In the Hamiltonian (Kogut--Susskind) formulation, it can be written as
\begin{equation}
H^{\rm KS}_g
= \sum_{\bold{x}\in \Omega, i} \frac{g^2}{a}\,\mathrm{Tr}\,\hat{L}_i^2(\bold{x})
-\sum_{\bold{x}\in \Omega}\frac{2}{g^2 a}\,\mathrm{Re}\,\mathrm{Tr}\,\hat{P}_{ij}(\bold{x}),
\label{eq:gaugeham}
\end{equation}
where $a$ is the lattice spacing, and $g$ denote the gauge coupling constant. This Hamiltonian can also be obtained by taking the continuous-time limit of the Wilson lattice action~\cite{Creutz:1976ch}.

The first term in Eq.~\eqref{eq:gaugeham} is the electric Hamiltonian. It is diagonal in the representation (electric flux) basis, with eigenvalues determined by the quadratic Casimir operator $\hat{L}_i^2(\bold{x})$ acting on a link. For a non-Abelian gauge group $G = SU(N)$,
\begin{equation}
\hat{L}_i^2(\bold{x}) \ket{R a b}
= C_2(R)\,\ket{R a b},
\end{equation}
where $R$ labels an irreducible representation (irrep) of $SU(N)$ and $C_2(R)$ is its quadratic Casimir eigenvalue.

For $SU(2)$, irreducible representations are labeled by a single quantum number $j$, with
\[
\dim(j) = 2j+1,
\qquad
C_2(j) = j(j+1).
\]

For $SU(3)$, irreducible representations are labeled by two Dynkin indices $(p,q)$, with
\[
\dim(p,q) = \frac{(p+1)(q+1)(p+q+2)}{2},
\qquad
C_2(p,q) = \frac{p^2+q^2+pq+3p+3q}{3}.
\]

The second term in Eq.~\eqref{eq:gaugeham} is the magnetic Hamiltonian, constructed from the plaquette operator
\begin{equation}
\hat{P}_{ij}(\bold{x})
=
\hat{U}(\bold{x}, \bold{x} + a \vec{i})
\hat{U}(\bold{x} + a \vec{i}, \bold{x} + a \vec{i} + a \vec{j})
\hat{U}(\bold{x} + a \vec{i} + a \vec{j}, \bold{x} + a \vec{j})
\hat{U}(\bold{x} + a \vec{j}, \bold{x}),
\end{equation}
which represents the ordered product of link operators around an elementary plaquette in the $ij$-plane.

The link operator $\hat{U}$ is diagonal in the magnetic (group-element) basis $\ket{U}$, defined by
\[
\hat{U}\ket{U} = U \ket{U},
\]
where $U \in G$ is the group-valued link variable associated with the corresponding lattice link.

The magnetic basis $\{\ket{U}\}$ and the representation basis $\{\ket{R a b}\}$ are related through a generalized non-Abelian Fourier transform (Peter--Weyl decomposition),
\begin{equation}
\braket{U|R a b}
=
\sqrt{\frac{\dim(R)}{|G|}}
\, D^R_{ab}(U),
\end{equation}
where $D^R_{ab}(U)$ denotes the $(a,b)$ matrix element of the unitary representation matrix of the group element $U$ in the irreducible representation $R$. For compact continuous groups, $|G|$ is understood as normalization with respect to the Haar measure.

\begin{figure}[htbp]
  \centering
  \begin{subfigure}[b]{0.14\textwidth}
    \raisebox{1.20cm}{\includegraphics[width=\textwidth]{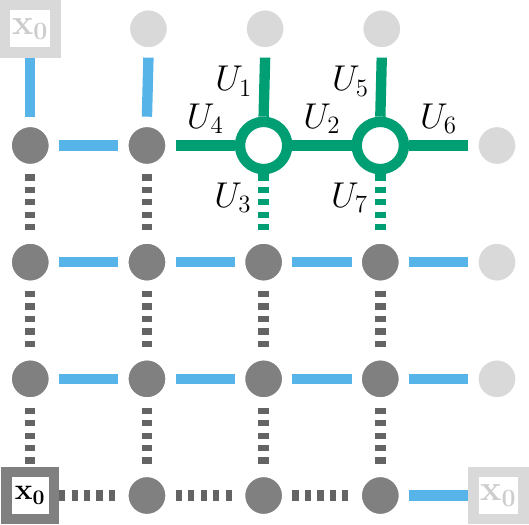}}\caption{}
  \end{subfigure}
  \hspace{0.4cm}
  \begin{subfigure}[b]{0.21\textwidth}
    \raisebox{0.8cm}{\includegraphics[width=\textwidth]{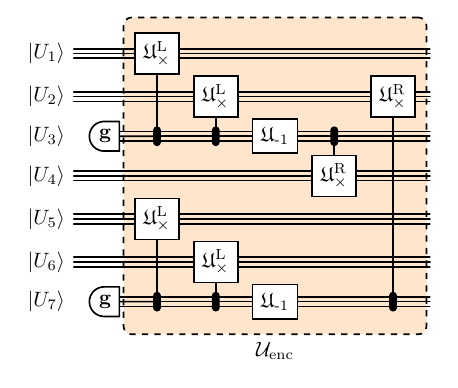}}\caption{}
  \end{subfigure}
  \begin{subfigure}[b]{0.3\textwidth}
    \includegraphics[width=\textwidth]{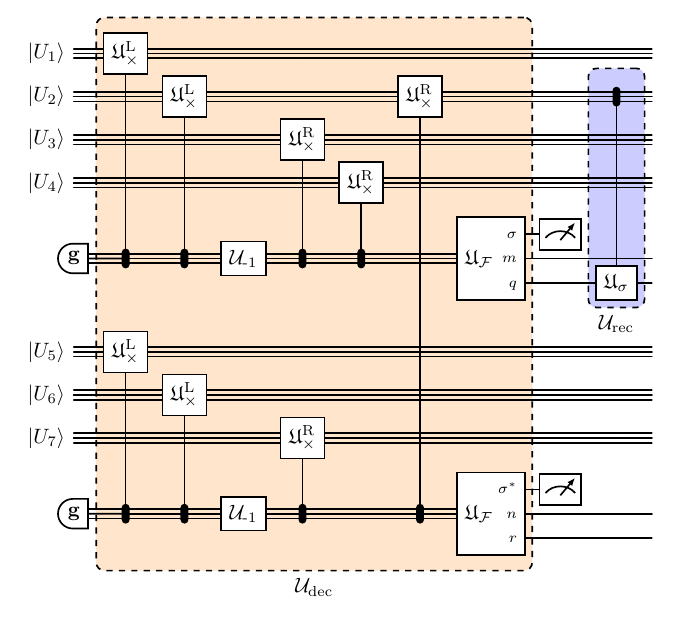}
    \caption{}
  \end{subfigure}
  \begin{subfigure}[b]{0.26\textwidth}
    \includegraphics[width=\textwidth]{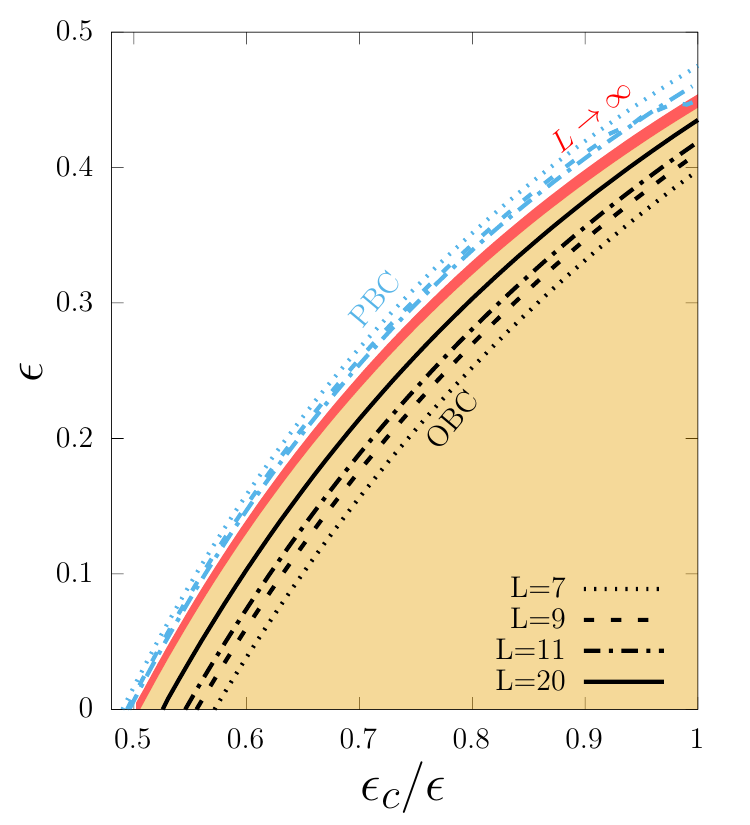}
    \caption{}
  \end{subfigure}
  \caption{(a) Maximal tree (dashed lattice links) of lattices for periodic boundary conditions. (b) Quantum circuits to impose Gauss's law from the redundant Hilbert space. (c) Gauss's law measurement and recovery circuit, where measuring a pair of charges $\sigma, \bar{\sigma}$ on the two ancillae indicates a gauge-violating error on $\ket {U_2}$. (d) Error threshold lines in the $\epsilon -\epsilon_c/\epsilon$ plane below which $F_{\mathrm {restored}}>F_{\mathrm{fixed}}$, indicating that using gauge redundancy to detect and correct the errors is advantageous, for 2D spatial lattices with local KL condition. The shaded region is the infinite volume limit. Adapted from Ref. \cite{Carena:2024dzu}, licensed under \textbf{CC BY 4.0}.}
  \label{fig:gauge-ec}
\end{figure}

The gluonic degrees of freedom are described by the Hilbert space 
$\mathcal{H}_{\rm full}$, defined as the tensor product over all lattice links of the corresponding single-link Hilbert spaces. The left panel of Fig.~\ref{fig:gauge-ec} illustrates this structure for a two-dimensional spatial lattice. For a continuous compact Lie group such as $SU(N)$, the Hilbert space associated with each link is infinite-dimensional; consequently, $\mathcal{H}_{\rm full}$ remains infinite-dimensional even on a finite lattice. 

The infinite dimensionality of each link Hilbert space for continuous gauge groups must be regularized for practical implementations. In the representation basis, this can be achieved by truncating to irreducible representations satisfying $R \le R_{\rm max}$, thereby rendering the link Hilbert space finite-dimensional. For a compact gauge group, the dimension of the truncated link space is
\begin{equation}
\sum_{R \le R_{\rm max}} \dim(R)^2,
\end{equation}
as follows from the Peter--Weyl decomposition. The predictions of the lattice gauge theory for the continuous group are then systematically recovered as $R_{\rm max}$ is increased. For example, convergence of the mass gap and of the vacuum expectation value of the Hermitian magnetic plaquette operator over a range of couplings has been demonstrated numerically for $SU(3)$ in Ref.~\cite{Ciavarella:2021nmj}.

The rate of convergence depends sensitively on the gauge coupling. In the strong-coupling regime, where the electric term in $H^{\rm KS}_g$ dominates, low-dimensional representations provide a good approximation and moderate values of $R_{\rm max}$ are typically sufficient. As the coupling $g$ decreases and the magnetic term becomes increasingly important, the ground-state wavefunction spreads over higher representations in representation space. Achieving a fixed target precision in this regime therefore requires progressively larger truncation thresholds. A systematic quantitative characterization of the scaling of observables with $R_{\rm max}$ and the gauge coupling $g$, particularly toward the weak-coupling regime, remains an open question.

Alternatively, one may work in a magnetic-field basis and truncate the continuous gauge variables to elements of a finite discrete subgroup. As the center of the gauge group plays a central role in the confinement mechanism of non-abelian gauge theories~\cite{Greensite:2011zz}, preserving the $\mathbb{Z}_N$ center is essential when constructing discrete subgroup truncations intended to reproduce the confinement properties of the continuous theory.

In the case of $SU(2)$, the binary tetrahedral (BT), binary octahedral (BO), and binary icosahedral (BI) groups are finite subgroups that contain the nontrivial central element and thus reproduce the $\mathbb{Z}_2$ center of the continuous theory. Moreover, owing to their high degree of symmetry, their elements are distributed over $S^3 \cong SU(2)$ in a manner that samples the group manifold more uniformly than those of binary dihedral subgroups $2D_n$ of comparable order.
For $SU(3)$, suitably chosen exceptional $\Sigma(n)$ subgroups such as $\Sigma(360)$, $\Sigma(648)$ and $\Sigma(1080)$ can likewise be constructed to contain the $\mathbb{Z}_3$ center. Although any finite subgroup provides only a discrete sampling of the eight-dimensional group manifold, these non-abelian subgroups offer a comparatively symmetric and structurally rich truncation that retains essential features of the full $SU(3)$ theory.

Such digitizations provide a rigid, finite approximation of the gauge degrees of freedom while preserving center symmetry and non-abelian structure, and they have enabled the development of dedicated quantum algorithms for truncated gauge theories~\cite{Gustafson:2022xdt, Gustafson:2023kvd, Lamm:2024jnl, Perez:2025cxl, Gustafson:2024kym}. However, because these subgroups are finite, establishing a systematic and controlled procedure by which observables converge to those of the continuous gauge theory remains an open and actively investigated problem.

The space $\mathcal{H}_{\rm full}$ contains gauge-equivalent states related by local gauge transformations and therefore includes unphysical redundancies. Strategies also needed to be pursued in quantum simulation to address this redundancy. The first is to explicitly solve Gauss’s law and restrict the theory to the gauge-invariant (physical) Hilbert space. While this eliminates redundant degrees of freedom at the outset, it generally leads to a more complicated form of the Hamiltonian in \eq{gaugeham} when expressed solely in terms of physical states. Concrete realizations of this approach include formulations in terms of plaquette variables \cite{Zohar:2013zla, Kaplan18_GaussLaw, Bender20_compactQED, Yamamoto:2020eqi, Bauer:2021gek, Gallimore:2022hai, Kane:2022ejm, PRXQuantum.2.030334}, states outside a maximal tree \cite{bauer2023new, Carena:2024dzu}, and local color-singlet multiplet bases \cite{Klco:2019evd, Ciavarella:2021nmj}.

The second strategy retains the full Hilbert space but enforces gauge invariance operationally, for instance by preparing gauge-invariant initial states and ensuring that time evolution preserves Gauss’s law. A commonly employed procedure is the maximal tree construction \cite{Gattringer:2010zz}, which fixes the gauge up to global transformations by explicitly solving the local Gauss’s law constraints. In this approach, the link variables along a chosen maximal tree (indicated by dashed lines in Fig.~\ref{fig:gauge-ec}(a)) are expressed as functions of the remaining links attached to each vertex. A quantum circuit implementing this construction in the magnetic basis for the preparation of gauge-invariant initial states was presented in Ref.~\cite{Carena:2024dzu} and is illustrated in Fig.~\ref{fig:gauge-ec}(b).

In contrast to scalar field theories, where systematic procedures exist for enlarging Hilbert-space truncations so as to approach the continuum limit while preserving structured and extensible circuit constructions, a comparably general and scalable framework has not yet been established for non-Abelian gauge theories such as $SU(N)$. This applies both to digitization in the electric (representation) basis—where increasing the truncation alters the representation content and circuit structure in ways that are not yet systematically characterized—and to discrete subgroup formulations in the magnetic basis, for which a controlled procedure for recovering the continuous gauge-group predictions remains to be developed. Establishing such extensible constructions, together with a quantitative understanding of their resource scaling, is essential for assessing the long-term feasibility of quantum simulations of lattice gauge theories.

\textbf{Fermion fields --- }
The lattice Hamiltonian for fermions can be written in terms of a one-component fermionic operator $\psi(\bold{x})$ at lattice site $\bold{x}$ as
\begin{equation}
    H_f = \sum_{\bold{x}\in \Omega, i} \bigg[\psi^\dagger(\bold{x})U(\bold{x}, i) \psi(\bold{x} + a \hat{i}) + h.c.\bigg] + m (-1)^\bold{x}\psi^\dagger (\bold{x})\psi(\bold{x})
    \label{eq:fermionham}
\end{equation}
where $(-1)^\bold{x} = (-1)^{\sum_i x_i}$.
This discretization follows from the staggered fermion formulation developed by Susskind in \cite{PhysRevD.16.3031} which removes partially the fermion doublings in $3+1$D theories by distributing the Dirac spin components over neighboring lattice sites. Other discretization methods for removing doublers include Wilson fermions \cite{wilson1977, Zache_2018}, which add a Wilson term that gives the doublers masses of order $1/a$. $\psi(\bold{x})$ satisfies the anti-commutation relation $\{\psi^\dagger(\bold{x}), \psi(\bold{x'})\} =\delta_{\bold{x, x'}}$. The local Hilbert space of a single fermionic mode is two-dimensional, expanded in the fermionic occupation basis $\ket{0}$ and $\ket{1}$ with the relation $\psi^\dagger(\bold{x}) \ket{0} = \ket{1}$. This has the same dimension of the Hilbert space of a spin-1/2 degree of freedom with basis $\ket{\downarrow}, \ket{\uparrow}$, described by Pauli operator $\sigma^i (\bold{x})$, which commute on different lattice sites.  
To map the fermionic system to a spin system in quantum simulations, efforts are needed to construct the anticommuting operator $\psi(\bold{x})$ out of the operator $\sigma^i (\bold{x})$ while preserving locality of the Hamiltonian. 

For a 1D spatial lattice, $\psi(\bold{x}\equiv n)$ can be built as chains of Pauli matrices using the Jordan-Wigner (JW) transformation \cite{Jordan:1928wi}
\begin{eqnarray}
    \psi(n) = \bigg(\underset{k<n}{\Pi} \sigma^z(k)\bigg)\sigma^+(n)\,\,, \psi^\dagger(n) = \bigg(\underset{k<n}{\Pi}
    -i\sigma^z(k)\bigg)\sigma^-(n).
\end{eqnarray} 
The mass term $\psi^+(n)\psi(n)$ transforms into local spin interaction $\sigma^-(n)\sigma^+(n)$. For open boundaries, the hopping term $\psi^\dagger(n)U(n, 1)\psi(n+1)$ maps to a local spin interaction $\sigma^-(n)U(n, 1)\sigma^+(n+1)$ under the JW transformation. 
In contrast, with periodic boundary conditions, the single hopping term that wraps around the lattice will map to a non-local spin interaction.
Furthermore, this non-locality is not restricted to boundary terms; in higher dimensions, the Jordan–Wigner transformation introduces non-local strings even in hopping terms within the bulk.
A major focus of generalizing the JW transformation to higher dimensions has been to maintain operator locality with a reduced size of the JW strings. Key strategies to achieve this include the use of auxiliary fermions \cite{Verstraete:2005pn}, symmetries \cite{po2021symmetric}, and multi-level computing systems \cite{carobene2024}.

Whereas JW transformation stores fermionic occupation numbers locally, necessitating non-local operator strings whose length scales as $\mathcal{O}(m)$ with the number of qubits $m$ in the chain to enforce fermionic anti-commutation relations, the Bravyi–Kitaev scheme \cite{Bravyi:2000vfj} instead distributes both occupation and parity information across qubits using a binary tree structure. This more sophisticated encoding allows both occupation and parity to be determined by accessing only $\mathcal{O}(\log m)$ qubits. Consequently, fermionic creation and annihilation operators transform into qubit operators with logarithmic—rather than linear—scaling in the number of local fermionic modes. This improved scaling directly reduces the circuit complexity for simulating fermionic Hamiltonians. It is also worthwhile to mention quantum devices built upon fermionic architectures. A universal gate set for such architectures was established by Bravyi and Kitaev \cite{Bravyi:2000vfj} and subsequent work has outlined promising routes for their experimental realizations \cite{OBrien:2018hki, Gonzalez-Cuadra:2023rex, Vilkelis:2023lbi}.

\subsubsection{Preparation of initial state}
\label{sec:state-prep}
The preparation of initial states is a central challenge in simulating real-time dynamics across diverse areas of particle physics. Depending on the physical process under study, these states can take various forms—ranging from localized particle wave packets in scattering processes, to bound states relevant for studying the dynamical properties of QCD, to thermal states characterizing the high-temperature environments of the early universe. Each class of initial state captures distinct aspects of underlying quantum field dynamics and may require specialized quantum algorithms for accurate and efficient preparation.

In quantum simulations of scattering phenomena, the relevant initial states consist of well-separated particle wave packets in the interaction picture. These are typically constructed from non-interacting wave packets and then evolved by adiabatically turning on the interactions, with the number of quantum gates for adiabatic evolution scales as $\mathcal{O}(1/\epsilon^{1+d/2})$ in terms of the state preparation error $\epsilon$ \cite{Jordan:2011ci} for scalar field theory. The non-interacting wave packets can usually be efficiently prepared. As an example, consider the preparation of non-interacting fermionic wave packets for scattering \cite{Chai:2023qpq}. For fermion field theories without interactions, we can define two sets of creation operators, $c^\dagger_k$ and $d^\dagger_k$ with momenta $k\in \frac{2\pi}{N}\times \bigg\{-[\frac{N}{4}], ..., [\frac{N}{4} -1]\bigg\}$, corresponding to creating particles and antiparticles, respectively. These operators are related to the original annihilation operator $\psi(n)$ and creation operators $\psi^\dagger(n)$ in position space by
\begin{eqnarray}
    c^\dagger_k &=& \frac{1}{\sqrt{N}}\sqrt{\frac{m+\omega_k}{\omega_k}} \sum_n e^{ikn}(\Pi_{n0} +v_k \Pi_{n1})\psi^\dagger(n)\notag\\
    d^\dagger_k &=& \frac{1}{\sqrt{N}}\sqrt{\frac{m+\omega_k}{\omega_k}} \sum_n e^{ikn}(\Pi_{n1} +v_k \Pi_{n0})\psi(n)
    \label{eq:p-basis}
\end{eqnarray}
where $v_k = \sin k/ (m+\omega_k)$,$\omega_k = \sqrt{m^2 + \sin^2 k}$, and $\Pi_{nl} = (1+(-1)^{n+1})/2$. Gaussian wave packets for particle and antiparticles with a mean momentum $\mu^{c(d)}_k$ and width $\sigma_k$, located around $\mu^{c(d)}_n$ in the position space can be created by the following operators:
\begin{eqnarray}
    C^\dagger(\phi^c) = \sum_k \phi^c_k c^\dagger_k, \,\,D^\dagger(\phi^d) = \sum_k \phi^d_k d^\dagger_k,
\end{eqnarray}
with $\phi^{c(d)}_k = \frac{1}{\sqrt{\mathcal{N}_k}}e^{-i k \mu^{c(d)}_n}e^{-(k - \mu^{c(d)}_k)^2/4\sigma^2_k}$ and $\mathcal{N}_k$ being the normalization factor. Using the relation in \eq{p-basis}, we can express the operator to create particle and antiparticle wave packet as:
\begin{eqnarray}
    C^\dagger(\phi^c) = \sum_n \tilde{\phi}^c_n \psi^\dagger(n), \, D^\dagger(\phi^d) = \sum_n \tilde{\phi}^d_n \psi(n)
\end{eqnarray}
The simulation of fermion--antifermion scattering involves preparing an initial state consisting of a pair of fermion and antifermion wave packets with appropriately chosen momenta, spatially separated by a finite distance. This state can be expressed as
\begin{equation}
    \ket{\Psi} = D^\dagger(\phi^c) C^\dagger(\phi^c) \ket{\Omega},
\end{equation}
where $\ket{\Omega}$ denotes the ground state of the non-interacting system. The wave packet creation operators $C^\dagger(\phi^c)$ and $D^\dagger(\phi^d)$, each expressed as a linear combination of fermionic operators, can be implemented in a quantum circuit with depth scaling linearly with the system
size $N$ using Givens rotations \cite{PhysRevA.92.062318}.

Yet, high-energy scattering of wave packets of bound states (such as hadrons and nuclei) is substantially more complex due to their composite structure of quarks and gluons governed by the nonperturbative dynamics of QCD. Algorithms have been developed to construct scattering states directly in the interacting theory based on Haag--Ruelle scattering theory \cite{Turco:2023rmx, Turco:2025jot}, as well as variational methods \cite{Liu:2021otn, Davoudi:2024wyv, Cao:2025shc}, where various ans\"atze are being investigated. When only observables associated with the asymptotic region are required, one may bypass explicit wave-packet preparation using methods such as the Lehmann--Symanzik--Zimmermann reduction formalism \cite{Li:2023kex, Briceno:2023xcm, Briceno:2020rar} to calculate scattering amplitudes, or the optical theorem \cite{Ciavarella:2020vqm} to extract decay rates.

Simulations of thermal and out-of-equilibrium dynamics—such as those relevant to the quark–gluon plasma in heavy-ion collisions or the early universe—require the preparation of Gibbs (thermal) states at finite temperature. These states describe systems in thermal equilibrium and serve as starting points for studying transport properties, phase transitions, and the nonequilibrium processes essential for explaining the origin of matter. Developing robust quantum methods for thermal-state preparation is thus a key step toward realistic simulations of QCD thermodynamics and cosmological dynamics. As a mixed state, thermal state preparation is non-trivial. There have been proposals such as quantum Metropolis sampling \cite{Temme:2009wa} which are somewhat formidable for near-term quantum simulators.
An increasing number of proposals based on hybrid quantum-classical variational methods have been developed to prepare purified thermal states \cite{Wu:2018nrn}, or a mixture of pure states with a classical probability distribution \cite{Martyn:2018wli}. Variational parameters are typically trained to minimize the free energy $F = E - T S$ of the system, where $E$ is the energy of the system, $T$ is the temperature at which the thermal state is prepared, and $S$ is the von Neumann entropy. Current methods for preparing thermal states as mixtures of pure states using variational quantum algorithms have been applied to study the $1+1$ dimensional Schwinger model at finite temperature \cite{Xie:2022jgj}, as well as the thermal states of the SYK model \cite{ Araz:2024xkw}, etc.

\subsubsection{Time evolution}
The time evolution of a quantum system over a duration $t$ is governed by the unitary operator
\[
U(t) = e^{-i H t},
\]
where we set $\hbar = 1$. Quantum algorithms seek to construct quantum circuits $\mathcal{U}(t)$ that approximate $U(t)$ efficiently with error $\epsilon_{\mathcal{U}} = |\mathcal{U}(t)- U(t)|$. Prominent approaches include product-formula methods, originally employed in the first explicit quantum algorithm for simulating $k$-local Hamiltonians by Lloyd \cite{Lloyd:1996aai}, as well as the linear combination of unitaries (LCU) method \cite{LCU} and quantum signal processing (QSP) \cite{Low_2017}. Among these, product-formula methods based on the Trotter--Suzuki decomposition are the most widely used in practical quantum simulations and are discussed in detail below.

Suppose the Hamiltonian can be decomposed into $N_l$ non-commuting terms,
\[
H = \sum_{l=1}^{N_l} H_l,
\]
then, at first order, the Trotter-Suzuki decomposition approximates the time evolution as
\[
\mathcal{U}(t) \approx \left( \prod_{l=0}^{N_l-1} e^{-i \delta t H_l} \right)^{N_t}, \quad N_t = t/\delta t,
\]
where each unitary $e^{i \delta t H_l}$ can be implemented efficiently by a quantum circuit over a universal gate set. Trotterization at higher orders $2p$ can be recursively defined as \cite{Suzuki:1991jtk}
\begin{eqnarray}
   \mathcal{U}_{p=1}(\delta t) &\equiv& (\prod^{N_l}_{l=1} e^{-i\frac{\delta t}{2} H_l}) (\prod^{1}_{l=N_l} e^{-i\frac{\delta t}{2} H_l}) \notag\\
   \mathcal{U}_p(\delta t) &\equiv& \bigg(\mathcal{U}_{p-1}(s_{p-1}\delta t)\bigg)^2 \mathcal{U}_{p-1}(\delta t-4 s_{p-1}\delta t) \bigg(\mathcal{U}_{p-1}(s_{p-1}\delta t)\bigg)^2
\end{eqnarray}
with $s_{p} = (4-4^{1/(2p+1)})^{-1}$, where the error $|\mathcal{U}_p(\delta t)- U(t)|$ scales as $\mathcal{O}(t \delta t^{2p})$ \cite{Childs_2021} which is polynomial in circuit depth. To reduce the quantum resources for reaching the continuous spacetime limit, renormalization techniques for the Trotter-Suzuki decomposition \cite{PhysRevD.104.094519} have recently been explored. These methods aim to suppress Trotter errors without significantly increasing the circuit depth when approaching the continuous spacetime limit. The key idea is to relate the Trotter step $\delta t$ to a physical temporal lattice spacing $a_t$, which can be determined via scale setting. By maintaining a fixed ratio between the spatial and temporal lattice spacings, predictions in the continuous spacetime limit can be systematically approached while controlling both Trotter errors and circuit depth. Such approaches provide a practical route to balance simulation accuracy and resource efficiency in quantum algorithms for Hamiltonian dynamics. Alternative approaches to implementing time evolution, such as quantum signal processing, can achieve polylogarithmic scaling with respect to the target accuracy. This scaling allows more efficient control of $\epsilon_{\mathcal{U}}$ and may be adopted in approaching the continuum limit within the recently proposed statistically bounded time evolution (SBTE) framework \cite{Kane:2025ybw}.

\subsubsection{Measurements} 
In quantum simulations of QFT, measurement typically represents the final stage of computation, providing the means to extract the desired physical observables. Physical quantities are represented by \emph{Hermitian operators} (observables), each of which naturally defines a corresponding projective measurement, with outcomes given by the operator’s eigenvalues. The choice of measurement basis—such as the Pauli-$X$, $Y$, or $Z$ basis—determines which aspect of the quantum state is revealed, thereby linking the abstract simulation to experimentally accessible quantities.

Quantum state tomography, the complete reconstruction of an unknown quantum state $\rho$, is infamously resource-intensive, requiring a number of measurements that scales exponentially with the number of qubits for generic quantum states. This fundamental bottleneck renders it impractical for characterizing large-scale quantum systems. An efficient method called a ``classical shadow'' \cite{shadow-measurement} has emerged as a powerful framework to predict $M$ different properties like quantum fidelities, entanglement entropies and two-point correlation functions with high success probability, requiring only $\mathcal{O}(\log M)$ measurements, a number independent of the system size. The technique involves repeatedly preparing the quantum state, measuring it in a randomly selected basis, and then processing the resulting classical outcomes—the ``classical shadows''—using classical algorithms. 

Subtleties can arise when attempting to measure the expectation value of an operator that is not an observable in the standard sense, due to its non-Hermitian nature. A prime example is the $n$-time correlator $\mathcal{O} = O_n(t_n)\cdots O_1(t_1) O_0(t_0)$, where each $O_j(t_j)$ can be a Hermitian operator. Such correlators are crucial for extracting observables that characterize the dynamics of QCD \cite{Lamm:2019uyc, Li:2021kcs, Lee:2024jnt}. To access the generally complex quantity $\bra{f} \mathcal{O} \ket{f}$, one may employ techniques such as the Hadamard test \cite{Lin:2022vrd}, or ancilla-free approaches based on combined real-time and quantum imaginary-time evolution \cite{Wang:2025ojn}. 

To implement the Hadamard test, the operator $\mathcal{O}$ is embedded into a unitary operator $U_{\mathcal{O}}$ via methods such as block encoding \cite{Camps:2022qce, Camps:2024siam, Gilyen:2019stoc, Kane:2024odt} or the linear combination of unitaries (LCU) approach \cite{Childs:2012aa}. In block encoding, $\mathcal{O}/\|\mathcal{O}\|$ is realized as a sub-block of a larger unitary acting on an extended Hilbert space that includes an ancilla qubit $\ket{q_a}$. Concretely, one constructs $U_{\mathcal{O}}$ as: 
\begin{eqnarray}
U_\mathcal{O} =
\begin{pmatrix}
\frac{\mathcal{O}}{\|\mathcal{O}\|} & -\sqrt{1 - \frac{\mathcal{O}^2}{\|\mathcal{O}\|^2}} \\
\sqrt{1 - \frac{\mathcal{O}^2}{\|\mathcal{O}\|^2}} & \frac{\mathcal{O}}{\|\mathcal{O}\|}
\end{pmatrix}.
\end{eqnarray}
When the ancilla is initialized in $\ket{q_a} = \ket{0}$, a post-selection on the outcome $\ket{0}$ effectively applies $\mathcal{O}/\|\mathcal{O}\|$ to the state $\ket{f}$.

A second ancilla qubit is then introduced to perform the Hadamard test. The probability of measuring this ancilla in the state $\ket{0}$ is
\begin{eqnarray}
P(0) = \frac{1 + \mathrm{Re}\left[\bra{q_a}\bra{f} U_{\mathcal{O}} \ket{f}\ket{q_a}\right]}{2}.
\end{eqnarray}
For $\ket{q_a} = \ket{0}$, this reduces to
\begin{eqnarray}
P(0) = \frac{1 + \mathrm{Re}\left[\bra{f} \mathcal{O} \ket{f} / \|\mathcal{O}\|\right]}{2}.
\end{eqnarray}
The imaginary part can be accessed by inserting an $S^\dagger$ gate on the ancilla prior to the final Hadamard gate.

\subsubsection{Quantum Error Correction and Mitigation (QEC and QEM)} 
Among the most widely used QEC schemes are stabilizer codes, which exploit symmetry constraints to detect and correct errors that drive quantum states outside the subspace defined by those symmetries. In the current NISQ (Noisy Intermediate-Scale Quantum) era, however, stabilizer codes require an enlarged, redundant Hilbert space that extends beyond the symmetry-constrained subspace. 

Following the logic underlying stabilizer codes, the redundancies inherent in gauge theories---when retained in the full Hilbert space---can similarly be leveraged to correct errors. In this framework, quantum simulations begin with a gauge-invariant state prepared within the full Hilbert space by enforcing Gauss’s law as a local consistency condition at each lattice site, as we have discussed in Sec. \ref{sec:Ham-digitization}. Since the Hamiltonian is gauge invariant, the time-evolution preserves the gauge symmetry and therefore should respect Gauss's law in the absence of quantum errors. Any observed violations of this constraint indicate the presence of errors.

Detecting and correcting such gauge-violating errors is particularly challenging for non-Abelian gauge theories. Checking Gauss’s law, \emph{i.e.}, computing the net flux at a given lattice site, involves evaluating the group’s Clebsch--Gordan coefficients, which are not diagonal in the electric basis for non-Abelian groups. However, this difficulty can be circumvented by performing an ``effective Clebsch--Gordan sum'' in the group-element basis, as demonstrated in~\cite{Carena:2024dzu}. 

This insight enables the construction of the quantum circuits ~\cite{Carena:2024dzu} shown in \fig{gauge-ec}(c), which are designed to detect quantum errors that break gauge symmetry and to correct those that satisfy the necessary and sufficient conditions for error correction, as specified by the Knill--Laflamme criteria. It is further shown that, once the hardware error rate falls below a certain threshold~$\epsilon$ \fig{gauge-ec}(d), this procedure yields higher fidelity compared to approaches that simply eliminate gauge redundancies from the Hilbert space.

On the other hand, QEM techniques (for a survey of recently developed methods, see \cite{Cai_2023}), such as Pauli twirling \cite{PhysRevA.94.052325} and zero-noise extrapolation (ZNE) \cite{Temme:2016vkz, Li:2016vmf}, are commonly employed to benchmark physics problems on NISQ devices.
Pauli Twirling mitigates the coherent accumulation of errors across multiple gates, such as CNOTs, by randomly surrounding each gate with Pauli operators $\{\mathbb{I}, X, Y, Z\}$ that satisfy
\begin{equation} \label{eq:paulicomp} 
\scalebox{0.95}[1]{$\left(\prod_{i}(\sigma^{b_i}_i)^{\otimes}\right)\text{CNOT}\otimes\mathbb{I} \left(\prod_{i} (\sigma^{a_i}_i)^{\otimes}\right) = \text{CNOT}\otimes\mathbb{I}$}, 
\end{equation}
where the $i$-th qubit is rotated by $\sigma_i^{a_i}$ before the CNOT and by $\sigma_i^{b_i}$ afterward. Although the number of independent circuits is enormous, $\mathcal{O}(10)$ randomly sampled circuits have been found to be sufficient for error mitigation~\cite{Erhard_2019}.

ZNE employs the strategy of amplifying hardware noise by a known factor $\mathcal{N}_s$ and extrapolating to the zero-noise limit, $\mathcal{N}_s = 0$. This technique has been widely adopted and shown to be effective. Noise amplification can typically be achieved through several strategies, such as Pulse stretching, Probabilistic noise application and Subcircuit repetition. Pulse stretching increases the duration of quantum gates, thereby amplifying decoherence and control errors~\cite{Temme:2016vkz, Kandala:2018kwe}. Probabilistic noise application amplifies the effect of the actual noise channel by applying it multiple times or with specified probabilities. For example, IBM~\cite{IBM-100} applied a learned noise model to a 127-qubit system with 60-CNOT layers, enabling noise amplification and allowing ZNE to produce estimates of the average magnetization—demonstrating reliable results for circuits beyond classical simulation capabilities.

Subcircuit repetition inserts noisy, identity-equivalent operations into the circuit, thereby increasing exposure to noise~\cite{Dumitrescu:2018njn, He:2020udd}. To be more specific, consider using the subcircuit repetition method to mitigate errors in a quantum circuit containing operations $U$. The noise level $\mathcal{N}_s$ for the original circuit is defined as $\mathcal{N}_s = 1$, whereas replacing $U$ with $U (U U^\dagger)^n$ increases the effective noise level to $\mathcal{N}_s = 1 + 2n$.
To amplify circuits with $\mathcal{N}_s < 3$, a random subset of $U$ gates in the original circuit can be chosen to be replaced. By performing simulations at different $\mathcal{N}_s$, the simulation results in the zero error limit ($\mathcal{N}_s = 0$) can be reached from extrapolations. Using this method, the CNOT-gate noise in preparing quantum states for a 20-qubit system with a 38-CNOT circuit depth on the \texttt{Zuchongzhi} chip is shown to be successfully mitigated \cite{Cao:2025shc}. With access to noise models native to the \texttt{Zuchongzhi} chip, even more specialized ZNE procedures can be carried out, enabling reliable results for larger systems.

\subsection{Benchmarks}
Leveraging the ingredients we have discussed, we illustrate two benchmark studies that exemplify state-of-the-art proof-of-principle simulations. One study investigates parton evolution effects within effective field theory \cite{Bauer_2021}, while the other examines particle–bubble collisions pertinent to baryon asymmetry in the early universe, targeting observables associated with symmetry violation \cite{Carena:2024peb}. Both cases employ simplified toy models.
\subsubsection{Fragmentation at colliders}
As a benchmark for testing algorithms, a massless scalar field theory in $1+1d$ is employed to investigate the showering processes of energetic particles \cite{Bauer_2021} that couples to the massless scalar through a Yukawa coupling $g$. Within the soft-collinear effective theory (SCET) \cite{scet} framework, the energetic state can be represented as a static object (Wilson line) moving along a light-like direction, since the soft scalar radiations do not change its direction. The Wilson line is defined as
\begin{equation}
Y_n = P \exp\bigg[i g \int^\infty_0 ds \,\hat{\phi}(x^\mu = n^\mu s)\bigg],
\end{equation}
where $ n^\mu = (1, \mathbf{n})$ denotes the light-like direction. Accordingly, the low-energy Hamiltonian is given solely by the scalar field theory in \eq{scalarham} for $d=1$, with $m_0 = 0$ and $\lambda_0 = 0$.

Considering soft radiation in a process involving a pair of energetic particle-antiparticle moving back-to-back, two Wilson lines on the lattice can be defined: $Y_n$ for the particle moving in the $n$ direction and $Y_{\bar{n}}^\dagger$ for the antiparticle moving in the opposite direction:
\begin{eqnarray}
Y_n &=& P \exp\bigg[i g \delta x \sum^{2n_0}{i=n_0} \hat{\phi}(t = x_i - n_0)\bigg], \notag\\
Y_{\bar{n}}^\dagger &=& P \exp\bigg[-i g \delta x \sum^{n_0}_{i=0} \hat{\phi}(t = n_0 - x_i)\bigg],
\end{eqnarray}
where $n_0 = (N-1)/2$ is the center of the lattice.

With this setup, one can compute the transition from the ground state $\ket{\Omega}$ to a final state $\ket{X}$ containing soft scalar radiations induced by the evolution of energetic particles
\begin{equation}
\bra{X} T[Y_n Y_{\bar{n}}^\dagger] \ket{\Omega},
\end{equation}
where $T$ denotes the time ordered product of the two Wilson lines. This formulation is particularly relevant in the regime of large coupling constant $g$ or when the final state $\ket{X}$ involves a large number of particles as in parton showers, where perturbative methods may fail.

\begin{figure}
    \centering
\includegraphics[width=0.5\linewidth]{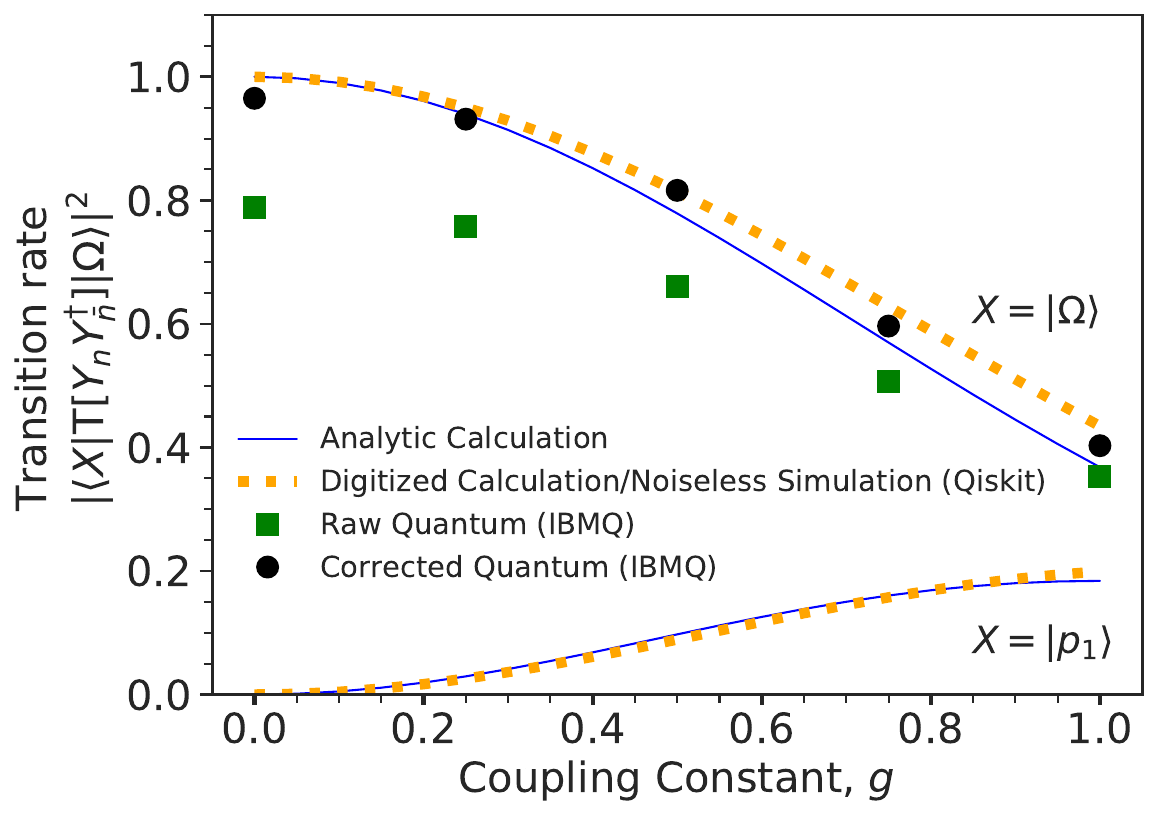}
    \caption{The transition rates from the vacuum to the lowest-lying single excited state induced by a pair of energetic particle-antiparticle moving back-to-back, for a system of three lattice sites and $n_q= 2$ qubits per site. Adapted from Ref. \cite{Bauer_2021}, licensed under \textbf{CC BY 4.0}.}
    \label{fig:SCET}
\end{figure}

For numerical demonstration in the NISQ era, a lattice with $N = 3$ sites, $\delta x = 1$, and $n_q = 2$ qubits per site is used. The transition rate from the vacuum to the lowest-lying single particle state (with nonzero lattice momentum) is computed on both a noiseless quantum simulator and the 65-qubit IBM Manhattan quantum computer without quantum error correction. The results are shown in \fig{SCET}, taken from \cite{Bauer_2021}.

\subsubsection{Out-of-equilibrium}

\begin{figure}[htbp]
  \centering
  \begin{subfigure}[b]{0.44\textwidth}
    \raisebox{0.10cm}{\includegraphics[width=\textwidth]{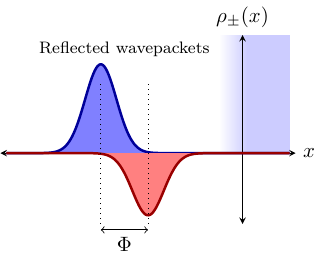}}
    \caption{}
    \label{subfig:asym-phase} 
  \end{subfigure}
  \hspace{0.4cm}
  \begin{subfigure}[b]{0.41\textwidth}
    \raisebox{0.10cm}{\includegraphics[width=\textwidth]{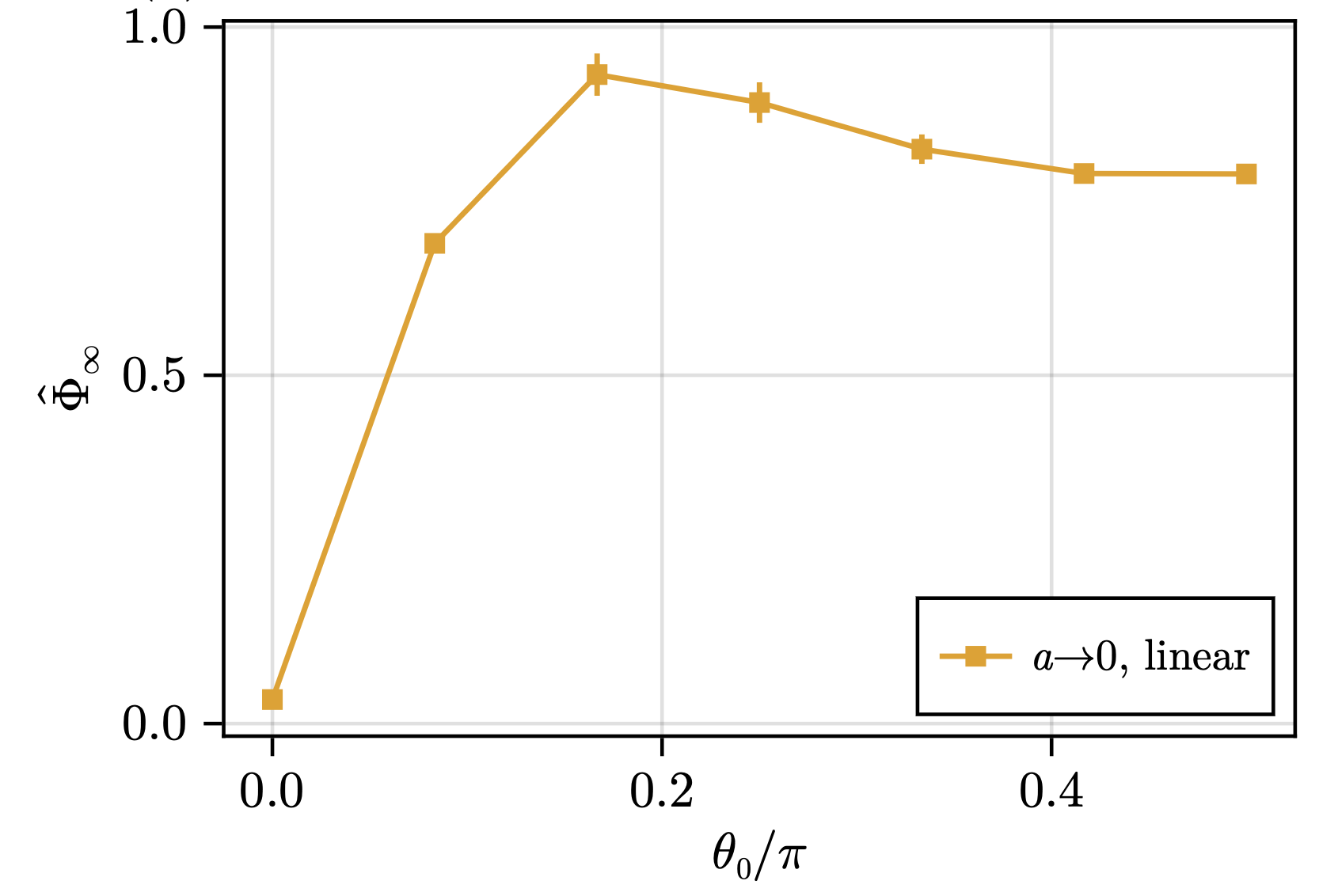}}
    \caption{}
    \label{subfig:PhaseAsymmetry}
  \end{subfigure}
  \caption{(a) The phase asymmetry $\Phi$ between the reflected conjugate pairs, which is defined as the distance between the peaks of the reflected Gaussian wave packets of the conjugate pair (blue for particle wave packet, red for anti-particle wave packet). $\rho_{\pm}(x)$ is the local charge densities for the wave packets. The magnitude asymmetry is the total charge difference in the symmetric phase between the conjugat pair. (b) The simulation results obtained using tensor network method for the phase asymmetry observable $\hat{\Phi}_\infty$ —the lattice counterpart of $\Phi$— are shown in the asymptotic regime $(t \to \infty)$, once the reflection process is complete. Adapted from Ref. \cite{Carena:2024peb}, licensed under \textbf{CC BY 4.0}.}
  \label{fig:asymmetry}
\end{figure}
In the early universe, out-of-equilibrium dynamics can arise when the cosmos undergoes a potential first-order phase transition (FOPT), a process relevant to the generation of baryon asymmetry, hadronic dynamics, and related phenomena. In the scenario of electroweak baryogenesis (EWBG), bubbles of the broken phase can nucleate during a FOPT, and dynamical processes—such as collisions between particle and the bubble—can generate chiral asymmetries. To model these dynamics, a $1+1d$ toy framework has been employed in \cite{Carena:2024peb}, as a prototype of particle-bubble scattering to generate chiral asymmetry necessary for baryon asymmetry generation.

In the setup of \cite{Carena:2024peb}, a Dirac fermion $\psi$ is coupled to a complex scalar field that undergoes FOPT, during which bubbles can form. Outside the bubble, the scalar field has a vanishing vacuum expectation value (vev) (phase $s$), while inside the bubble it acquires a non-vanishing vev (phase $b$). In the rest frame of the bubble wall, the effect of vev profile on the fermion can be encoded in a complex mass term
\begin{equation}
m(x) = |m(x)| e^{i \theta(x) \gamma^5},
\end{equation}
where $|m(x)| = 0$ outside the bubble and $|m(x)| \neq 0$ inside. The spatial dependence of the phase $\theta(x)$ explicitly breaks chiral symmetry. The Hamiltonian density in the continuous spacetime for the fermion is given by:
\begin{eqnarray}
    \mathcal{H} &
= -i\Bar{\psi}\gamma^i\partial_i\psi
    +|m(x)|\bar{\psi}\left[\cos\theta(x)+ i\sin\theta(x)\gamma^5\right] \psi,
    \label{eq:hcon}
\end{eqnarray}
with the index $i$ running over the spatial dimensions. The mapping of fermions onto qubits, once the system is discretized on a lattice, can be performed in the same manner as for the lattice fermionic Hamiltonian in \eq{fermionham}.

The scattering process relevant for generating chiral asymmetry near the bubble wall is as follows: a fermion (or its charge-conjugate antifermion) located outside the bubble moves toward the bubble wall and scatters, either being reflected or transmitted. To simulate this process, a non-interacting wave packet of massless Weyl fermion $\ket{\Psi(0)}_\pm$ with a central momentum $k$ towards the bubble needs to be prepared following the techniques of wave packet preparation in \Sec{state-prep}. Here $\pm$ refers to the conjugate pairs, \textit{i.e.} $+$ is for fermion, and $-$ for anti-fermion.

The quantum simulation framework for this process leverages established techniques—including staggered fermion discretization, Jordan-Wigner encoding, and tailored wave packet initialization for massless fermions—followed by the time evolution under the mass-varying Hamiltonian. To characterize symmetry-violating signatures in the scattering dynamics, one can measure comparative metrics such as the phase asymmetry $\Phi$ between reflected conjugate pairs and the magnitude asymmetry. These quantities are illustrated in \fig{asymmetry}(a), along with results obtained by tensor network methods, which are further extrapolated to the continuum lattice in \fig{asymmetry}(b) from \cite{Carena:2024peb}.

\subsection{Outlook}
The prospects for quantum simulation of QFTs in HEP are viewed with long-term optimism, as progress critically depends on synergistic advances in both quantum hardware and quantum algorithms. Although the development of fault-tolerant quantum computers remains a long-term objective, substantial breakthroughs are still required in theoretical studies of the computational framework and quantum algorithmic methods to ultimately enable the solving of HEP problems with quantum computers. At the same time, classical simulation techniques—especially those employing tensor networks, see \textit{e.g.} Refs.\cite{Orus:2019, Carena:2024peb, Milsted:2020jmf, Magnifico:2024eiy,
Barata:2025rjb, Barata:2026aoq} and quantum-inspired algorithms—serve a dual purpose: they not only deliver immediate physical understanding and benchmarking capabilities for small-scale systems but also help identify inherent limitations in simulating real-time dynamics and high-entropy regimes of interacting field theories. These combined efforts will clarify precisely which problems will benefit most from future quantum computational advantages. Moreover, the algorithmic toolkit emerging from quantum simulation research—including variational quantum algorithms and Hamiltonian learning techniques—strongly intersects with quantum machine learning (QML). These QML approaches are already being investigated for HEP data analysis tasks \cite{Fang:2024ple}, such as event classification, anomaly detection, and pattern recognition in complex detector data, offering potential improvements in data processing even on early-stage, non-fault-tolerant quantum devices.

%% file: sections/quantum-cavity.tex
\section{Cavity and Circuit-Based Searches for Ultralight Bosons and High-Frequency Gravitational Waves}
\label{sec:cavity}

Electromagnetic (EM) detectors, such as resonant cavities and circuits, have been widely employed and proposed to probe ultralight dark matter (DM) candidates, including axions~\cite{Preskill:1982cy, Abbott:1982af, Dine:1982ah} and dark photons~\cite{Abel:2008ai,Goodsell:2009xc,Nelson:2011sf}, as well as high-frequency gravitational waves (HFGWs)~\cite{Aggarwal:2020olq,Aggarwal:2025noe}. These detection platforms are at the forefront of experimental efforts to explore physics beyond the Standard Model.

This section reviews the fundamental principles, recent advancements, and future prospects in these directions. We begin by outlining the detection principles underlying EM detectors. All three sources under consideration, axions, dark photons, and HFGWs, can generate effective electric currents through their interactions with EM photons. These effective currents produce measurable EM signals. Such signals can be detected either within resonant cavities~\cite{Sikivie:1983ip, Sikivie:1985yu,Berlin:2021txa} or in shielded environments where circuits register the induced magnetic flux~\cite{Sikivie:2013laa,Chaudhuri:2014dla,Domcke:2022rgu}. For axion and HFGW detection, the presence of an external EM field is essential, which can be either static~\cite{Sikivie:1983ip,Sikivie:1985yu,Sikivie:2013laa,Berlin:2021txa} or oscillatory~\cite{Goryachev:2018vjt,Berlin:2019ahk,Berlin:2023grv}.

EM detectors can operate in either a resonant~\cite{Sikivie:1983ip,Sikivie:1985yu,Sikivie:2013laa} or broadband configuration~\cite{Kahn:2016aff,Berlin:2020vrk,Bourhill:2022alm}. Resonant detectors typically require frequency tuning across multiple steps to scan a target frequency range. We review the figure-of-merit for the scan rate and discuss avenues for improvement, including increasing cavity volume, enhancing the quality factor, and utilizing stronger magnetic fields. In single-mode resonant detectors, the scan rate and noise level are often limited by the standard quantum limit (SQL)~\cite{Chaudhuri:2018rqn}. However, various techniques, both demonstrated and proposed, aim to surpass this fundamental limit.

Most EM detection strategies target the standard halo model for DM, where the signal bandwidth is approximately $10^{-6}$ of the central frequency, dictated by the virialization of DM in the Galactic halo~\cite{Foster:2017hbq}. However, potential signals from an ultralight boson fields or HFGWs may exhibit deviations from this description, such as broader spectra, anisotropies, or macroscopic polarizations. To extract such information, advanced analysis techniques are essential. These include exploiting the Earth’s rotation to observe diurnal signal modulations and employing detector networks to analyze correlations across multiple detectors~\cite{Foster:2020fln,Chen:2021bdr,SHANHE:2024tpr}.

This section is organized as follows. We begin with the effective currents induced by axions, dark photons, and HFGWs in Sec.~\ref{sec:effective_current}, followed by their detection with resonant systems (cavities and LC circuits) in Sec.~\ref{sec:resonant_systems}. Heterodyne upconversion is discussed in Sec.~\ref{sec:heterodyne}, including EM and mechanical couplings of HFGWs, as well as cavity-based light-shining-through-wall searches in Sec.~\ref{sec:lsw}. We then present the sensitivity framework in Sec.~\ref{sec:sensitivity}, together with the main experimental ingredients: cavity volume (Sec.~\ref{sec:volume}), quality factor (Sec.~\ref{sec:quality}), and external magnetic field (Sec.~\ref{sec:bfield}). Quantum-enhanced strategies beyond the SQL are reviewed in Sec.~\ref{sec:beyond_sql}, including squeezing (Sec.~\ref{sec:squeezing}), multi-mode systems (Sec.~\ref{sec:multimode}), and quantum non-demolition (QND) readout (Sec.~\ref{sec:qnd}). Finally, we discuss the characterization of wave-like fields in Sec.~\ref{sec:characterization}, focusing on detector networks (Sec.~\ref{sec:network}) and diurnal modulation (Sec.~\ref{sec:diurnal}), before concluding in Sec.~\ref{sec:EMoutlook}.

\subsection{Electromagnetic Detectors for Axion, Dark Photon, and High-Frequency Gravitational Waves}

\subsubsection{Effective Currents Induced by Wave-like Fields}\label{sec:effective_current}

Wave-like fields, such as ultralight axions, dark photons, and HFGWs, induce effective electric currents in resonant systems through their interactions with EM photons. These effective currents act as sources that excite EM signals, which can be amplified by resonant detectors specifically designed for this purpose. In the following, we derive the expressions for these effective currents in each case, starting from their respective Lagrangians.

\paragraph{Axion}
The axion is a well-motivated particle beyond the Standard Model, originally proposed to explain the absence of a neutron electric dipole moment and thereby providing an elegant solution to the strong CP problem~\cite{Preskill:1982cy,Abbott:1982af,Dine:1982ah}. In addition, theories with extra dimensions, such as string theory, predict a landscape of axion-like particles with a broad range of masses and couplings~\cite{Svrcek:2006yi, Arvanitaki:2009fg}. Among the various portals connecting axions to the Standard Model, the coupling to the EM field has received particular experimental focus due to its straightforward laboratory accessibility, and it is further supported by strong theoretical motivation. For example, this interaction arises naturally through mixing with the neutral pion.

The Lagrangian describing the axion field $a$ and its interaction with the EM field is given by:
\begin{equation}
\mathcal{L}_a = \frac{1}{2} \partial_\mu a \partial^\mu a - \frac{1}{2} m_a^2 a^2 - \frac{g_{a\gamma}}{4} a F_{\mu\nu} \tilde{F}^{\mu\nu},
\label{eq:La}
\end{equation}
where $F_{\mu\nu}$ is the EM field tensor, $\tilde{F}^{\mu\nu}$ is its dual, $g_{a\gamma}$ is the axion-photon coupling constant, and $m_a$ is the axion mass. 

Performing an integration by parts on the interaction term in Eq.~(\ref{eq:La}) leads to a form proportional to $A_\mu J^\mu_{\rm eff}$, where $A_\mu$ represents the signal EM field and $J^\mu_{\rm eff}$ is the effective current induced by the axion field in the presence of an EM field $F_0^{\mu \nu}$. The effective current is given by:
\begin{equation}
J_{\text{eff}}^{a\,\mu} = g_{a\gamma} \tilde{F}_0^{\mu \nu} \partial_{\nu} a.
\label{eq:aJ}
\end{equation}
For non-relativistic axion DM, the time derivative of the axion field dominates over its spatial derivative. Consequently, an external magnetic field $\vec{B}_0$ is typically used to generate a spatial effective current, which takes the form:
\begin{equation}
\vec{J}^a_{\rm eff} = g_{a\gamma} \omega_a a \vec{B}_0,
\label{jaxion}
\end{equation}
where $\omega_a$ is the frequency of the axion field, approximately equal to its mass $m_a$ for non-relativistic DM. 

For DM detection, the axion field is typically modeled using the standard halo model, which assumes a thermal velocity distribution of approximately $10^{-3}$ times the speed of light. This distribution results in a frequency bandwidth of approximately $10^{-6}$ of the central frequency. These parameters further determine the correlation length and correlation time of the axion field, which are typically $10^3$ times the Compton wavelength and $10^6$ times the axion oscillation period, respectively~\cite{Foster:2017hbq}. 
The local DM density is typically inferred to be 
$\rho_{\rm DM} \sim 0.3-0.6~\mathrm{GeV/cm^3}$ 
from dynamical measurements based on astrometric observations~\cite{deSalas:2020hbh}.

Axions may also exist in relativistic form, produced through a variety of cosmological and astrophysical mechanisms~\cite{Baumann:2016wac,Dror:2021nyr}. Their total energy density is expected to be much smaller than that of cold dark matter, and their spectral and angular distributions are more uncertain and depend sensitively on their production history.

\paragraph{Dark Photon}
Dark photons, $A'_\mu$, are well-motivated candidates for ultralight bosonic DM. They arise naturally in extensions of the Standard Model that include an additional $U(1)$ gauge symmetry, which is also commonly predicted in string theory~\cite{Abel:2008ai, Goodsell:2009xc}. The simplest interaction with the Standard Model occurs through kinetic mixing, described by the term $\propto{\epsilon} F_{\mu\nu} F'^{\mu\nu}$, where $\epsilon$ is the kinetic mixing coefficient, and $F'_{\mu\nu}$ is the the field strength tensor of the dark photon.

A convenient framework for describing dark photons in a resonant cavity or a shielded environment is the interaction basis. In this basis, the relevant Lagrangian is given by~\cite{Chaudhuri:2014dla}:
\begin{equation}
\mathcal{L}_{A'} = -\frac{1}{4} \tilde{F}'_{\mu\nu} \tilde{F}'^{\mu\nu}  + \frac{1}{2} m_{\gamma'}^2 \tilde{A}'_\mu \tilde{A}'^\mu + \epsilon m_{\gamma'}^2 \tilde{A}_\mu \tilde{A}'^\mu,
\label{eq:LdpIB}
\end{equation}
where the tilde ($\,\tilde{}\,$) denotes quantities in the interaction basis, and $m_{\gamma'}$ is the mass of the dark photon. In this basis, the dark photon does not couple directly to the EM current. However, the interaction term $\epsilon m_{\gamma'}^2 \tilde{A}_\mu \tilde{A}'^\mu$ contributes to an effective current, expressed as:
\begin{equation}
J^{A'\,\mu}_{\text{eff}} = \epsilon m_{\gamma'}^2 \tilde{A}'^\mu.
\label{eq:dp_current}
\end{equation}
Unlike axions, detecting dark photons does not require an external EM field. The effective current is directly proportional to the dark photon wavefunction, with its spatial component aligned parallel to the dark photon's vector field.

As a DM candidate, the dark photon possesses polarization degrees of freedom, unlike the axion. In the standard halo model, the virialization process of DM can wash out macroscopic polarization during production. As a result, a randomly polarized distribution is typically assumed, where the projections of the dark photon wavefunction onto three orthogonal axes are expected to be equivalent on average~\cite{Guo:2019ker,Chen:2021bdr}.

\paragraph{High-Frequency Gravitational Wave}
HFGWs refer to gravitational waves (GWs) with frequencies significantly higher than the kHz range, which makes them challenging to probe using traditional laser interferometers. Unlike lower-frequency GWs with known astrophysical origins, such as binary mergers, HFGWs are typically associated with processes in new physics scenarios. Potential production mechanisms include early Universe phenomena, exotic compact objects like primordial black holes, and ultralight bosons surrounding black holes~\cite{Aggarwal:2020olq,Aggarwal:2025noe}. Consequently, detecting HFGWs could provide insights into fundamental physics beyond the Standard Model.

GWs interact with EM fields through the Einstein–Maxwell action,
\begin{equation}
\mathcal{L}_{h} =  \sqrt{-g}\left(-\frac{1}{4} g^{\mu\alpha} g^{\nu\beta} F_{\mu\nu}F_{\alpha\beta}\right),
\label{eq:Lh}
\end{equation}
with $g_{\mu\nu}=\eta_{\mu\nu}+h_{\mu\nu}$, where $h_{\mu\nu}$ denotes the GW strain. 
Linearizing in $h_{\mu\nu}$ and decomposing the EM field strength as 
$F^{\mu\nu}=F_0^{\mu\nu}+\delta F^{\mu\nu}$, where $F_0^{\mu\nu}$ is an external EM field, 
the GW effect can be recast as a source for $\delta F^{\mu\nu}$. 
It is then convenient to express this source in terms of an effective current constructed from $h_{\mu\nu}$ and $F_0^{\mu\nu}$. A representative bulk expression is
\begin{equation}
J^{h\,\mu}_{\text{eff}} = \partial_\nu \!\left(\frac{1}{2} h\, F_0^{\mu\nu}
+ h^{\nu}{}_{\alpha}\, F_0^{\alpha\mu}
- h^{\mu}{}_{\alpha}\, F_0^{\alpha\nu}\right),
\label{eq:GW_current}
\end{equation}
where $h\equiv h^\alpha{}_\alpha$. This formulation makes clear that, as in the axion case, a nonzero external EM field is necessary to generate a nonvanishing source.

While the effective current provides a useful description for scaling estimates, it is not itself a gauge-invariant observable. Physical predictions must ultimately be formulated in terms of quantities measured by the detector, namely the EM fields evaluated in the detector frame and projected onto cavity or circuit eigenmodes. Recent work has clarified that apparent ambiguities associated with gauge or frame choice are not physical, and cancel once all relevant contributions, including boundary and inertial effects, are treated consistently at the level of observable detector responses~\cite{Gue:2026kga}.

For GW wavelengths small compared to the detector size, it is often convenient to start from TT-gauge expressions for the incident wave~\cite{Ratzinger:2024spd,Gue:2026kga}. For more general configurations, however, a careful treatment of the detector frame and boundary effects is required. A fully covariant formulation that makes the gauge and frame invariance of the observable signal manifest is presented in Ref.~\cite{Gue:2026kga}.

With this understood, one may still use compact scaling estimates for intuition. For a static external magnetic field of characteristic strength $B_0$ extending over a scale $V^{1/3}$, the effective current can be approximately parameterized as~\cite{Berlin:2021txa, Domcke:2022rgu}:
\begin{equation}
\vec{J}^{\,h}_{\text{eff}} \simeq \omega_h^2 V^{\frac{1}{3}} B_0\, h_0\, \hat{j} (\vec{r\,}),
\label{SCGW}
\end{equation}
where $\hat{j} (\vec{r}\,)$ is a spatially dependent, dimensionless vector encoding the direction and polarization of the incoming GW.

In addition to the direct EM coupling described by Eq.~(\ref{eq:Lh}), HFGWs can also induce mechanical deformations of a cavity. These deformations, in the presence of an external EM field, can lead to additional EM signals~\cite{Berlin:2023grv}. This mechanism will be discussed in further detail in Sec.~\ref{sec:HFGWM}.

Unlike DM, HFGWs are inherently relativistic, with their frequencies and spectra determined by their sources. Depending on the production mechanism, these waves can exhibit various bandwidths, making them coherent or stochastic in nature. Moreover, HFGWs carry polarization-specific information, which can reveal unique features of their origins.

\subsubsection{Resonant Systems: Cavities and LC Circuits}\label{sec:resonant_systems}

EM resonant detectors, such as resonant cavities or LC circuits, are highly efficient tools for detecting weak signals generated by effective currents, including those induced by ultralight dark matter or HFGWs. These systems leverage their resonant properties to amplify faint signals, enabling the extraction of critical information about their sources. This section introduces the fundamental principles governing resonant cavities and LC circuits, with a focus on how they capture and respond to effective currents.

\paragraph{Resonant Cavities}

A resonant cavity is a hollow structure enclosed by walls, typically made of metal or reflective materials, designed to trap EM waves. These waves reflect back and forth between the walls, reinforcing each other at specific frequencies determined primarily by the cavity's size and shape. When an effective current oscillates at the cavity's resonant frequency, it efficiently excites an EM cavity mode.

While the effective current $\vec{J}_{\rm eff}$ is the primary source of excitation, the time component of $J^\mu_{\rm eff}$, representing the effective charge density, can also influence electrodynamics. However, it does not excite the resonant modes of interest in this context~\cite{hill2009electromagnetic} and will not be discussed further.

The dynamics of the EM fields within the cavity are governed by Maxwell's equations, subject to appropriate boundary conditions. The effective current $\vec{J}_{\rm eff}$ acts as a source term for the electric field $\vec{E}$:
\begin{equation}
 \Box \vec{E}(t,\vec{x}) = -\frac{\partial \vec{J}_{\rm eff}}{\partial t},
\label{eq:maxwell_cavity}
\end{equation}
where $\Box$ is the d'Alembert operator. The boundary conditions of the cavity allow the electric field to be decomposed into a discrete sum of orthogonal cavity modes, labeled by $n$:
\begin{equation}
 \vec{E} (t,\vec{x}) = \sum_n e_n(t) \vec{\epsilon}_n (\vec{x}),
\label{Edecom}
\end{equation}
where $e_n(t)$ describes the time evolution of each mode, and $\vec{\epsilon}_n (\vec{x})$ forms a complete and orthogonal basis within the cavity. These modes satisfy:
\begin{equation}
 \nabla^2 \vec{\epsilon}_n + \omega_n^2 \vec{\epsilon}_n = 0, \qquad \int_V \vec{\epsilon}_n \cdot \vec{\epsilon}_m^{\,*} \, \D V = \delta_{mn},
\end{equation}
where $\omega_n \equiv 2\pi f_n$ is the resonant frequency of the $n$-th mode.

Substituting Eq.~(\ref{Edecom}) into Eq.~(\ref{eq:maxwell_cavity}) and projecting onto a specific mode $\vec{\epsilon}_n^*$ yields the equation of motion for the mode function $e_n(t)$:
\begin{equation}
 \frac{\D^2 e_n}{\D t^2} + \frac{\omega_n}{Q_n} \frac{\D e_n}{\D t} + \omega_n^2 e_n = \int_V \vec{\epsilon}_n^{\,*} \cdot \frac{\partial \vec{J}_{\rm eff}}{\partial t} \, \D V,
\label{eq:mode_eom}
\end{equation}
where $Q_n$ is the quality factor of the cavity mode, and $\omega_n / Q_n$ quantifies its dissipation rate.

When the frequency of the effective current matches the cavity mode frequency $\omega_n$, the mode is resonantly excited, and its amplitude is enhanced by the quality factor $Q_n$. Specifically, the mode reaches a steady-state amplitude:
\begin{equation}
 e_n(t) \sim \frac{Q_n}{\omega_n} \int_V \vec{\epsilon}_n^{\,*} \cdot \vec{J}_{\rm eff} \, \D V,
\end{equation}
which is proportional to the overlap between the cavity mode and the effective current. The power stored in the cavity due to the steady-state mode field is given by:
\begin{equation}
 P_{\rm sig} = \frac{Q_n}{2\omega_n}  V |\bar{J}_{\rm eff}\, \eta_n|^2,
\label{eq:power_signal}
\end{equation}
where $\bar{J}_{\rm eff}$ is the spatially averaged effective current density, and $\eta_n$ is the form factor,
\begin{equation}
 \bar{J}_{\rm eff} \equiv \sqrt{ \frac{1}{V}\int_V |\vec{J}_{\rm eff}|^2 \, \D V}, \qquad \eta_n \equiv \frac{\int_V \vec{\epsilon}_n^{\,*} \cdot \vec{J}_{\rm eff} \, \D V}{\sqrt{V}\bar{J}_{\rm eff}}.
 \label{eq:Jeffeta}
\end{equation}
The form factor $\eta_n$ quantifies the geometric overlap between the effective current and the cavity mode.

The resonant frequencies of a cavity are typically inversely proportional to its characteristic size and, for practical haloscope designs, generally fall within the $0.1$–$10$~GHz range.

For axion detection, the TM$_{010}$ cavity mode is typically chosen because its electric field component is predominantly oriented in the $\hat{z}$ direction, which maximizes the overlap with an external magnetic field~\cite{Sikivie:1983ip,Sikivie:1985yu}. A similar choice is often made for dark photon detection, where the goal is to detect the projection of the randomly polarized dark photon wavefunction onto the electric field component of the cavity mode~\cite{Ghosh:2021ard,Caputo:2021eaa}. In contrast, HFGWs exhibit more complex dependence cavity modes, depending on both the incoming wave direction and polarization~\cite{Berlin:2021txa,Navarro:2023eii}. Higher-order cavity modes, although generally suboptimal for DM searches due to reduced geometric overlap, can in certain cases exhibit enhanced sensitivity to HFGWs arriving from specific directions~\cite{Berlin:2021txa}.

\paragraph{LC Circuits}

Unlike resonant cavities, whose resonant frequency is constrained by their physical dimensions, LC circuits offer a flexible and tunable approach for detecting signals across a broad frequency range, particularly below the GHz regime. An LC circuit comprises an inductor $L$ and a capacitor $C$, forming a resonant system with a frequency:
\begin{equation}
\omega_{\rm rf} = \frac{1}{\sqrt{LC}}.
\label{eq:LC_freq}
\end{equation}
By varying either $L$ or $C$, the resonant frequency can be tuned over a broad range, making LC circuits well suited for scanning searches.

In this system, the capacitor stores energy in the form of an electric field associated with a charge $Q$ on its plates, while the inductor stores energy in the form of a magnetic field associated with a magnetic flux $\Phi$ through its coil. The total EM energy in the circuit is
\begin{equation}
H_0 = \frac{\Phi^2}{2L} + \frac{Q^2}{2C},
\label{eq:H_LC}
\end{equation}
which takes the familiar form of a harmonic oscillator. Energy oscillates between the electric field of the capacitor and the magnetic field of the inductor at frequency $\omega_{\rm rf}$. When driven by an external signal near resonance, the circuit undergoes coherent amplification, leading to enhanced oscillations in $Q$ and $\Phi$ that can be extracted using a sensitive current or flux readout.

In practical implementations, the LC resonator is coupled to a pickup loop that intercepts magnetic flux sourced by an effective current in the detector volume. The induced flux drives the circuit and builds up resonant energy, which is typically read out by a low-noise magnetometer such as a superconducting quantum interference device (SQUID).

The LC circuit is typically placed inside an EM shielded room, where the effective current is expected to generate EM signals. Since the operating frequency is typically well below the GHz regime, the shielded room can be approximated as an off-resonant cavity. In this limit, the effective current primarily generates a magnetic field, as opposed to an electric field. While the electric field source is proportional to the time derivative of $\vec{J}_{\rm eff}$, and thus scales with frequency, the magnetic field source is instead set by the characteristic length scale of the system, determined by the spatial extent of the external EM field and the geometric separation to the pickup loop of the LC circuit.
This results in the magnetic field dominating at frequencies much below the GHz range.

The magnetic field satisfies the inhomogeneous Maxwell equation:
\begin{equation}
\Box \vec{B}(t, \vec{x}) = \vec{\nabla} \times \vec{J}_{\rm eff},
\label{eq:maxwell_B}
\end{equation}
Compared to Eq.~(\ref{eq:maxwell_cavity}), it is evident that the induced magnetic field is stronger than the electric field by a factor of $1/(\omega V^{1/3})$. 

A precise calculation of the magnetic field requires summing over contributions from various cavity modes, with the lowest quantum numbers typically providing the dominant contribution~\cite{Jiang:2023jhl}. An approximate estimate of the magnetic field induced by the effective current can be expressed as:
\begin{equation}
\vec{B} (\vec{r}\,) \approx \int \frac{\vec{J}_{\rm eff} (\vec{r}\,’) \times (\vec{r} - \vec{r}\,’)}{4\pi |\vec{r} - \vec{r}\,’|^3} \D^3 \vec{r}\,’.
\label{eq:B_LC}
\end{equation}
Consequently, the induced magnetic flux through the pickup loop is approximately:
\begin{equation}
\Phi = \eta V \bar{J}_{\rm eff}.
\label{eq:PhiPsi}
\end{equation}
Here, $V$ represents the spatial volume relevant to the experimental configuration, including the size of the effective current region, the shield room, and the pickup loop area. The geometric factor $\eta \sim \mathcal{O}(0.1)$ accounts for the spatial overlap and configuration.

For axion detection, the pickup loop is typically oriented perpendicular to the external magnetic field $\vec{B}_0$ to maximize the signal~\cite{Sikivie:2013laa,Kahn:2016aff}. For dark photon detection, the randomly polarized nature of the dark photon field results in effective currents in all three spatial directions, allowing the pickup loop to be placed close to the shield wall and oriented perpendicularly to maximize flux capture~\cite{Chaudhuri:2014dla,Jiang:2023jhl}. 

However, HFGW detection requires specially designed pickup-loop geometries due to the spatial structure of the induced signal. The GW-induced effective current generates magnetic fields with a nontrivial angular dependence that integrates to zero for symmetric configurations such as a single circular loop. As a result, the leading contribution cancels, leaving only subleading effects. Differential geometries, such as a figure-eight loop composed of oppositely oriented semicircles, avoid this cancellation by sampling regions with opposite phase, thereby restoring the leading-order signal and enhancing sensitivity~\cite{Domcke:2022rgu,Domcke:2023bat}.

However, HFGW detection requires specially designed pickup loop configurations to account for the quadrupolar nature of GWs. The GW-induced effective current changes sign across the detector, so a single-loop pickup tends to average out the signal. Differential configurations, such as figure-eight (gradiometric) loops, are therefore used to avoid this cancellation by sampling regions of opposite sign. Detailed designs for HFGW-sensitive setups can be found in Refs.~\cite{Domcke:2022rgu,Domcke:2023bat}.

\subsubsection{Heterodyne Upconversion within Cavities}\label{sec:heterodyne}

\paragraph{Axion and HFGW EM Coupling}

Heterodyne upconversion in resonant cavities provides a powerful method for detecting axion DM or HFGWs at frequencies below the GHz regime~\cite{Goryachev:2018vjt,Berlin:2019ahk,Berlin:2020vrk,Thomson:2021zvq,Berlin:2022hfx,Bourhill:2022alm,Berlin:2023grv,Thomson:2023moc,Li:2025pyi,Fischer:2024msc,Marconatoetal,Crew:2026waz}. The key idea is to use a strongly driven cavity mode (the pump mode) to upconvert a low-frequency signal into a higher-frequency mode (the signal mode) that can be efficiently detected. This enables sensitivity to wave-like fields whose frequencies lie well below the intrinsic resonance frequency of the cavity.

This mechanism relies on transition between two nearly degenerate cavity modes induced by the effective current generated by axions or HFGWs. In contrast to conventional cavity or circuit haloscopes, where a static external magnetic field is used, here the external EM field is provided by the pump mode, oscillating at frequency $\omega_0$. The effective current thus takes the same geometric form as in Eq.~(\ref{eq:mode_eom}), but with the replacement of $\vec{B}_0$ to a cavity mode.

As an example, consider axion DM. The pump mode is driven at frequency $\omega_0$ and couples to a nearly degenerate signal mode at frequency $\omega_1 = \omega_0 \pm m_a$, where $m_a$ is the axion mass and $m_a \ll \omega_0 \sim \omega_1$. The axion field induces energy transfer between these two modes, generating a signal at $\omega_1$.

The signal power generated through heterodyne upconversion can be estimated as~\cite{Berlin:2019ahk}
\begin{equation}
P_{\rm sig} \sim \min \left( \frac{Q_a}{m_a}, \frac{Q_1}{\omega_1}  \right)V |\bar{J}_{\rm eff}\, \eta_n|^2,
\label{eq:signal_power}
\end{equation}
where $Q_a \sim 10^6$ is the coherence quality factor of the axion field, and $Q_1$ and $\omega_1$ are the quality factor and resonant frequency of the signal mode. The overlap factor $\eta_n$ and averaged effective current $\bar{J}_{\rm eff}$ are defined analogously to Eq.~(\ref{eq:Jeffeta}), but with the external EM field now provided by the pump mode rather than a static magnetic field. For axion masses well below the GHz range, the sensitivity benefits from the exceptionally high quality factors achievable in superconducting radio-frequency (SRF) cavities, typically $Q_1 \sim 10^{11}$. However, operating SRF cavities at such high $Q$ imposes a constraint on the maximum sustainable external EM field, usually limited to $B_0 \lesssim 0.2~\text{T}$, which introduces a trade-off between achievable $Q$ and pump field strength.

By tuning the frequency splitting between the pump and signal modes, heterodyne upconversion enables scanning over axion or HFGW frequencies that are far below the GHz resonant frequency of the cavity. If the initial mode separation is already small, such tuning can be achieved with minimal cavity deformation by mechanically squeezing the cavity volume, exploiting the fact that different modes respond differently to geometric perturbations. At very low target frequencies, particularly below the kHz scale, mechanical vibrations of the cavity structure and phase noise from the pump injection line can induce unwanted transitions between the modes, generating excess noise that dominates the low-frequency sensitivity.

The upconversion concept can also be adapted for broadband searches. One simple approach is to tune the pump and signal modes to be exactly degenerate, $\omega_0 = \omega_1$. In this configuration, very light axion DM can still induce mode conversion as long as its mass lies within the bandwidth of the signal mode, allowing sensitivity to all axion masses below that bandwidth~\cite{Berlin:2020vrk}. In addition, cavity designs can be engineered so that a single mode possesses both electric and magnetic field overlap, enabling self-transition. An example is the recently proposed twisted axion cavity~\cite{Bourhill:2022alm}, where the axion field induces a sideband of the pump mode without requiring a separate signal mode. Broadband search strategies have the advantage of avoiding frequency tuning.

\paragraph{HFGW Mechanical Coupling}\label{sec:HFGWM}

The detection sensitivity to HFGWs using cavities can be significantly enhanced by exploiting the mechanical coupling between the cavity and the GW~\cite{Berlin:2023grv}. In this approach, the tidal forces induced by the GW generate small mechanical displacements of the cavity walls. These deformations break the exact orthogonality of the cavity modes and enable energy transfer from the pump mode to the signal mode.

In this setup, a GW of frequency $\omega_g$ couples to the mechanical vibrations of the cavity and resonantly excites the signal mode when the frequency difference between the two EM modes matches the GW frequency,
$\omega_1 - \omega_0 = \omega_g$.

Compared to pure EM coupling, mechanical coupling becomes advantageous below the MHz regime, since mechanical eigenmodes of realistic cavities typically lie in the kHz range. Near these resonances, the mechanical response is strongly enhanced, particularly when the GW frequency falls within the mechanical mode bandwidth.

\subsubsection{Light Shining Through Wall}\label{sec:lsw}

The light-shining-through-a-wall (LSW) technique is a purely laboratory-based method for searching for new light and weakly coupled particles, such as axions and dark photons~\cite{VanBibber:1987rq,Ehret:2010mh,Redondo:2010dp,Arias:2010bh,Ortiz:2020tgs}. Unlike DM searches, LSW experiments do not rely on an ambient energy density of these fields; instead, they are produced and detected entirely within the laboratory. In a traditional LSW experiment, an intense laser beam is resonantly enhanced inside a high-finesse optical cavity in the presence of a strong magnetic field, enabling photon–axion conversion via the Primakoff effect. An opaque wall blocks the laser photons but is transparent to axions due to their extremely weak coupling to matter. A second magnetic field region and optical cavity on the far side of the wall then attempt to reconvert axions into photons, yielding a background-free signal of axion regeneration. Experiments based on this approach, such as ALPS~\cite{Ehret:2010mh,Bahre:2013ywa} and OSQAR~\cite{OSQAR:2015qdv}, have set mass-independent constraints on the axion–photon coupling $g_{a\gamma}$ for axion masses well below the laser photon energy scale ($m_a \ll \text{meV}$). Recent proposals aim to improve ALP production by replacing the static magnetic field region with ultra-strong EM fields generated using laser wakefield acceleration techniques~\cite{An:2025lll,An:2025gax}.

The LSW concept has been extended to the microwave domain using SRF cavities for dark photons~\cite{Jaeckel:2007ch,Graham:2014sha}, leading to what is often referred to as Dark SRF~\cite{Romanenko:2023irv,Kalia:2025afc}. In this approach, a high-$Q$ SRF cavity is used to coherently generate dark photons from a pump mode. The superconducting barrier automatically separates this cavity from a second empty SRF cavity used for particle regeneration and detection. Such setups benefit from extraordinarily large quality factors $Q \sim 10^{11}$ and extremely low noise achievable at cryogenic temperatures. Unlike optical LSW experiments, they do not require large-scale magnetic fields and are particularly sensitive to dark photon kinetic mixing.

A related SRF implementation applies the LSW concept specifically to axion dark matter by exploiting axion–photon conversion inside resonant EM structures~\cite{Gao:2020anb}. Unlike the Dark SRF setup for dark photon searches, axion LSW with SRF cavities requires two pump modes with significant spatial overlap of their electric and magnetic fields. The interaction between these two modes generates an effective axion current that sources an axion flux propagating toward the detection cavity. The axion signal generated in the production cavity is then converted back into EM photons in the detection cavity using a pump field, in close analogy to heterodyne-based DM detection schemes.

A practical challenge in SRF-based LSW experiments is the requirement of precise frequency matching between the generation and detection cavities. The very large quality factors of SRF cavities imply extremely narrow resonant bandwidths, so even small detuning between the two cavities can significantly suppress the regenerated signal power. Recent analyses have studied the impact of frequency instabilities arising from microphonics, fast fluctuations of the cavity resonance induced by mechanical vibrations or environmental perturbations, and have shown that such rapid fluctuations do not strongly degrade sensitivity when appropriately averaged~\cite{Cui:2025kxk,Kalia:2025afc}. However, slow long-term drifts in the cavity resonance frequency remain a limiting factor, as they restrict the maximum achievable coherent integration time unless active frequency stabilization techniques are employed.

\subsection{Sensitivity and Figure of Merit}\label{sec:sensitivity}

In most existing and proposed resonant detection schemes, the observable used to search for weak signals is the power deposited in a detector mode, or equivalently the power spectral density (PSD) of the readout variable, such as the voltage at a cavity readout port or the current or magnetic flux in an LC circuit. Operationally, this corresponds to measuring a two-point correlation function of the detector output and integrating it over time to extract excess power at a given frequency.

This power-based detection strategy is optimal for stationary and incoherent signals and forms the basis of standard radiometer techniques~\cite{Dicke:1946glx}. However, because it relies on linear amplification and second-order correlations, it is intrinsically subject to quantum fluctuations of the measured field. In the quantum-limited regime, vacuum fluctuations enter on equal footing with the signal and cannot be eliminated by linear amplification alone. 
The balance between measurement imprecision and quantum backaction in such power measurements leads directly to the Standard Quantum Limit (SQL), which sets a benchmark for the minimum detectable signal power within a fixed integration time, corresponding to resolving energy at the level of order one quantum in the detector mode over its coherence time~\cite{Chaudhuri:2019ntz}.

We define the detected PSD of the readout variable $v_r(t)$ using the symmetrized correlator
\begin{equation}
\langle { v_r(t)\, v_r(0) } \rangle
= \int_{-\infty}^{\infty} \frac{\text{d}\omega}{2\pi}\,
S_{v_r}(\omega)\, e^{-i\omega t},
\end{equation}
where $S_{v_r}(\omega)$ is a two-sided, symmetrized PSD. It can be decomposed into signal and noise contributions as
\begin{equation}
S_{v_r}(\omega) = S_{\text{sig}}(\omega) + S_{\text{noise}}(\omega).
\end{equation}

The signal term is sourced by the effective current associated with the bosonic field or GW background and can be written as
\begin{equation}
S_{\text{sig}}(\omega)
= |S_{0r}(\omega)|^2 \frac{\alpha^2}{4\Gamma}\, S_\Psi(\omega),
\end{equation}
where $\alpha$ characterizes the coupling strength between the detector mode and the external source, $\Gamma$ is the total damping rate of the mode, and $S_\Psi(\omega)$ is the source PSD. For a wave background with spectral energy density $\mathrm{d}\rho_\Psi/\mathrm{d}\omega$, we adopt the one-sided PSD convention ($\omega>0$),
\begin{equation}
S_\Psi(\omega)
= \frac{2\pi}{\omega^2}\,\frac{\mathrm{d}\rho_\Psi}{\mathrm{d}\omega}.
\end{equation}

The noise contribution includes intrinsic thermal fluctuations, quantum vacuum noise, and amplifier noise,
\begin{equation}
S_{\text{noise}}(\omega)
= |S_{0r}|^2 n_{\rm occ}
	+	|S_{rr}|^2 \frac{1}{2}
	+	\frac{1}{2},
\label{eq:Snoise}
\end{equation}
where $n_{\rm occ}= \tfrac12 + (e^{\omega/T}-1)^{-1}$ is the symmetrized thermal occupation number of the resonant mode. In the quantum-limited limit $T\to 0$, unitarity of the scattering matrix ensures that the total noise reduces to $S_{\text{noise}}=1$, explicitly realizing the SQL.

To assess sensitivity, the scan strategy aims to achieve an $\mathcal{O}(1)$ signal-to-noise ratio (SNR) over the source bandwidth. The SNR is given by the Dicke radiometer formula~\cite{Dicke:1946glx}:
\begin{equation}
\text{SNR}^2 = \frac{t_\text{int}}{2\pi} \int \left( \frac{S_{\text{sig}}}{S_{\text{noise}}} \right)^2 \mathrm{d}\omega,
\label{eq:SNR}
\end{equation}
where $t_\text{int}$ is the integration time per frequency step. The integrand naturally factorizes into the PSD of the background field and a detector response function.

The target signal field is characterized by an effective central frequency and bandwidth,
\begin{align}
\overline{\omega}_\Psi &= \frac{\int \omega^2 S_\Psi \alpha^2 / \Gamma\, \mathrm{d}\omega}{\int \omega S_\Psi \alpha^2 / \Gamma\, \mathrm{d}\omega}\,, \qquad
\Delta\omega_\Psi = \frac{\int \omega^2 S_\Psi \alpha^2 / \Gamma\, \mathrm{d}\omega}{\overline{\omega}_\Psi^2 S_\Psi(\overline{\omega}_\Psi) \alpha^2(\overline{\omega}_\Psi)/\Gamma(\overline{\omega}_\Psi)}.
\end{align}

The effective bandwidth for each scan is set by the detector response window,
\begin{equation}
\Delta\omega_r = \int \left( \frac{|S_{0r}|^2 n_\text{occ}}{S_{\text{noise}}} \right)^2 \mathrm{d}\omega,
\label{eq:Deltaomegar}
\end{equation}
which quantifies the spectral range over which the detector maintains near-optimal sensitivity.

At a fixed frequency setting, the SNR integral in Eq.~(\ref{eq:SNR}) receives support only over the overlap of the detector response and the background field PSD, and is therefore controlled by the smaller of $\Delta\omega_r$ and $\Delta\omega_\Psi$.

If a total time $t_e$ is allocated to cover one $e$-fold in frequency or mass parameter space, the integration time per step is set by the number of statistically independent measurements required to span that interval. Scanning one $e$-fold corresponds to a frequency range of width $\sim \overline{\omega}_\Psi$. 

However, statistical independence between neighboring frequency settings is lost if the step size is much smaller than the larger of the two characteristic widths. 
If $\Delta\omega_r \gg \Delta\omega_\Psi$, the detector response is nearly unchanged as the tuning is moved within a range $\lesssim \Delta\omega_r$, so measurements taken with $\delta\omega_{\rm step} \ll \Delta\omega_r$ largely reuse the same detector weighting of the spectrum. 
If instead $\Delta\omega_\Psi \gg \Delta\omega_r$, the signal spectrum varies only on scales $\sim \Delta\omega_\Psi$, and tuning steps with $\delta\omega_{\rm step} \ll \Delta\omega_\Psi$ probe essentially the same portion of the signal realization. 
In either limit, taking steps smaller than the broader of the two scales does not provide independent information, so the effective step size is
\begin{equation}
\delta\omega_{\rm step} \sim \max(\Delta\omega_r,\Delta\omega_\Psi).
\end{equation}
Distributing the total scan time $t_e$ uniformly over these independent steps yields
\begin{equation}
t_\text{int} \simeq t_e \cdot \frac{\max(\Delta\omega_r, \Delta\omega_\Psi)}{\overline{\omega}_\Psi}.
\end{equation}

Substituting the factorized form of the signal and noise PSDs, together with the approximations discussed above, into the SNR expression in Eq.~(\ref{eq:SNR}) yields
\begin{equation}
\text{SNR}^2(\overline{\omega}_\Psi) \simeq \frac{t_e}{\overline{\omega}_\Psi} \Delta\omega_r \Delta\omega_\Psi \left. \frac{\alpha^4 S_\Psi^2}{32 \pi \Gamma^2 n_\text{occ}^2} \right|_{\omega = \overline{\omega}_\Psi}.
\label{snrS}
\end{equation}
This expression highlights two key principles for optimizing scan-based searches: maximizing the detector response width $\Delta\omega_r$ and enhancing the coupling strength $\alpha$ both directly increase the achievable scan rate and sensitivity reach. 

The detector response width $\Delta\omega_r$ is controlled by the coupling to the readout port and can be optimized by tuning the readout damping rate $\Gamma_r$ to balance thermal noise and amplifier noise. In the zero-temperature limit ($n_\text{occ}=1/2$), the optimal scan rate is achieved for $\Gamma_r=2\Gamma$, which yields $\Delta\omega_r\simeq3\Gamma$~\cite{Krauss:1985ub}. At higher temperatures, where $n_\text{occ}\gg1$, the optimal readout coupling becomes $\Gamma_r\simeq2n_\text{occ}\Gamma$, corresponding to a response width $\Delta\omega_r\simeq2n_\text{occ}\Gamma$~\cite{Chaudhuri:2018rqn}. These relations specify the conditions under which the detector saturates the Standard Quantum Limit while maximizing scan efficiency. In practice, achieving this regime requires careful impedance matching, cryogenic operation, and quantum-limited amplification.

The signal coupling $\alpha$, which quantifies how the external source excites the resonant mode, enters quadratically in the SNR and therefore plays a central role in determining the ultimate sensitivity reach. We summarize the expressions for $\alpha$ across different detection schemes and target sources in Table~\ref{tab:alpha_summary}. These couplings follow directly from the effective currents in Eqs.~(\ref{jaxion}, \ref{eq:dp_current}, \ref{eq:GW_current}) and encapsulate the relevant parameters, including the external magnetic field strength $B_0$, detector volume $V$, and the geometric overlap factor $\eta$ between the induced signal and the detector mode.

\begin{table}[h!]
\centering
\renewcommand{\arraystretch}{1.4}
\begin{tabular}{lccc}
\hline\hline
\textbf{Scheme} & \textbf{Axion} & \textbf{Dark Photon} & \textbf{HFGW} \\
\hline
LC Circuit & 
$g_{a\gamma} \eta B_0 V^{5/6} \omega_a \sqrt{\omega_{\text{rf}}}$ &
$\epsilon \eta V^{5/6} m_{A'}^2 \sqrt{\omega_{\text{rf}}}$ &
$\eta B_0 V^{7/6} \omega_h^2 \sqrt{\omega_{\text{rf}}}$ \\

Cavity & 
$g_{a\gamma} \eta B_0 V^{1/2} \omega_a / \sqrt{\omega_{\text{rf}}}$ &
$\epsilon \eta V^{1/2} m_{A'}^2 / \sqrt{\omega_{\text{rf}}}$ &
$\eta B_0 V^{5/6} \omega_h^2 / \sqrt{\omega_{\text{rf}}}$ \\

SRF$^{\text{EM}}$ &
$g_{a\gamma} \eta B_0 V^{1/2} \omega_a / \sqrt{2\omega_{\text{rf}}}$ &
$\epsilon \eta V^{1/2} m_{A'}^2 / \sqrt{\omega_{\text{rf}}}$ &
$\eta B_0 V^{7/6} \omega_h^2 \sqrt{\omega_{\text{rf}}/2}$\\
\hline\hline
\end{tabular}
\caption{Effective signal couplings $\alpha$ to resonant sensor modes for axions, dark photons, and HFGWs across three haloscope schemes: LC circuit, EM cavity, and SRF cavity with EM upconversion. Couplings are defined through the effective currents in Eq.~(\ref{jaxion},\ref{eq:dp_current},\ref{eq:GW_current}). The SRF$^{\text{EM}}$ scheme rely on heterodyne upconversion using a strong external EM mode.}
\label{tab:alpha_summary}
\end{table}

Overall, this analysis shows that large detection volumes, strong external EM fields, and high quality factors are common and crucial ingredients for achieving optimal sensitivity in resonant searches. In the following, we discuss these factors in more detail, focusing on cavity-based axion detection as a representative and well-developed framework, noting that searches for dark photons and HFGWs often reuse or reinterpret experimental setups and analysis techniques developed for axion haloscopes.

\subsubsection{Cavity Volume}\label{sec:volume}

In a cavity haloscope the signal power is proportional to the effective detection volume $V$. 
For cavity-based axion detection the detector response scales approximately as $\alpha \propto V^{1/2}$, and can scale even more strongly in other detection schemes, as summarized in Table~\ref{tab:alpha_summary}. Increasing the cavity volume is therefore one of the most direct ways to enhance sensitivity.

However, the cavity volume is closely tied to the resonant frequency. 
In a typical cylindrical cavity the resonant frequency is determined by the cavity size. 
For example, the TM$_{0nl}$ modes satisfy
\begin{equation}
\omega_{{\rm TM}_{0nl}} =
\sqrt{\left(\frac{\xi_n}{R}\right)^2 + \left(\frac{l\pi}{L}\right)^2},
\end{equation}
where $R$ and $L$ are the cavity radius and height and $\xi_n$ is the $n$-th zero of the Bessel function. 
Since the cavity dimensions scale roughly with the photon wavelength, higher-frequency searches naturally require smaller cavities. 
This creates a fundamental tension between increasing $V$ and accessing higher axion masses.

This volume-frequency tension has become one of the central design challenges of modern axion haloscopes. 
Low-frequency experiments such as ADMX~\cite{ADMX:2009iij,ADMX:2018gho,ADMX:2018ogs,ADMX:2019uok,ADMX:2021nhd,ADMX:2021mio,ADMX:2024xbv,ADMX:2025vom} can exploit large cylindrical cavities inside large-bore solenoids, whereas searches at higher frequencies increasingly rely on alternative geometries, modular arrays of smaller resonators, or elongated and multicell cavity structures in order to recover the effective volume lost at large axion masses.

Several experimental strategies have been developed to increase the effective detection volume:

\begin{itemize}

\item \textbf{Large-volume cavities in existing magnets.}  
For low axion masses the resonant frequency can be matched using very large cavities, thereby maximizing the detection volume. 
Examples include WISPDMX, which reuses a large cavity from the HERA accelerator~\cite{Nguyen:2015ktw}, 
and the KLASH proposal, which plans to install a $\sim 50~{\rm m}^3$ cavity in the KLOE magnet~\cite{Alesini:2017ifp}. 
Related efforts also explore alternative magnet geometries. 
For example, toroidal cavities can provide large magnetic volumes while significantly reducing the fringe magnetic fields typical of solenoidal configurations~\cite{Choi:2017hjy}. 

\item \textbf{Combining several cavities.}  
Instead of constructing a single large resonator, one can combine the outputs of multiple cavities operating at the same frequency~\cite{Goryachev:2017wpw}. 
This strategy increases the effective detection volume while keeping the resonance frequency fixed. 
Early demonstrations of phase-matched multi-cavity readout were performed in ADMX R\&D~\cite{Kinion:2001fp}. 
More recently, the ADMX collaboration has implemented a four-cavity configuration to extend sensitivity toward higher frequencies~\cite{Yang:2020xsc}, 
and the ADMX Extended Frequency Range (EFR) program explores arrays of smaller cavities to probe axion masses beyond the reach of a single large cavity~\cite{2023APS..APRC01002K}. 
Similar concepts are also being pursued in the CAPP program and other high-frequency searches~\cite{Jeong:2017xqz}. 

\item \textbf{Multicell and mode-engineered cavities.}  
When the cavity size becomes much larger than the axion Compton wavelength, the axion tends to couple to higher-order modes whose fields oscillate many times across the cavity volume, leading to partial cancellations in the overlap integral. 
This can be mitigated by engineering the EM mode structure so that the electric field remains approximately phase aligned with the axion-induced current over a large volume. 
Examples include dielectric-loaded cavities~\cite{Morris:1984nu} and photonic band-gap cavities designed to control mode localization at high frequencies~\cite{Seviour:2014dqa,McAllister:2016fux, Jeong:2020cwz}. 

Another widely explored realization is to divide the resonator into multiple coupled cavity cells whose fields are phase matched by design. 
This idea is used in the RADES program, where rectangular cavities are connected through irises to form a multicell structure~\cite{Melcon:2018dba}. 
A related concept is the ``pizza-slice'' cavity developed at CAPP, in which a cylindrical cavity is divided into several azimuthal sectors that can be combined coherently~\cite{Jeong:2017hqs}. 
Note that multicell cavities are also widely used in accelerator physics, where superconducting SRF cavities employ several coupled cells to increase the effective accelerating gradient and efficiency~\cite{Aune:2000gb,Belomestnykh:2006wd,Blas:2026ybh}.

\item \textbf{Long cavities.}  
In rectangular or waveguide-like cavities the resonant frequency is mainly determined by the transverse dimensions, so the cavity length can be increased to enlarge the detection volume. 
This idea was proposed in Ref.~\cite{Baker:2011na} and implemented by the CAST–CAPP effort in the CAST magnet~\cite{Desch:2221945}. 

\end{itemize}

\subsubsection{Quality Factor}\label{sec:quality}

Another key parameter that determines the sensitivity of resonant detectors is the quality factor $Q$ of the cavity. 
The signal power stored in the resonator increases with the photon dwell time, which scales with $Q$, so higher-$Q$ resonators enhance the signal response. 
Importantly, increasing $Q$ also improves the scan rate over a wide range of practical values, even when $Q$ is well below $10^6$~\cite{Chaudhuri:2018rqn,Berlin:2019ahk,Chen:2021bgy,Cervantes:2022gtv,SHANHE:2023kxz,Chen:2023ryb}. 
Achieving higher quality factors has therefore become an important experimental direction for modern haloscopes.

Several experimental strategies have been developed to increase $Q$:

\begin{itemize}

\item \textbf{Superconducting resonators.}  
SRF cavities can achieve extremely high quality factors, often exceeding $Q\gtrsim10^9$ in accelerator applications~\cite{Padamsee:2017ohf}. 
Adapting SRF technology to axion haloscopes offers the possibility of dramatically extending photon storage times and improving detector sensitivity. 
However, conventional SRF cavities are generally incompatible with the strong static magnetic fields required for axion–photon conversion. 
As a result, SRF cavities are currently better suited for dark photon detection~\cite{Cervantes:2022gtv,SHANHE:2023kxz} or for heterodyne upconversion schemes in axion searches, where the EM pump fields injected into the cavity are much weaker than the strong static magnetic fields used in traditional haloscopes~\cite{Berlin:2019ahk,Berlin:2020vrk}.

\item \textbf{Surface engineering and superconducting coatings.}  
Another approach is to reduce resistive losses by improving the surface properties of the cavity. 
Techniques include improved copper surface treatment, electroplating, and thin-film or high-temperature-superconducting (HTS) coatings with lower surface resistance. 
In particular, several efforts have explored HTS cavities using materials such as YBCO or REBCO to maintain high $Q$ even in strong magnetic fields. 
Such developments have been actively pursued within the IBS-CAPP program and later the IBS-DMAG program~\cite{Jeong:2017hqs,Ahn:2019gbo,Lee:2019mfy,Lee:2020cfj,CAPP:2020utb,Yi:2022fmn,Adair:2022rtw,Yi:2023jrq,Kim:2023vpo,Youn:2024omj,CAPP:2024dtx,Bae:2024kmy,Bae:2025bqr,Ahn:2025via,Ahn:2026ssw,Lee:2026xxx}, where HTS cavities and coated resonators have been extensively investigated for axion haloscope applications. 
Related HTS cavity developments are also explored in the RADES collaboration~\cite{Ahyoune:2024klt}.

\item \textbf{Cavity geometry and coupling optimization.}  
In addition to material improvements, the EM design of the cavity can be optimized to reduce dissipation. 
By shaping the cavity geometry or tuning the coupling configuration, one can reduce the surface electric fields responsible for ohmic losses or distribute the fields more uniformly over the cavity walls. 
Examples include re-entrant cavity geometries and other mode-engineered structures designed to minimize surface losses while maintaining strong overlap with the axion-induced current~\cite{McAllister:2017lkb,Quiskamp:2022pks}. 

Such geometric optimization is particularly important in high-frequency searches, where the cavity dimensions become comparable to the microwave wavelength and surface losses increase. 
For example, the ORGAN experiment employs optimized cavity geometries and higher-order modes while maintaining reasonable quality factors and form factors~\cite{McAllister:2017lkb,McAllister:2018ndu,Quiskamp:2022pks,Quiskamp:2024oet,Quiskamp:2025wme}. 

\end{itemize}

\subsubsection{External Magnetic Field}\label{sec:bfield}

Axion and HFGW detection with microwave cavities or circuits both rely on the presence of an external magnetic field $B_0$. 
In such detectors the signal power scales as $B_0^2$, so increasing the external magnetic field is one of the most direct ways to enhance the experimental sensitivity.

In practice the achievable magnetic field is limited by magnet technology, cryogenic constraints, and the need to integrate resonant cavities and readout electronics within the magnet bore. 
Most existing haloscope experiments operate in superconducting magnets with fields of a few to several Tesla, while several programs are actively developing higher-field systems to extend the experimental reach. 
Different experimental efforts therefore pursue complementary strategies for generating strong magnetic fields.

\begin{itemize}

\item \textbf{Conventional superconducting solenoids.}  
Most haloscope experiments use superconducting solenoids based on NbTi or Nb$_3$Sn technology, which provide stable magnetic fields in the range of roughly $5$--$10~{\rm T}$. 
Examples include ADMX, which operates in an 8~T NbTi solenoid optimized for large cylindrical cavities~\cite{Du:2018uak}, and HAYSTAC, which uses a 9~T magnet designed for smaller high-frequency resonators~\cite{Kenany:2016tta,Brubaker:2016ktl}. 
Such magnets provide relatively large and uniform magnetic fields and remain the most mature technology for cavity-based searches.

\item \textbf{Ultra-high-field superconducting magnets.}  
To further enhance sensitivity, several programs are developing magnets capable of significantly higher fields. 
In particular, the CAPP program has carried out extensive research on high-field magnet systems for axion searches, including hybrid superconducting designs and high-temperature-superconductor (HTS) technologies targeting magnetic fields in the $18$-$25~{\rm T}$ range~\cite{Woohyun:2016hkn,Lee:2020cfj}. 

\end{itemize}

\subsection{Beyond the Standard Quantum Limit}\label{sec:beyond_sql}

In conventional resonant detection, the SQL arises from vacuum fluctuations in both signal quadratures. Achieving the SQL represents the fundamental sensitivity bound for any single-mode linear amplifier readout operating with quantum-limited noise. However, recent advances have demonstrated several techniques that can surpass the SQL, either by reshaping the quantum state of the EM field, engineering the readout process to squeeze one quadrature, or exploiting the structure of multi-mode systems. These strategies open new directions for enhancing the reach of resonant EM detectors searching for ultralight bosonic DM and HFGWs.

In the following subsections, we review three approaches that can enhance the scan rate beyond the limit imposed by the SQL in conventional resonant detection schemes. The first is the use of squeezed quantum states, which reduce the uncertainty in the measured quadrature at the cost of increased uncertainty in the conjugate one, thereby lowering the effective vacuum noise in the signal port~\cite{Zheng:2016qjv, Malnou:2018dxn, HAYSTAC:2020kwv, Lehnert:2021gbj}. We then discuss multi-mode resonator generalizations, including entangled mode configurations, state-swapping protocols, and engineered non-Hermitian networks that enable broadband detection while preserving narrowband sensitivity. Finally, we describe photon-counting measurement schemes based on quantum non-demolition (QND) readout with superconducting qubits, which enable projective detection of signal-induced excitations without disturbing the resonator quantum state.
Together, these methods illustrate a growing toolbox of quantum sensing techniques that extend beyond the limits of standard amplifier-based haloscope experiments.

\subsubsection{Squeezing}\label{sec:squeezing}

A widely explored quantum strategy for improving the performance of resonant detectors is the use of squeezed states. The basic idea is to reduce quantum fluctuations in the measured quadrature of the output field below the vacuum level, at the expense of increased fluctuations in the conjugate quadrature. When the signal predominantly occupies a single quadrature of the cavity output field, suppressing noise in that quadrature can improve the measurement sensitivity. This concept was proposed as a way to accelerate weak-signal searches in resonant detectors by reshaping the quantum noise entering the measurement chain~\cite{Zheng:2016qjv,Malnou:2018dxn,Lehnert:2021gbj}.

The impact of squeezing can be understood directly from the noise decomposition introduced in Eq.~(\ref{eq:Snoise}). In the absence of squeezing, the noise PSD contains three contributions: fluctuations associated with the internal loss channel of the resonator, vacuum fluctuations entering through the readout port, and the added noise of the phase-preserving amplifier. In the quantum-limited and zero-temperature limit these combine to give a total noise level corresponding to one photon, realizing the SQL.

Injecting a squeezed vacuum state into the readout port modifies only the vacuum fluctuations entering through that port. If the incoming field is squeezed by a factor $r$ and the measurement quadrature is aligned with the squeezed direction, the vacuum contribution associated with the readout port is reduced by a factor $e^{-2r}$. The internal fluctuations of the resonator and the amplifier-added noise remain unchanged. The noise PSD therefore becomes
\begin{equation}
S_{\rm noise}^{\rm sq}(\omega)
=
|S_{0r}|^2 n_{\rm occ}
+
|S_{rr}|^2 \frac{e^{-2r}}{2}
+
\frac{1}{2},
\label{eq:noise_sq}
\end{equation}
which should be compared directly with Eq.~(\ref{eq:Snoise}). Here the first term corresponds to intrinsic fluctuations associated with the resonator loss channel, the second term represents vacuum noise entering from the readout port and is the part suppressed by squeezing, and the final term is the irreducible noise added by a phase-preserving amplifier.

Because the amplifier contribution cannot be squeezed in a linear measurement chain, the total noise can be reduced below the SQL value but cannot fall below the ultimate limit of one half quantum. Consequently, the main benefit of squeezing in resonant searches is not an arbitrarily large reduction of the on-resonance noise floor. Instead, suppressing vacuum noise from the readout channel allows the detector to operate with stronger readout coupling without incurring a large vacuum-noise penalty. This enables operation in a more strongly overcoupled regime, thereby increasing the effective response bandwidth $\Delta\omega_r$ (defined in Eq.~(\ref{eq:Deltaomegar})) and improving the scan rate in frequency-scanning searches.

Experimentally, this strategy has been realized in axion haloscope searches such as HAYSTAC~\cite{HAYSTAC:2020kwv,Malnou:2018dxn,HAYSTAC:2023cam}. In these experiments, squeezed microwave states generated by Josephson parametric amplifiers are injected into the detector chain, reducing vacuum fluctuations entering through the readout port and enabling operation closer to the quantum limit over a wider bandwidth. The resulting squeezed-state receiver demonstrated quantum-enhanced axion searches and achieved an improved scan rate over a broader frequency interval around resonance.

\subsubsection{Multi-Mode Resonator Generalizations}\label{sec:multimode}

Another strategy to improve the scan rate is to exploit multiple resonant modes and construct a multi-mode detection system. 
If $N$ modes couple coherently to the same signal field with comparable response, their outputs can be combined with calibrated phases to form a collective observable. In this case the signal contributions add coherently, leading to a signal PSD enhanced by a factor of $N^2$, while the noise from independent channels adds incoherently and therefore grows only as $N$~\cite{Chen:2021bgy,Brady:2022bus}. 
Consequently, the signal-to-noise ratio in Eq.~(\ref{snrS}) can scale approximately as $N$.

Beyond simply combining independent resonators, one can also engineer coherently coupled multi-mode systems whose collective response to an external signal is significantly enhanced~\cite{Li:2020cwh,Chen:2021bgy,Wurtz:2021cnm,Jiang:2022vpm,Chen:2023ryb}. 
In conventional single-mode resonators, high sensitivity is achieved only within a narrow bandwidth around the resonance frequency, requiring a large number of tuning steps to scan a broad mass or frequency range. 
By contrast, properly designed multi-mode systems can distribute sensitivity across several nearby modes, effectively broadening the usable detection bandwidth while maintaining comparable peak sensitivity within the band.

\paragraph{Collective Readout of Multi-Mode Resonator Arrays}

One can combine the outputs of multiple resonator modes or sensors through coherent collective readout. In such schemes, several resonators or cavity modes probe the same external signal field, and their outputs are combined with appropriate phase calibration and weighting. 
If the signal field is coherent across the detector array, the measured observables from different sensors can be combined to form a collective observable constructed as a weighted linear combination of individual readouts.

For independent noise sources, the signal adds coherently while the noise from independent channels adds incoherently, leading to the familiar scaling
\begin{equation}
\mathrm{SNR} \propto \sqrt{N},
\end{equation}
for $N$ identical detectors~\cite{Chen:2021bgy,Brady:2022bus}. 
Such collective measurements effectively increase the detection volume and therefore improve the scan rate.

Beyond classical coherent combination, more advanced protocols have been proposed in which quantum correlations are engineered among the readout channels~\cite{Brady:2022bus}. 
In these schemes the outputs of different resonators are coupled through parametric interactions or joint measurements that generate correlations between their quantum fluctuations.
By constructing suitable collective observables, these correlations can partially suppress vacuum noise contributions that would otherwise appear independently in each sensor. In this sense, entanglement or two-mode squeezing can act as a quantum resource that suppresses measurement noise in the collective observable, enabling further sensitivity improvements beyond the classical scaling achievable with coherent signal addition alone~\cite{Chen:2025tgj,Fukuda:2025afi}.

\paragraph{Simultaneous Resonant and Broadband Detection via Parametrically Coupled Multi-Mode Resonators}

A fundamental limitation of traditional haloscope designs is the narrow bandwidth of high-$Q$ resonators, which restricts the efficiency of scanning a broad signal frequency range. To address this limitation, a class of approaches has been proposed that broadens the detector response by engineering multi-mode resonant systems~\cite{Li:2020cwh,Chen:2021bgy,Wurtz:2021cnm,Jiang:2022vpm,Chen:2023ryb}.

These schemes employ parametric couplings between the detector mode and auxiliary modes, which can realize beam-splitter (state-swapping) interactions or two-mode squeezing interactions. For example, an interaction Hamiltonian can be written as
\begin{equation}
\mathcal{H}_{\text{int}} = g\, a^\dagger b + G\, a^\dagger b^\dagger + \text{h.c.},
\label{eq:Hpara}
\end{equation}
where $a$ and $b$ are the annihilation operators of the sensor and auxiliary (readout) modes, respectively. The first term corresponds to a beam-splitter or state-swapping interaction, while the second term describes a two-mode squeezing interaction that generates correlated or entangled excitations.
Experimental realization of these couplings can be achieved using Josephson ring modulators~\cite{Bergeal:2010}.

When the coupling strengths $g$ and $G$ are comparable, one quadrature of the signal in the sensing mode $a$ can be coherently transferred to the auxiliary mode $b$ and back, thereby extending the resonant response over a broader frequency interval without sacrificing peak sensitivity, as demonstrated in a prototype experiment reported in~\cite{Jiang:2022vpm}.

This mechanism arises from coherent interference between the two parametric couplings. By optimizing both couplings and the readout coupling $\Gamma_r$, the effective response bandwidth $\Delta\omega_r$ in Eq.~(\ref{eq:Deltaomegar}) can be enhanced by a factor of $(g/\Gamma n_{\rm occ})^{2/3}$~\cite{Chen:2021bgy}. 

One can further introduce additional auxiliary modes and connect them through the same two types of couplings in Eq.~(\ref{eq:Hpara}), leading to a progressive broadening of the effective bandwidth $\Delta\omega_r$. In the limit of many coupled modes, the achievable bandwidth can approach the order of the cavity mode resonance frequency itself~\cite{Chen:2021bgy,Chen:2023ryb}. In this generalized setup, Eq.~(\ref{eq:Hpara}) effectively forms a chain-like structure in which one quadrature of the signal is sequentially enhanced along the chain toward the readout, while the orthogonal quadrature is progressively suppressed.

An alternative design introduces additional auxiliary modes that separate the two couplings in Eq.~(\ref{eq:Hpara}) into different modes, forming a binary-tree structure~\cite{Chen:2021bgy,Chen:2023ryb}. This architecture is more robust because both signal quadratures can be sequentially amplified along different branches. As a result, it avoids phase-calibration requirements and reduces nonlinear effects when implementing the couplings in Eq.~(\ref{eq:Hpara}) using Josephson junction circuits~\cite{Chen:2023ryb}.

The bandwidth enhancement becomes particularly significant for high-$Q$ resonators such as SRF cavities, since the dissipation rate scales as $\Gamma \propto 1/Q$. In heterodyne upconversion schemes, once the effective bandwidth $\Delta \omega_r$ approaches the order of the resonant frequency, one no longer needs to tune the frequency difference between modes, and the detector can simultaneously probe wave-like fields over several orders of magnitude in frequency, extending up to the GHz scale~\cite{Chen:2023ryb}.

\subsubsection{Photon-Number Readout via Quantum Non-demolition Superconducting Qubits}\label{sec:qnd}

A conceptually distinct route beyond conventional SQL-limited readout is to change the measured observable itself. 
In standard haloscope detection, one measures a linear signal such as a field quadrature or output voltage, and the sensitivity is therefore limited by vacuum fluctuations of that quadrature. 
By contrast, superconducting-qubit-based readout enables measurements of the photon number stored in the resonator, converting the search into a QND measurement of energy quanta rather than field amplitude~\cite{Dixit:2020ymh,Chen:2022quj,Agrawal:2023umy,Chen:2024aya,Nakazono:2025tak,Zheng:2025qgv}.

This distinction is important because the vacuum state has a definite photon number, namely $n=0$, even though it exhibits fluctuations in field quadratures. 
A photon-number measurement is therefore insensitive to vacuum fluctuations in the same way as linear amplification. 
In particular, if the search targets transitions from the vacuum state to the one-photon Fock state or higher excited states, the irreducible vacuum noise that limits linear readout no longer appears as a background. 
Instead, the dominant backgrounds arise from thermal photons, detector inefficiencies, and spurious excitations of the resonator or qubit.

The basic implementation couples the signal mode of a high-$Q$ cavity to a superconducting qubit operated in the dispersive regime.
For a transmon qubit, the effective Hamiltonian takes the form
\begin{equation}
H_{\rm eff}
=
\omega_r a^\dagger a
+
\frac{\omega_q}{2}\sigma_z
+
\chi a^\dagger a \sigma_z ,
\end{equation}
where $\omega_r$ and $\omega_q$ are the resonator and qubit frequencies and $\chi$ is the dispersive shift. 
The qubit transition frequency therefore depends on the photon number in the resonator, allowing repeated QND interrogation of the cavity occupation without destroying the stored excitation. 
In a DM search, a weak axion- or dark-photon-induced excitation populates the resonator, and the qubit is then used as a number-resolving probe to detect whether the cavity has transitioned from the vacuum to an excited Fock state.

In this framework the impact of thermal noise becomes especially transparent. 
If the resonator has a nonzero thermal occupation, thermal photons can mimic the signal and therefore set the dominant background once vacuum fluctuations are removed. 
Since the noise floor is then determined primarily by the thermal photon population and by the fidelity of qubit readout, the advantage of QND readout is greatest in the regime where the thermal occupation is $\ll 1$. 
This principle has been demonstrated experimentally in Ref.~\cite{Dixit:2020ymh}, which implemented photon-number-resolving detection in a microwave cavity and set stringent constraints on dark photons near the cavity resonance.

The basic vacuum-versus-one-photon discrimination can be further generalized using engineered quantum states of the resonator. 
Preparing the cavity in these states can enhance the response of the system to weak perturbations, such as in higher Fock states~\cite{Agrawal:2023umy}, Schr\"odinger cat states~\cite{Zheng:2025qgv} and entangled Fock states~\cite{Freiman:2025tse}.
In such schemes the signal effectively rotates or displaces the prepared state in phase space, enabling improved discrimination relative to a vacuum initial state.

A practical limitation of photon-number-based detection is that it does not directly preserve the phase information of the signal, making interferometric combination across multiple detectors less straightforward. 
Recent work therefore explores arrays of qubit-coupled resonators together with joint measurement protocols that recover collective advantages through quantum correlations~\cite{Chen:2023swh,Shu:2024nmc}. 
In these schemes, measurements from multiple sensors are combined through entangled readout channels or collective observables, enabling signal enhancement analogous to classical coherent combination while maintaining the low-noise benefits of QND detection. 
Such architectures point toward scalable quantum sensor networks that extend resonant dark matter searches beyond SQL-limited linear amplification.

Another direction is to use superconducting qubits themselves as the sensing elements rather than only as readout devices. 
In this case the DM field couples directly to the qubit degrees of freedom and induces transitions between the qubit energy levels. 
Because superconducting qubits are sub-wavelength detectors whose resonance frequencies can be tuned in situ, multiple qubits with different frequencies can be integrated on a single chip to probe different mass windows simultaneously~\cite{Kang:2025kaf}.

A particular challenge arises when applying qubit-based detection to axion searches, since conventional axion haloscopes require a strong static magnetic field, whereas superconducting qubits are typically fragile in magnetic environments, which can degrade coherence times or even suppress superconductivity in circuit elements. 
A dedicated line of work has therefore focused on developing magnetic-field-compatible qubit architectures and circuit designs~\cite{Kroll:2018hqa,Krause:2021llk}. 
These studies show that appropriate choices of junction materials, device geometry, and cavity design can substantially improve magnetic resilience, even for otherwise conventional superconducting circuits. 
Overcoming this challenge remains an active area of experimental development and will be essential for extending qubit-based QND detection schemes to axion dark matter searches.

\subsection{Characterizing Properties of Wave-like Signals}\label{sec:characterization}

Beyond setting bounds or discovering ultralight bosonic fields, EM resonant detectors can also be used to characterize the properties of the underlying wave-like signal. Owing to the coherent wave nature of these fields, such detectors can probe not only the signal amplitude and frequency, but also spatial and temporal correlations of the wavefront. This capability enables studies of the DM velocity distribution and potential deviations from the standard halo model~\cite{Foster:2017hbq}. 

Possible wave-like fields arise from a variety of sources. These include stochastic cosmological backgrounds~\cite{Baumann:2016wac, Dror:2021nyr}, as well as localized or directional sources such as dark matter streams~\cite{OHare:2017yze, Foster:2017hbq, Knirck:2018knd, OHare:2023rtm}, bound bosonic clumps~\cite{JacksonKimball:2017qgk, Budker:2023sex}, or radiation from compact astrophysical objects~\cite{Krause:1994ar, Dror:2021wrl, Hou:2021suj, Baryakhtar:2020gao, Dailey:2020sxa, East:2022ppo, Duque:2023seg, Khamis:2024oqa, Gavilan-Martin:2026zzw,Li:2026dls}. 
Such wave-like signals may exhibit characteristic observables including directionality, polarization, and spectral structure. These properties encode information about both the microscopic nature and the macroscopic structure of the wave-like field~\cite{Chen:2021bdr}. For HFGWs, the range of possible sources is even broader. Although detecting a stochastic cosmological HFGW background is challenging due to constraints from Big Bang nucleosynthesis (BBN)~\cite{Planck:2018vyg}, nearby coherent or transient sources, such as primordial black hole binaries, may still produce detectable signals~\cite{Aggarwal:2020olq, Aggarwal:2025noe}.

The ability to extract these observables depends crucially on the detector response. In resonant haloscope experiments, signals arise from the overlap between the induced effective currents and the EM eigenmodes of the detector, with analogous geometric projections appearing in circuit-based detectors. As a result, the detector response is inherently directional and depends on the relative orientation between the detector and the incoming wave, as well as on the EM configuration of the resonator. For instance, vector fields such as dark photons with a particular polarization can preferentially excite cavity modes whose electric field profiles overlap with the polarization of the incoming wave.

These properties can be probed experimentally by comparing signal variations across detectors with different response patterns. A network of detectors with distinct orientations or locations can exploit their different responses to infer properties of the incoming signal~\cite{Foster:2020fln, Chen:2021bdr, Jiang:2023jhl, Sulai:2023zqw, SHANHE:2024tpr, Jiang:2024boi, Gavilan-Martin:2024nlo, Reina-Valero:2025rul, Arza:2025kuh, Wilson:2025lhq}. 
If the detectors are additionally separated by long baselines, interferometric timing and phase information can further improve source localization~\cite{Foster:2020fln,Chen:2021bdr,Jiang:2023jhl,Gavilan-Martin:2024nlo}. Alternatively, a single detector can extract directional information through diurnal modulation induced by the Earth's rotation~\cite{SHANHE:2024tpr}. Furthermore, multiple EM modes within a single cavity can effectively function as a detector network when the incoming signal spans several resonant modes~\cite{Blas:2026ybh}. Below we discuss several representative approaches.

\subsubsection{Cavity and Circuit Networks}\label{sec:network}

A coherent network of spatially separated, phase-sensitive detectors, such as LC circuits or resonant cavities, can be employed to probe the coherence and propagation properties of wave fields~\cite{Derevianko:2016vpm,Foster:2020fln}. This approach relies on cross-correlations between pairs of detectors and is closely analogous to techniques used in radio astronomy and GW observatories~\cite{Finn:2008vh,Romano:2016dpx}.

As a simple toy model, consider a classical coherent field 
$\phi(t,\vec{x})=\phi_0\cos(\omega t-\vec{k}\cdot\vec{x})$, 
which induces time-dependent signals (e.g., effective currents) in multiple detectors. When two detectors are separated by a baseline $\vec{L}$, the signal measured at the two locations differs by a phase shift $\delta=\vec{k}\cdot\vec{L}$ due to the finite wavevector $\vec{k}$. For DM described by the standard halo model, the typical momentum is $k\sim10^{-3}m_\phi$, reflecting the virial velocity $v\sim10^{-3}$. Consequently, detectors separated by distances of order $1/k$ can exhibit interferometric signatures of the wave field~\cite{Foster:2020fln,Chen:2021bdr}.

By comparing cross-correlations across multiple detector pairs in a network, one can effectively perform a Fourier analysis over the set of baseline vectors $\vec{L}$ and infer the momentum distribution $\vec{k}$ of the incoming wave. This allows the reconstruction of the local DM velocity distribution and provides a way to distinguish the standard halo model from nonstandard scenarios such as streams, local clumps, or relativistic wave backgrounds. The achievable angular resolution follows a diffraction-like scaling,
$\Delta\theta \sim 1/({kL\times\mathrm{SNR}})$,
where $L$ denotes the characteristic baseline length. Even for an isotropic wave background, correlations between detectors persist over a coherence length set by the inverse momentum dispersion of the field.

Besides long-baseline correlations, one can also exploit the distinct responses of individual detectors within the network to extract wave properties. The response of each sensor depends on the projection of the induced effective current onto the local detector mode, making it sensitive to both the propagation direction and the polarization of the field. For instance, the signal from a dark photon background depends directly on the polarization vector of the hidden electric field. By deploying detectors with different orientations but separations smaller than the coherence length, it is possible to reconstruct the properties of the incoming dark photon wave, including its propagation direction and polarization structure~\cite{Chen:2021bdr}. These observables probe both the microscopic nature and the macroscopic distribution of the wave-like fields.

The same interferometric concepts can also be extended to the detection of HFGWs. For stochastic backgrounds, correlations between detectors are primarily determined by their response functions, leading to overlap-reduction functions analogous to those used in conventional GW interferometry~\cite{Chen:2024xzw}. For relativistic waves such as GWs, the characteristic correlation length and time scale are typically set by the inverse of the signal frequency.

Beyond networks of spatially separated detectors, similar principles can be applied using multiple resonant modes within a single detector. For example, a cavity resonator supports higher EM modes characterized by mode indices determined by its geometric symmetry~\cite{hill2009electromagnetic,Navarro:2023eii,Navarro-Madrid:2025qtm}. If an incoming signal excites several nearby modes, the relative amplitudes and phases of these modes can be used to infer properties of the wave-like fields. As an illustrative example, a multi-cell cavity with at least five closely spaced modes can in principle reconstruct the full GW strain tensor, including propagation direction and polarization, for HFGWs emitted by compact binaries~\cite{Blas:2026ybh}.

Experimental realizations of sensor networks have already begun to emerge in the search for wave-like DM and other weak signals. 
One of the earliest examples is the Global Network of Optical Magnetometers for Exotic physics searches (GNOME) collaboration~\cite{Pospelov:2012mt,Pustelny:2013rza}. 
GNOME was originally proposed to search for nonstandard DM scenarios beyond the standard halo model, such as topological defect dark matter including domain walls or other transient field configurations passing through the Earth~\cite{Pospelov:2012mt,Pustelny:2013rza,GNOME:2023rpz}. 
By operating a worldwide array of synchronized optical magnetometers, the experiment can look for correlated transient signals propagating across the network.

More recently, long-range correlations between spatially separated sensors have been explored as a powerful way to suppress common-mode noise and improve sensitivity to wave-like fields~\cite{Jiang:2023jhl,Gavilan-Martin:2024nlo}. 
By comparing signals across detectors separated by large distances, local noise sources that are uncorrelated between sites can be strongly reduced, while correlated signals induced by a coherent wave-like field remain detectable.

The network concept has also been proposed for HFGW searches. 
For example, the GravNet collaboration proposes a network of synchronized resonant cavities operating as phase-sensitive detectors for HFGWs~\cite{Schmieden:2023fzn,Schneemann:2024qli,Amaral:2026bef}.

In addition to geographically separated detectors, network-like strategies can also be implemented within a single experiment using multiple resonant cavities. 
For instance, the ADMX collaboration has developed a four-cavity configuration in which the signals from several cavities are operated simultaneously and combined coherently to increase the effective detection volume~\cite{Yang:2020xsc,Chung:2025jdz}. 
More recently, the ADMX EFR program envisions arrays of $18$ small cavities measured simultaneously, forming a scalable cavity network for axion searches~\cite{2023APS..APRC01002K}.

\subsubsection{Diurnal Modulation}\label{sec:diurnal}

For wave-like fields with a preferred propagation direction, the Earth's rotation naturally induces periodic variations in the projection of the signal onto the detector response.
In the Galactic frame, the direction of an incoming wave field can remain approximately fixed over long timescales, while the orientation of a laboratory detector changes due to the Earth's daily rotation. 
Consequently, the projection of the signal field onto the sensitive mode of the detector varies in time, producing a characteristic diurnal modulation in the measured signal power. 
Such modulation patterns can be exploited to infer both the direction and polarization structure of the incoming wave field.

In detector networks with long baselines, directional information can also be extracted through cross-correlations between spatially separated sensors. 
As the Earth rotates, the projection of the Galactic-frame momentum onto the detector baseline changes periodically, leading to a time-dependent modulation of the cross-correlation signal~\cite{Foster:2020fln}.

Even for a single detector, directional sensitivity can arise from the anisotropic response of the detector mode. 
A recent example is the ADMX search for a relativistic anisotropic axion population originating from the Galactic center~\cite{ADMX:2023rsk}. 
For such relativistic axions with a preferred momentum direction, the effective current induced in the cavity depends on the angle between the axion propagation direction and the cavity field axis. 
As the Earth rotates, this angle changes periodically, producing a characteristic diurnal modulation that can be used both to set constraints and to distinguish potential signals from noise.

A similar directional effect can occur for dark photon field. 
Because the signal is determined by the projection of the hidden electric field onto the detector mode, the measured power depends on the relative orientation between the detector axis and the dark photon polarization. 
The Earth's rotation therefore induces a daily modulation in the signal amplitude, which can be used to probe the polarization structure of the dark photon field~\cite{SHANHE:2024tpr}.

\subsection{Outlook}\label{sec:EMoutlook}

Cavities and circuits have developed into a versatile experimental platform for searching for ultralight bosons and HFGWs across a wide range of frequencies. What originated as the axion haloscope concept has now expanded into a broader family of detectors, including LC circuits, microwave cavities, SRF-based heterodyne upconversion schemes, and light-shining-through-wall experiments. Continued progress will likely arise from both incremental improvements to established detector concepts and the development of new architectures that exploit resonant mode structure and quantum measurement techniques.

On the hardware side, the most direct path to improved sensitivity remains the optimization of the key parameters summarized in Table~\ref{tab:alpha_summary}, namely the detector volume, external magnetic field, and resonator quality factor, together with noise reduction. Different frequency regimes naturally favor different design strategies. Low-frequency searches benefit from large magnetic volumes and large cavities, whereas higher-frequency searches increasingly rely on modular cavity arrays, multicell resonators, and mode-engineered structures that recover effective volume while maintaining the target frequency. In parallel, advances in high-field magnet technology, superconducting cavity development, and HTS-compatible resonators will continue to expand the accessible parameter space.

A second important direction is the development of detection architectures that decouple the signal frequency from the intrinsic resonance frequency of the detector. Heterodyne upconversion and other multi-mode schemes enable low-frequency axion or GW signals to be probed using GHz-scale resonators, greatly extending the useful range of high-$Q$ cavities. These approaches naturally benefit from the extremely low losses achievable in SRF systems, but they also introduce new experimental challenges, including pump stability, mechanical vibrations, and mode mixing at very small frequency splittings.

Finally, quantum sensing techniques are opening new opportunities for resonant detectors. Squeezed-state receivers can reduce measurement noise and increase scan efficiency, while parametrically coupled resonator networks can broaden the effective detection bandwidth. Superconducting qubits further enable photon-number-based quantum non-demolition measurements, which evade vacuum fluctuations that limit conventional linear amplification. As these technologies mature, they may allow resonant detectors to move beyond single-instrument searches toward coordinated, quantum-enabled sensor networks capable not only of discovering new wave fields but also of characterizing their directionality, polarization, and temporal structure.

%% file: sections/quantum-magnetometer.tex
\section{Quantum Magnetometry Probes of New Physics}
\label{sec:magnetometry}
\subsection{Basics of Quantum Magnetometry}	\label{sec:spinbasics}
Quantum magnetometers rely on the quantum properties of matter in order to accurately measure magnetic fields. In practice, two primary technological approaches exist\footnote{At frequencies in the THz range and above, other quantum sensors sensitive to electromagnetic radiation might be useful for particle-physics searches~\cite{Berlin:2023ubt,Mitridate:2020kly,Chigusa:2020gfs}. Although such devices technically may also respond to the magnetic component of the field, they are not typically classified as magnetometers and rely on substantially different detection principles. Accordingly, they are not discussed here.}. The first, Superconducting Quantum Interference Devices (SQUIDs), use flux quantization in a superconductor, and are discussed in section~\ref{sec:superconducting}. The second approach is spin-based magnetometers, and they are the focus of this section. Spin-based magnetometers employ ensembles which possess a magnetic dipole through their spins in atomic, molecular, and solid state systems. A magnetic field applies a torque on the spins, causing them to tilt or precess. The primary method to read out this effect is either optically, or using an auxiliary magnetometer that measures the magnetization of the sample.

Most spin-based magnetometers can broadly be split into two groups. The first group consists of magnetometers that utilize samples with electron-based spins, and so their dipole is of the same order as the Bohr magneton $\mu_B\propto e/m_e$. The second group consists of magnetometers based on nuclear spins, and so their dipole is of the same order as the nuclear magneton $\mu_N\propto e/m_N$. Magnetometers that utilize multiple spin species are often referred to as comagnetometers. Comagnetometers are also sometimes used as gyroscopes, as non-inertial rotation couples to spins in a similar way to magnetic fields, and by having multiple spin species measuring spin-coupled effects, one can remove the sensitivity to magnetic fields. In the following, we elaborate on the Bloch equations, which describe the evolution of the expectation value of the average spin of an atomic sample. 

\subsubsection{The Bloch Equations}
For a spin $\vec{S}$, the Bloch equations are  
\begin{equation}
\frac{d \langle \vec{S}\rangle}{dt} 
= \langle \vec{S}\rangle \times\left(\gamma  \vec{B}+\vec{b}\right)
- \Gamma \left(\langle \vec{S}\rangle- \vec{S}_0\right).
\label{eq:Bloch1}
\end{equation}

Here, $\gamma$ is the gyromagnetic ratio, which determines the precession frequency of the spin about the local magnetic field $\vec{B}$. The applied magnetic field is typically chosen such that it is predominantly aligned with the steady-state polarization $\vec{S}_0$, which defines the longitudinal ($\hat{z}$) direction. The vector $\vec{S}_0$ denotes the equilibrium spin polarization established in the absence of transverse magnetic or anomalous fields, and relaxation drives the system toward this state. The anomalous field $\vec{b}$ represents a spin-coupled BSM interaction with Hamiltonian $H = -\vec{b}\cdot\vec{S}$\footnote{Authors sometimes use different notations, wherein, for example, ${\vec b}\to {\vec b}/\gamma$, so that ${\vec b}$ carries magnetic field units}. The tensor $\Gamma$ is usually assumed diagonal in the frame defined by $\vec{S}_0$, and cannot be described by a Hamiltonian. The longitudinal component of $\Gamma$ is the relaxation rate $\Gamma_1 \equiv 1/T_1$, while the two transverse components are equal and correspond to the decoherence rate $\Gamma_2 \equiv 1/T_2$. 

Before discussing the individual terms in further detail, we provide a simplified estimate of the spin response. In a standard transverse magnetometry scheme, a bias field $B_z$ is applied, and one measures either $\langle S_x \rangle$ or $\langle S_y \rangle$ to detect transverse fields $\gamma B_{x,y} + b_{x,y}$. Typically, the sensor is maximally sensitive to fields oscillating in the $xy$ plane at the Larmor frequency $|\gamma B_z|$, with sensitivity to either the co-rotating or counter-rotating component depending on the sign of $\gamma B_z$. The frequency response is Lorentzian, with a linewidth given by $2/T_2$. On resonance, the transverse spin amplitude scales as
\begin{equation}
\langle S_{x,y} \rangle \sim S_z\, T_2\, (\gamma B_{x,y} + b_{x,y}) .
\end{equation}

We now move to discuss each term in detail. The gyromagnetic ratio $\gamma$ characterizes the coupling of a spin to magnetic fields. For atoms with no net electronic spin, the gyromagnetic ratio arises solely from nuclear spin and is typically of order $|\gamma|\sim 10^7$--$10^8~{\rm rad}~{\rm s}^{-1}~{\rm T}^{-1}$. Atoms with unpaired electronic spins generally possess much larger gyromagnetic ratios, $|\gamma|\sim 10^{11}~{\rm rad}~{\rm s}^{-1}~{\rm T}^{-1}$. Such atoms may also possess nuclear spin, which cannot necessarily be ignored even when focusing on electronic spin dynamics, due to the hyperfine coupling. For alkali vapors in the rapid-spin-exchange, low-field regime, the coupled electron--nuclear system can often be treated as an effective spin-$1/2$ system with a reduced gyromagnetic ratio $\gamma = \gamma_e/Q$, where $Q$ is the slowing-down factor and is typically of order $4$--$20$, depending on the nuclear spin and degree of atomic polarization~\cite{PhysRevA.71.023405}.

The magnetic field $\vec{B}$ generally has several distinct contributions. First, magnetometers typically use calibrated coils to produce controlled magnetic fields. While initializing the spin ensemble may involve a sequence of applied field pulses, during the signal-sensitive period the field is often held fixed in a single direction. Second, although most experiments employ magnetic shielding to suppress environmental noise, such shielding is imperfect. Furthermore, the ${\vec B}$ induced by the shields often acts as a source of noise~\cite{kornack2007low}. Finally, in comagnetometers that utilize multiple collocated spin species~\cite{Bloch:2019lcy,Bloch:2022kjm,Bloch:2021vnn,Gavilan-Martin:2024nlo,GNOME:2023rpz,Wu:2019exd,Lee:2022vvb,Wei:2023rzs,Vasilakis:2008yn,VasilakisThesis,Kornack:2004cs,KornackThesis,Brown:2010dt,BrownThesis,Xu:2023vfn,Jiang:2021dby,Abel:2017rtm,Garcon:2019inh,Alonso:2018dxy,Wei:2022mra,Wei:2022ggs}, the different spin species generate mutual magnetic fields. These arise from two primary sources: (i) long-range magnetic dipole--dipole interactions, and (ii) short-range interactions during close collisions. One common example where the utilization of two spin species can be advantageous is when dark matter couples predominantly to one spin species, while a second species provides a high-sensitivity readout channel.

The anomalous precession term $\langle \vec{S} \rangle \times \vec{b}$ encodes possible new-physics effects. Typically, as previously mentioned, when the characteristic frequency of the anomalous field $\vec{b}$ is comparable to $|\gamma \vec{B}|$, the magnetometer becomes maximally sensitive to its effects. For most DM-induced anomalous fields, this frequency is approximately $m_{\rm DM}$, so probing higher dark matter masses requires larger applied magnetic fields. Given that laboratory magnetic fields are typically limited to $\mathcal{O}(10~{\rm T})$, systems based on nuclear spins are expected to probe wavelike dark matter up to $\mathcal{O}({\rm \mu eV})$ masses~\cite{Dror:2022xpi}, while those exploiting electronic spins can in principle reach masses up to $\mathcal{O}({\rm meV})$~\cite{Berlin:2023ubt}.

As introduced above, the diagonal matrix $\Gamma$ determines the rates at which the spin ensemble approaches its steady state. Although spin-$1/2$ systems are strictly two-level systems, even systems with total spin $s>1/2$ are often well described as effective two-level systems (see the later discussion of the density-matrix formalism). Within this picture, the $\Gamma$ longitudinal component, $\Gamma_1$ corresponds to the decay rate of the excited state, while $\Gamma_2$ arises from processes that randomly dephase the individual spins. In general $\Gamma_1\leq 2\Gamma_2$, though usually it is also true that $\Gamma_1\ll \Gamma_2$ as well.  In many systems, an additional timescale $T_2^*<T_2=1/\Gamma_2$, known as the effective coherence time, arises due to spatial inhomogeneities—most commonly in the applied magnetic field—which cause dephasing between different regions of the ensemble. In such cases, the sensitivity of the sensor can depend on both time scales depending on the exact regime operated at~\cite{Walter:2023edn}\footnote{For simplicity, throughout this review we assume $T_2=T_2^*$, though in most contexts, when $T_2>T_2^*$, $T_2^*$ is the timescale that should be used in the Bloch equations, and its solutions.}.

Without the repolarization term proportional to $\Gamma \vec{S}_0$, the spin ensemble would relax toward a completely randomized orientation, eliminating any useful measurement signal. Therefore, a polarization mechanism is required to achieve a nonzero steady-state (or temporary, for pulsed schemes) spin alignment. A variety of such mechanisms exist, and the choice of polarization technique is typically dictated by the experimental platform and measurement requirements, since not all methods are universally applicable. Below, we briefly describe four of the primary polarization methods employed in particle-physics searches.
\begin{itemize}
    \item {\textit {Thermal Polarization}.}
    Given a constant magnetic field $B_{\rm ext}$, an energy splitting is generated between spins aligned and anti-aligned with it. Therefore, if the system is cold enough and $B_{\rm ext}$ is sufficiently large, the spin ensemble will be thermally polarized. In thermal equilibrium, a spin-$1/2$ ensemble has polarization given by
    \begin{equation}
        P=\tanh(\mu B_{\rm ext}/k_B T),
        \label{eq:polarization}
    \end{equation}
    where $\mu$ is the magnetic moment of an individual spin. In practice, this is mostly used when a system is cryogenically cooled~\cite{JacksonKimball:2017elr,Dror:2022xpi}, and a large $B_{\rm ext}$ is temporarily applied to polarize the spins, followed by a data-taking period $T\lesssim T_1$. For example, for $B_{\rm ext}\sim\mathcal{O}({\rm T})$, nuclear spins (electron spins) achieve $\mathcal{O}(1)$ polarization at $\mathcal{O}({\rm mK})$ ($\mathcal{O}({\rm K})$) temperatures.
    \item {\textit{Spontaneous Magnetization.}} Certain materials exhibit spontaneous spin polarization arising from magnetic ordering~\cite{frenkel1930magnetisation,QUAX:2020adt,Crescini:2018qrz,Chigusa:2023szl}. Ferromagnets are the most common example: spontaneous symmetry breaking causes the system to become polarized even without an external magnetic field, along an axis determined by atomic interactions.
    \item {\textit {Optical Pumping.}} Optical pumping employs circularly polarized light tuned near an atomic transition of the magnetometer atoms~\cite{HAPPER:1972ars}. The light carries angular momentum along its direction of propagation, which is transferred to the atomic system. Optical pumping is commonly used in alkali vapor magnetometers (e.g.\ Rb, Cs, or K vapors). Importantly, unlike thermal polarization, optical pumping can easily achieve $\mathcal{O}(1)$ polarization in room-temperature (or heated) magnetometers.  
    \item \textit{Spin-Exchange Polarization}. Spin-exchange processes are relevant for magnetometers that contain two different spin species~\cite{Walker:1997zzc}. The first, the ``donor'', is polarized by some other method (often optical pumping), and through its interaction with the second species, the ``recipient'', it polarizes the latter. This method is commonly used to polarize nuclear spins in comagnetometers~\cite{Bloch:2019lcy,Bloch:2022kjm,Bloch:2021vnn,Gavilan-Martin:2024nlo,Lee:2022vvb,Wei:2023rzs,Vasilakis:2008yn,VasilakisThesis,Kornack:2004cs,KornackThesis,Brown:2010dt,BrownThesis,Xu:2023vfn,Jiang:2021dby,Alonso:2018dxy}, typically achieving up to $\mathcal{O}(10\%)$ nuclear polarization.
\end{itemize}
Other approaches, such as Dynamical Nuclear Polarization (DNP)~\cite{Atsarkin_2011}, are also possible. Furthermore, multiple methods are sometimes combined to optimize spin polarization and thereby experimental sensitivities in detecting BSM signals.

\subsubsection{Beyond the Bloch Equations and Quantum Noise}

The Bloch equations are phenomenological and do not provide the full description of a system. For example, both the spatial dependence of the spins and the inherent quantum uncertainty in a finite number of spins (see also the discussion below) are not captured by the Bloch equations, which only describe the averaged spin of the sample. However, for many applications, the Bloch equations (or slight variations thereof, such as the Landau-Lifshitz-Bloch equation~\cite{Atxitia_2017}) are sufficient. A more detailed description of spin dynamics can be achieved using density-matrix-based formalisms. Many effects captured by these approaches—such as spin projection noise (SPN) arising from the quantized nature of the spin degrees of freedom—can also be effectively incorporated into appropriately modified Bloch equations. In practice, many parameters of the Bloch equations must be determined experimentally, as they depend sensitively on the technical details of the apparatus and cannot generally be derived from first principles.

The SPN, also known as the atomic shot noise, is closely related to the Standard Quantum Limit. A finite number of spins presents a fundamental uncertainty when measuring a spin component for which the state is not an eigenstate. Assuming the measurement is weak, the induced SPN can be characterized to leading order by the two-point correlation function of the measured spins. We briefly describe the Keldysh formalism to compute the SPN~\cite{Braun_2007} in the following. Consider a measurement operator $\hat{\mathcal{O}}$ (e.g. $S_x$, the average spin in the ${\hat{x}}$ direction). For simplicity, assume the operator does not depend explicitly on time. The corresponding expectation value is
\begin{equation}
\mathcal{O}(t)={\rm Tr}(\hat{\mathcal{O}}\rho (t)).
\end{equation}
We can define the symmetrized two-point temporal correlation as
\begin{equation}
S(t)=\frac{1}{2}{\rm Tr}\left({\overset{\leftrightarrow}{\mathcal{O}}}{\Pi}(t) \overset{\leftrightarrow}{\mathcal{O}} \rho(0)\right),
\end{equation}
where $\rho(0)$ is the initial density matrix, while for any matrix $M$, $\overset{\leftrightarrow}{\mathcal{O}} M\equiv \hat{\mathcal{O}}M+M\hat{\mathcal{O}}$, and ${\Pi}(t)$ is the time evolution operator for a density matrix. For simplicity we chose the time separated correlation to be evaluated between initial time $0$ and $t$, though generally it should be evaluated between any two times $t$ and $t'$. 
For a linear evolution and a density matrix of size $n\times n$, ${\Pi}(t)$ can be written as an ${n^2}\times n^2$ matrix, operating on a vectorized version of $\rho$. ${\Pi }(t)$ can be derived by computing the time evolution for arbitrary initial conditions, with the additional requirement that ${\Pi}(t)$ preserves the trace of the density matrix. 

For the simple case of a fully polarized system of $N$ spins, a large axial magnetic field ${\vec B}=B{\hat z}$ with no applied transverse magnetic fields, and a measurement of the average $S_x$, the amplitude spectral density at $\omega=\omega_0\equiv \gamma B$ scales as
\begin{equation}
\delta S_x\sim\sqrt{\frac{T_2}{N}},
\end{equation}
with a full width at half maximum of $2/T_2$~\cite{Aybas:2021cdk}.  We implicitly took the rotating wave approximation (i.e. neglecting a counter-rotating term oscillating at frequency $-\omega_0$).

As mentioned previously, the SPN discussed above can be incorporated into the Bloch equations as an effective stochastic magnetic field (see Ref.~\cite{VasilakisThesis} for a detailed semi-classical treatment). Much like an anomalous field, this noise term couples only to one spin species (for the SPN it is the one which generates it). Its mean vanishes, but it possesses a non-vanishing power spectral density in frequency space, reflecting its stochastic nature. In this formulation, for a single spin species, the amplitude spectral density of the effective magnetic noise at the Larmor resonance frequency is given by
\begin{equation}
\delta B\sim \tfrac{1}{\gamma}\sqrt{\frac{1}{NT_2^*}},
\end{equation}
where the field can be taken to be in either the $x$ or $y$ direction, as both have similar effects on resonance (within the rotating wave approximation).

In addition to the inherent quantum noise in the spins, the readout method also introduces noise which is highly system-dependent. For optically read magnetometers, the fundamental quantum noise sources are the Photon Shot Noise (PSN), which limits the optical readout, and Light Shift Noise (LSN), in which fluctuations of the probe polarization act back on the spins. Both will be expanded on later. If instead the spins are read out via a secondary magnetometer (e.g., a SQUID), the intrinsic noise of the secondary magnetometer should also be taken into account.

\subsection{Particle Physics Applications}

Spin-based magnetometers have found several applications in particle physics. The two primary directions are searches for ultralight wavelike dark matter that couples to spins and searches for novel spin-dependent forces. A third avenue, somewhat less active in recent years, involves tests of Lorentz symmetry~\cite{Brown:2010dt,Kornack:2004cs}. 

Another distinct approach employs a geographically distributed network of synchronized magnetometers to search for correlated, often transient signals~\cite{Pospelov:2012mt}. The Global Network of Optical Magnetometers (GNOME) pioneered this approach for searches for correlated exotic spin signals with its international network of magnetometers~\cite{Khamis:2024oqa,Afach:2018eze,Afach:2021pfd,GNOME:2023rpz,Masia-Roig:2019hsy,JacksonKimball:2017qgk,Wlodarczyk:2013ria,Pustelny:2013rza}, while other groups have since established similar networks~\cite{Wang:2026rwy}. These efforts target a variety of Standard Model extensions, including Axion Stars, Dark Matter Q-balls, and more (see Ref.~\cite{GNOME:2023rpz}).

We will now discuss searches employing magnetometers to probe wavelike dark matter and signatures of new interactions.

\subsubsection{Ultralight Wave-like Dark Matter}

In the context of quantum magnetometers, the primary dark matter candidate that can be searched for is the ultralight ALP. The anomalous field $\vec{b}$, arising from the direct coupling of ALPs to a fermion species $\psi$ with spin 1/2, can be written as: 
\begin{equation}
{\vec b}=\epsilon_\psi g_{a\psi\psi}\sqrt{{2\rho_{\rm DM}}}{\vec v}_{\rm DM} \cos(E_{\rm DM} t+\theta_0).
\end{equation}

where $g_{a\psi\psi}$ is the dimensionful coupling strength between ALP and the fermion species\footnote{The notation used here has these couplings carrying mass dimension $-1$, but common alternatives have dimensionless couplings, with $g_{a\psi\psi}^{\rm dimensionless}=2m_{\psi}g_{a\psi\psi}^{\rm dimensionful}$ (or more rarely $g_{a\psi\psi}^{\rm dimensionless}=m_{\psi}g_{a\psi\psi}^{\rm dimensionful}$).}. Currently, the majority of quantum magnetometer experiments looking for the above cast their bounds in terms of the coupling of ALPs to electrons, protons, or neutrons ($g_{aee},g_{app},g_{ann},$ respectively). The form factor $\epsilon_\psi$\footnote{This form factor, $\epsilon$ has no agreed-upon symbol across the community.} accounts for the fact that experimental measurements probe composite systems rather than bare spin-1/2 fermions. For instance, in setting ALP-neutron limits from ALP-$^{3}$He interactions, $\epsilon_\psi$ quantifies the effective contribution of the neutron spins to the $^3$He nuclear spin (see e.g. Refs.~\cite{JacksonKimball:2014vsz,Hu:2021awl} for further discussion and estimations of the form factor for various atoms).

The variables ${\vec v}_{\rm DM}$ and $\theta_0$ are sampled from a statistical distribution. The distribution of ${\vec v}_{\rm DM}$ is usually taken from the Standard Halo Model, with the averaged velocity magnitude $v_{\rm DM}\sim {10}^{-3}$, and variation of $ {\delta v}_{\rm DM} \sim {10}^{-3}$. The energy of dark matter is given by $E_{\rm DM}= m_{a}+\tfrac{1}{2}m_a \vec{v}_{\rm DM}^2$, with the center signal frequency $\langle \omega\rangle \sim \langle {E_{\rm DM}}\rangle \sim m_a$, and a width of $\delta \omega \sim \delta E_{\rm DM}\sim 10^{-6}m_a$. The appropriate way to analyze the stochastic nature of the DM field can be found in e.g.~\cite{Bloch:2021vnn,Lee:2022vvb,Centers:2019dyn}. However, a simplified treatment is often sufficient for estimating projected sensitivities. For measurement times $T$ shorter than the ALP coherence time ($\tau_a = 2\pi/\delta E \approx 2\pi \times 10^{6}/m_a$), the field may be approximated as monochromatic with a well-defined phase. In this coherent regime ($T < \tau_a$), the signal adds coherently while broadband background noise does not, and the sensitivity to the coupling constant therefore improves as $T^{1/2}$ (assuming $T > T_2$). For measurement times longer than the coherence time ($T > \tau_a$), the phase of the field undergoes stochastic drift. Nevertheless, by fitting the data to an assumed ALP velocity distribution, continued improvement remains possible. In this regime, longer integration times provide improved spectral resolution of the ALP lineshape, and the sensitivity to the coupling constant scales more slowly, as $T^{1/4}$ (see also Refs.~\cite{Dror:2022xpi,Zhang:2023eya} for more details on different scaling regimes).

The primary alternative DM detection effort to the direct ALP-spin interaction targets ALP-DM induced oscillating electric dipole moments (EDMs)~\cite{Graham:2013gfa,Budker:2013hfa,JacksonKimball:2017elr,Abel:2017rtm,JEDI:2022hxa,Fan:2024pxs}. A background axion field can induce a time-dependent EDM,
\begin{equation}
d_n=g_d a.
\end{equation}
In particular, if axions couple to gluons as $\tfrac{\alpha_s}{8\pi}\tfrac{a}{f_a} G_{\mu\nu}\tilde{G}_{\mu\nu}$, then for neutrons, the $g_d$ induced by that coupling is
\begin{equation}
g_d=\frac{2.4\times 10^{-16}{\rm e}\cdot{\rm cm}}{f_a}.
\end{equation}

Because an EDM couples fermion spins to electric fields, these searches require either an applied external field or, more commonly, the use of strong internal ones. In the presence of such fields and the oscillating EDM, the atoms experience a spin-dependent splitting, leading to an anomalous spin-coupled field.

While most spin-based magnetometer searches for dark matter concentrate on ALPs through the linear spin couplings discussed above, alternative scenarios have also been explored. Some studies have considered quadratic spin couplings to ALPs (see, e.g.,~\cite{Wu:2019exd,Bloch:2021vnn}), and recent works have pointed out that quadratic couplings to the Earth's matter density can substantially enhance the induced field gradients, thereby amplifying the effective linear spin-coupled anomalous field~\cite{Banerjee:2025dlo,delCastillo:2025rbr} (see Ref.~\cite{Huang:2026tgv} for a dedicated search).  Others have noted that sensitivity to ALP--photon interactions can arise naturally within spin-based searches~\cite{Beadle:2025dgy}. Additional efforts have extended beyond pseudoscalar couplings altogether. For instance, magnetometer-based searches have been performed for dark photons, where the observable signal can depend sensitively on the presence of nearby conducting surfaces, such as the ground and ionosphere~\cite{kwyg-5v64}, or on surrounding conductive enclosures\footnote{In these scenarios, the interaction creates an ordinary magnetic field, and therefore does not intrinsically require the use of spin-based magnetometers, compared to other magnetometry schemes.}. Proposals targeting scalar dark matter have also been put forward~\cite{Bloch:2023uis}.

\subsubsection{New Force Searches}

In addition to DM searches, spin-based magnetometers have a long history of probing novel forces and interactions across a variety of distance scales. See Ref.~\cite{Cong:2024qly} for a recent comprehensive review. The primary theoretical motivation for novel forces is the presence of new particles which mediate the interactions between SM particles. 
Unlike searches for a background DM field, detecting the effects of novel forces requires a sensor and (at least one) source. Depending on the interaction, the source may need to be in motion, spin-polarized, or simply have a finite density of SM fermions. In some cases, large bodies such as the Earth or Sun can serve as a source.

Several classification schemes for novel interactions have been proposed~\cite{Cong:2024qly,Fadeev:2018rfl}. The most widely used approach was introduced by Ref.~\cite{Dobrescu:2006au}, building upon the earlier work of Ref.~\cite{Moody:1984ba} (see Ref.~\cite{Fadeev:2018rfl} for refinements). Based on Lorentz-symmetry arguments, this framework identifies 16 distinct potentials describing long-range rotationally-invariant interactions between two non-relativistic spin-1/2 fermions. Fifteen of these depend on the spins of the source, the sensor, or both, rendering them accessible to spin-based sensors.

\subsection{Magnetometer Technologies}

\subsubsection{Optical Magnetometers}
The primary class of spin-based magnetometers that has demonstrated results in the context of particle physics searches is optical magnetometers (see e.g.~\cite{Bloch:2019lcy,Bloch:2022kjm,Bloch:2021vnn,Gavilan-Martin:2024nlo,GNOME:2023rpz,Lee:2022vvb,Wei:2023rzs,Vasilakis:2008yn,VasilakisThesis,Kornack:2004cs,KornackThesis,Brown:2010dt,BrownThesis,Xu:2023vfn,Jiang:2021dby,kwyg-5v64}). Optical magnetometers~\cite{BudkerRomalis_OM,BudkerJacksonBook} form a broad family of technologies that share a common principle: the use of light to measure atomic spins.
One common method is to pass a linearly polarized probe beam through the spin ensemble. The probe exits with a modified polarization due to the Faraday effect. Measuring this polarization change allows one to infer the orientation of the spins that interacted with the probe beam. This method is advantageous because it suppresses sensitivity to fluctuations in the intensity of the probe beam, which are typically much larger than the polarization instabilities of the probe beam. 
The PSN arises from photon-counting fluctuations in the probe beam. The LSN arises from fluctuations in its circular polarization, which act back on the spins through the AC Stark shift. The uncertainty in the measured polarization rotation is inversely correlated with the photon flux
$\Phi$ (photons per second) at the photodiodes as~\cite{BudkerRomalis_OM}
\begin{equation}
\delta \varphi \approx \frac{1}{2\sqrt{\Phi}},
\end{equation} 
with $\delta \varphi$ as the amplitude spectral density of the rotation angle, expressed in ${\rm rad}/\sqrt{\rm Hz}$. 
At fixed detuning, increasing the power of a continuous probe reduces the PSN but increases the LSN and the probe-induced relaxation rate. The latter shortens the spin coherence time, thereby increasing the field-referred SPN. At fixed power, larger detuning mitigates LSN and probe-induced relaxation but also reduces the optical response (the polarization rotation for a fixed transverse spin polarization). Thus, optimizing performance requires balancing PSN, LSN, SPN, and the optical response to maximize the magnetometer SNR. In an optimized system, the PSN and LSN contributions can be made comparable and subdominant to the SPN, which then dominates the total noise~\cite{PhysRevLett.95.063004}.

One of the primary types of optical magnetometers used is the alkali vapor magnetometer~\cite{Bloch:2019lcy,Bloch:2022kjm,Bloch:2021vnn,Gavilan-Martin:2024nlo,GNOME:2023rpz,Lee:2022vvb,Wei:2023rzs,Vasilakis:2008yn,VasilakisThesis,Kornack:2004cs,KornackThesis,Brown:2010dt,BrownThesis,Xu:2023vfn,Jiang:2021dby}, which typically use rubidium, potassium, or cesium (or combinations thereof). Their atomic structure allows the spins to be polarized and measured with near-IR lasers. Such devices often utilize axial magnetic fields in the $\mathcal{O}({\rm mG}-{\rm G})$ range. The response of alkali metals to magnetic fields is dominated by the electron in their outermost shell, making them possess a gyromagnetic ratio of similar magnitude to that of free electrons. Because this electron is coupled to the nuclear spin, spin-exchange collisions can move atoms between levels with different precession rates, shortening $T_2$ outside the SERF regime and making them relatively broadband magnetometers. The Spin-Exchange Relaxation-Free (SERF) regime~\cite{Allred2002} arises at high atomic densities and sufficiently low magnetic fields, where rapid collisions average over the different precession rates of these levels, suppressing this dephasing and allowing longer $T_2$s, sometimes of the order of milliseconds or longer.

To achieve high sensitivity to new physics effects, alkali magnetometers are often coupled to a second spin species, typically a noble-gas isotope with nonzero nuclear spin. The noble-gas magnetometer is often far narrower, far denser, and its spins possess a lower gyromagnetic ratio. These properties imply a far lower sensitivity to magnetic backgrounds, higher quality factors, and lower quantum noise. Such experiments are currently mostly limited by the magnetic backgrounds measured by the alkali metals (induced by imperfect shields), though some have reached the SPN of their alkali sample. Some quantum-enhanced strategies can in principle reduce the impact of SPN, but this has not yet been demonstrated in the context of particle physics searches.

Other types of optical magnetometers are also in use or have been proposed in the context of BSM physics. For example, it has been suggested that Nitrogen Vacancy center magnetometers could serve to search for axion-electron interactions~\cite{Chigusa:2023hms}. Furthermore, optical readout has been used in the measurement of ${}^{199}{\rm Hg}$, using the spin-dependent optical depth~\cite{Abel:2017rtm}. While technically also read out optically, spin-pendulums offer a very different sort of measurement: the interaction of DM with the spin is transduced to a physical torque which is measured to achieve sensitivity for axion-electron interactions~\cite{Terrano:2019clh}.

\subsubsection{Alternative Magnetometry Techniques}

Many additional technologies for magnetometers have been proposed, or have shown some results, and rely on different principles (even if they could technically be classified as optical magnetometers, or are read-out via SQUIDs). These include the usage of storage-rings~\cite{JEDI:2022hxa,Lenisaa:2025pxn} (with future projections for a dedicated experiment estimated in Ref.~\cite{Brandenstein:2022eif}), the use of interferometers~\cite{Crescini:2023zyl}, Levitated Ferromagnetic Torque Sensors~\cite{Ahrens:2024yzo,Vinante:2021fpj,Kalia:2024eml,Ji:2025yvn,Peng:2026ffa,Xu:2026jyl}, and more. 

 However, the primary alternative to optical readout is reading the spins with a secondary magnetometer~\cite{OldWind,Budker:2013hfa}. The usage of SQUIDs or other non-spin-based magnetometers to measure solid or liquid spin-ensembles which interact with DM was proposed over three decades ago~\cite{OldWind}, though implementations are far more recent~\cite{QUAX:2020adt,Walter:2025ldb}. As the density of solids and liquids is far larger than the vapors or gases used in many optical magnetometers, such avenues offer great promise, especially for frequencies above $\sim~{\rm kHz}$. In such cases, the first spin sample is used as a transducer that converts novel particle-physics signals into magnetic fields measured by the auxiliary magnetometer.

One notable example of an effort in this direction is the CASPEr collaboration. Split into two efforts, one focused on $g_{ann}$, and the other on $g_{d}$, the CASPEr collaboration is one of the largest collaborations in the field pushing in this direction. Their $g_d$ search, CASPEr-Electric, is one of the only existing proposals for probing QCD axions with masses below $10^{-10}{\rm eV}$~\cite{JacksonKimball:2017elr}. CASPEr-Electric~\cite{Graham:2013gfa,Budker:2013hfa}, plans to utilize spin-polarized ferroelectric crystals. The strong, effective internal electric field of such materials can be of the order of $100~{\rm MV/cm}$. However, the observable atomic EDM is suppressed relative to the nuclear EDM due to Schiff screening, whereby the rearrangement of the atomic electron cloud partially cancels the electric field generated by the nuclear EDM. As a result, the atomic EDM—particularly in light atoms—can be significantly smaller than the underlying nuclear EDM. The use of heavy nuclei helps mitigate this screening, increasing the observable atomic EDM relative to the nuclear EDM. However, even for heavy nuclei, CASPEr typically adopts a residual suppression factor of order $10^{-2}$ as a figure of merit in its sensitivity estimates.  In CASPEr-Electric, a magnetic field $B$ is applied, and the spins resonantly respond at a frequency $\gamma B$. Hence, by scanning the magnitude of the magnetic field, the DM search is planned to be sensitive between the range of $\mathcal{O}(100~{\rm Hz})$ and $\sim 100~{\rm MHz}$. The current prototype sensitivity can be found in Ref.~\cite{Aybas:2021nvn}, and has been shown at a frequency of $\sim 40~{\rm MHz}$, with a $1~{\rm MHz}$ band scanned around it. It is currently far less sensitive than the planned goals of the collaboration. The two most major changes that should improve the sensitivity are the utilization of a much larger sample and focusing on sub-MHz frequencies, where the $T_2$ of the sample is expected to improve compared to current performance~\cite{Aybas:2021nvn}. With all improvements implemented, CASPEr-Electric plans to target a QCD-axion DM scenario in the mass range of $\sim 10^{-12}{\rm eV}-10^{-9}~{\rm eV}$.

\subsection{Outlook}

Quantum magnetometry has become a versatile probe of particle physics, with sensitivity to ultralight dark matter, a broad class of spin-dependent interactions, and other beyond-the-Standard-Model scenarios. Although the underlying techniques have a long history in precision metrology, the past decade has seen significant growth in their application to fundamental physics, accompanied by a surge of innovative experimental concepts. Proposed strategies and novel detection schemes are expected to enhance sensitivities to new physics by orders of magnitude~\cite{Bloch:2024uqb,qin2024new,jiang2023enhanced,Boyers:2025qgc,Galanis:2025amc,Brandenstein:2022eif,JacksonKimball:2017elr,Kalia:2024eml}, promising access to highly motivated BSM models that are currently beyond reach.

%% file: sections/quantum-superconducting_updated.tex
\section{Superconducting quantum devices for light dark matter}
\label{sec:superconducting}
This chapter will discuss micro-fabricated superconducting devices and their application in particle physics, especially in dark matter searches. Nine decades have passed since the existence of dark matter was first hypothesized. Tremendous effort has gone into searching for the well-motivated classic candidates, weakly interacting massive particles (WIMPs) and QCD axions. While they are still being pursued, the community has also started to emphasize the direct search in a broader mass range, including light particle-like dark matter below $\SI{1}{\giga\electronvolt}$ and axion-like particles (ALPs) \cite{Ballmer:2022uxx}, which is boosted by recent advances in quantum sensing technologies\cite{Golwala_2022, Sikivie:2020zpn, Sushkov:2023fjw, Bass:2023hoi}. 

Direct light dark matter detection measures the small energy from dark matter scattering or absorption. Fermionic DM scatters with the target material through new mediators of unknown nature, and bosonic DM is directly absorbed by the target. The couplings can be to electron or nucleon numbers, mass, or spin \cite{Battaglieri:2017aum}. Detectors that are sensitive to the different types of interactions are favored. For this reason, athermal phonon sensors, which sense the energy deposition as phonons in target crystals \cite{Kahn:2021ttr, Knapen:2021bwg}, are one of the most developed technologies. Some crystals also have enhanced coupling to specific DM candidates. For example, polar crystals have optical phonon modes that resonantly couple to DM mediated through dark photons of the matched masses \cite{Knapen:2017ekk, Griffin:2018bjn, Griffin:2019mvc}. On the other hand, as another important signal channel, photon-sensitive detectors target the standard model photon converted from ALPs or dark photons \cite{Caputo:2021eaa, Sikivie:2020zpn}.  

Thermal noise sets the boundary between signal pulse detection and power measurement. Modern cryogenic system can easily achieve \SI{1}{\kelvin} (\SI{0.1}{\milli\electronvolt}) across cubic meter volumes, and as low as \SI{10}{\milli\kelvin} (\SI{1}{\micro\electronvolt}) across smaller volumes. Thus, in both phonon and photon signal channels, pulse identifying detectors should aim for energy threshold as low as \SI{0.1}{\milli\electronvolt}. This energy scale is sufficient to detect single optical phonon modes in crystals, accessing the benefit of enhanced energy transfer of scatterings through single-phonon processes \cite{Campbell-Deem:2022fqm}.
On the photon detection side, it covers the short wavelengths which significantly limit the effective volume of resonant cavities.  

Traditional ionizing or scintillating particle detectors do not respond in this condensed matter energy scale. Superconductors, on the other hand, have superconducting gaps of Cooper pair breaking on the scale of \SI{0.1}{\milli\electronvolt}, making them perfect sensor materials. Other novel detectors \cite{Chou:2023hcc} are also proposed, such as low-gap semiconductors \cite{Abbamonte:2025guf} and magnons \cite{Trickle:2019ovy}, which will not be covered here.

External energy breaks Cooper pairs and changes the quasiparticle (QP) density $n_\mathrm{qp}$ in superconducting devices. The changing $n_\mathrm{qp}$ results in perturbations in electrical properties in the superconducting circuits, which are used to convert the energy signal to an electrical signal. Depending on the type of detectors, the signal can be either sub-MHz (DC signal), or sub-MHz modulation signal on a GHz carrier frequency (RF signal), or as fast impulses with ps timing resolution. They correspond to transition edge sensors (TESs) \cite{Irwin2005}, kinetic inductance devices (KIDs) \cite{Zmuidzinas_2012}, and Qubits \cite{Ramanathan:2024hsf, Shaw_2009}, and superconducting nanowire single-photon detectors (SNSPDs) \cite{EsmaeilZadeh:2021gqi}, respectively. We will first give a short review of each type of device's operation principles and state-of-art development in section \ref{sec:SC-sensors}.

Pair-breaking is a thermal process that involves dissipation. Unfortunately, all the technologies are still far from resolving the energy levels of a single Cooper pair. Besides SNSPD, which is primarily only sensitive to photons, all the other devices are limited by the thermal noise of QPs. So, strictly speaking, they are not quantum-limited devices. However, their operation involves using amplifiers operating at the standard quantum limit, and these readout circuits utilize the nonlinearities in superconductors, including the kinetic inductance and the Josephson effects in superconducting weak links. We will discuss their applications as quantum-limited amplifiers and multiplexers in section \ref{sec:SC-amps}. 

Modern dilution refrigerators (DRs) can easily cool large samples to \SI{10}{\milli\kelvin} temperate and below, which should suppress the thermal noise in phonon and QP systems to negligible levels at signal energy scales, but many experiments have shown evidence that this is actually not the case. Various sources of excess QPs have been identified, including environmental and cosmic ionizing radiation, black body radiation from high temperature stages in the DR, and stress releasing from thermal contractions. We will discuss the limitations in realistic systems in section \ref{sec:SC_BKG}. 

\subsection{Pair-breaking superconducting sensors}
\label{sec:SC-sensors}

\begin{figure}
    \centering
    \includegraphics[width=\linewidth]{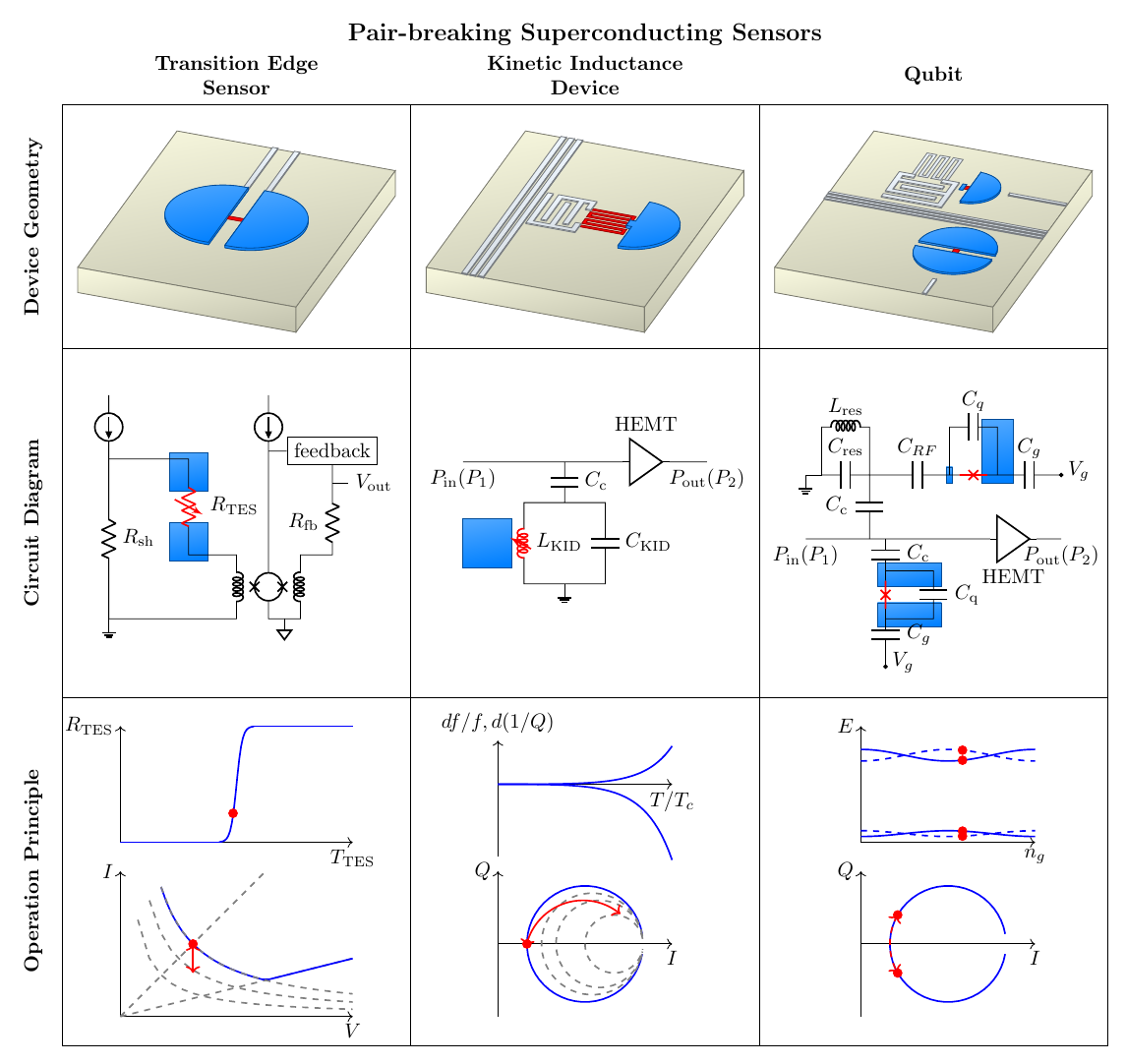}
    \caption{Three major phonon-sensitive pair-breaking superconducting sensors. The typical device geometries (top row), equivalent circuit diagram including the readout electronics (middle row), and the operation principles (bottom row) are shown for transition edge sensors (TESs, left column), kinetic inductance devices (KIDs, middle column), and Qubits (right column). The blue regions in the top and middle rows indicate the phonon-absorbing volume, red regions indicate the active sensing elements: a superconductor resistor biased on the transition edge in a TES; a superconducting inductor in a KID; a Josephson junction in a Qubit. The top diagrams in the operation principles demonstrate the operation status of the device, where the status of TESs and KIDs changes with respect to the device temperature and the status of Qubits jumps between the even and odd parity of $n_g$. Finally, The bottom diagrams demonstrate the measurement variables, current signals in the case of TESs and I-Q RF transmission signals in the case of KIDs and Qubits, and the corresponding behavior as an incident signal changes the sensor status. Details of each sensor are explained in the Sec.~\ref{sec:SC-sensors-TES} to \ref{sec:SC-sensors-Qubit}. Conceptual diagram based on \cite{Irwin2005, Zmuidzinas_2012, Ramanathan:2024hsf}.}
    \label{fig:SC-sensors}
\end{figure}

\subsubsection{Transition edge sensors}
\label{sec:SC-sensors-TES}
Transition edge sensors can be considered an extreme case of a thermistor, which operates on the steep normal-superconducting phase transition edge of a superconductor. The small temperature change due to a small energy injection is amplified into a significant change in the resistance. Biasing the TES with a constant voltage can create negative electrical-thermal feedback (ETF) that stabilizes the operation point on the transition edge. With the constant voltage bias, the resistance change is measured as a current signal, which requires low-impedance, low-noise current readout. This is satisfied by superconducting quantum interference device (SQUID) current amplifiers, which will be introduced later in Sec.~\ref{sec:SC-SQUID}. 

The negative ETF has an intuitive picture. Modern TESs are thin superconducting films patterned into micrometer-scale resistors on substrates. With a constant voltage bias, Joule heating from the bias raises the TES temperature to $T_0$ above the substrate temperature $T_b$. $T_0$ is in the transition region around the true critical temperature $T_c$ of the superconductor. When the active heating matches the heat dissipation, the system reaches a steady-state. Now, if an external energy signal $\delta E$ perturbs the TES temperature, it will increase TES resistance, causing the current through TES to decrease, and reducing the heating power. Heat dissipation exceeds the heating power, preventing the TES temperature from rising. Eventually, as the external signal is removed, the TES recovers the steady-state bias point. To the leading order, the time integral of the reduced heating power equals to $\delta E$. 

The constant voltage bias is realized by connecting a small cold resistor relative to the TES normal resistance in parallel with the TES, as shown in fig.~\ref{fig:SC-sensors}. 
A stable ETF requires the circuit's electrical time constant $L/R_0$ to be faster than the thermalization time with ETF $\tau_\mathrm{ETF} \sim C_\mathrm{TES}/G\alpha$, where $C_\mathrm{TES}$ is the heat capacity of the TES, proportional to TES volume $V_\mathrm{TES}$ and $T_0$, $G$ is the thermal conductance from TES to the thermal bath, and $\alpha \equiv \frac{T_0}{R_0}\frac{\partial R}{\partial T}|_{I_0}$ is the steepness of the transition edge under constant current. The exact stability criteria also depend on the steepness of the transition edge, a comprehensive derivation can be found in \cite{DeLucia:2024sxp}.

In CMB and astronomy applications, the TESs are exposed to IR photons from the sky, micro-Watt optical power prevents the detectors from running far below \SI{100}{\milli\kelvin}. Thus, materials with $T_c>\SI{100}{\milli\kelvin}$ are chosen, such as Al/Mn, Hf and Ir-Pt bilayers \cite{Singh:2022rck}. The TESs are fabricated on suspended SiN membranes to reduce $G$, slowing down the TES for stability, and reducing the TFN at the same time.

In particle dark matter detectors, the signals propagate as phonons in the substrate. TES has to be fabricated directly on the substrate to achieve high signal collection efficiency. Low-$T_c$ superconducting films are chosen to reduce $G$ and stabilize ETF. The heat dissipation is limited by electron-phonon coupling, which scales as $P=\Sigma V_\mathrm{TES} (T_0^5-T_b^5)$ \cite{Wellstood_1994}, where $\Sigma$ is a material-dependent constant. The thermal conductance at the bias point scales as $G\equiv \mathrm{d}P/\mathrm{d}T_0 \propto 5 \Sigma V_\mathrm{TES}T_0^4$. 

The fundamental noise source in a TES is the thermal fluctuation noise (TFN). It is an inevitable noise due to the fluctuation in the thermal conductance between $T_0$ and $T_b$. The noise power scales as 
\begin{equation}
    S_{P_\mathrm{TFN}} = 4k_\mathrm{B}T_0^2GF(T_0,T_b)
\end{equation}
where $F$ is a factor close to $1/2$ when $T_b\ll T_0$.
Other noise sources include the TES and shunt resistor current noise, and the amplifier noise. When TFN dominates, the energy resolution of the TES is proportional to the TFN power integrated over the signal bandwidth, i.e.
\begin{equation}
    \sigma_\mathrm{TES} \propto \left(S_{P_\mathrm{TFN}}\tau_\mathrm{ETF}\right)^{1/2} 
    \propto(T_0^2C_\mathrm{TES}/\alpha)^{1/2}
    \propto V_\mathrm{TES}^{1/2} T_c^{3/2} \alpha^{-1/2}
\end{equation}
Designing TESs with a small volume of low-$T_c$ sharp-transition material significantly improves the resolution, while the low bias power poses a challenging requirement for low parasitic noise power. 

Another challenge is to collect phonon signals efficiently. Athermal phonon decays quickly at high energies as the decay rate is proportional to $E_\mathrm{ph}^5$ \cite{Harrelson:2021fyd}. Signal energy will be invisible once phonons decay to energies below the superconducting gap. One may increase the phonon collection area by increasing the TES surface coverage, but $V_\mathrm{TES}$ and $G$ are increased simultaneously causing higher noise. The QP-trapping technique breaks the constraints \cite{Irwin:1995kk}, where a high $T_c$ phonon absorber, usually aluminum, collects phonon and convert the energy into excited QPs, then the QPs will be trapped into the lower $T_c$ TES, for example tungsten, which is in contact with the phonon absorber. The world-leading experiments all use similar structures \cite{TESSERACT:2025tfw, SuperCDMS:2020aus, CRESST:2022lqw}, and the best achieved energy threshold in phonon systems on a \SI{0.23}{\gram} detector is as low as \SI{1.5}{\electronvolt} \cite{TESSERACT:2025tfw}. The excellent threshold is achieved by optimizing the surface phonon absorber and TES surface coverage fraction such that the phonon collection time matches the TES bandwidth \cite{Fink:2022rme, Golwala_2022}. With the optimized bandwidth, meaning $\tau_\mathrm{ETF}$ is fixed to match the phonon collection time while $T_c$ changes, the energy resolution of the detector scales as
\begin{equation}
    \sigma \propto V_\mathrm{det}^{1/2} T_c^3 
\end{equation}
where $V_\mathrm{det}$ is the detector substrate volume instead of the TES volume. This scaling did not account for non-TFN noises. In the most recent development, as the TES is being optimized, phonon shot noise from low energy excess (LEE) events has become the limiting factor \cite{Anthony-Petersen:2024vdh, TESSERACT:2025odn}.

Other configurations use the TES material as the target for optical photons \cite{Chen:2025cvl, Schwemmbauer:2024jel} and $\beta$ particles\cite{PTOLEMY:2018jst}, or using gold wires to conduct the heat from targets to the TES\cite{Billard:2022cqd, Ricochet:2023yek}.

\subsubsection{Kinetic inductance devices}
\label{sec:SC-sensors-KID}
The conductance of superconductors diverges at zero frequency, while it remains finite at non-zero frequencies. Both the real and imaginary components depend on the density of QPs in the superconductor. The imaginary dependence, which can be effectively treated as an additional contribution to the inductance, is called the kinetic inductance effect. Superconducting micro resonators that exploit this effect can be engineered into highly sensitive sensors, which are referred to as kinetic inductance devices (KIDs), or more explicitly microwave kinetic inductance devices (MKIDs). 

Incident energy signals are detected by measuring the impedance change of KIDs. Usually, one use a probing radio-frequency (RF) tone, $f$, fixed to the resonant frequency $f_r$ at zero temperature, and monitor the amplitude and phase change in the RF transmission, $S_{21}$, or reflection, $S_{11}$. Tone tracking and feedback techniques are also developed \cite{Henderson:2018jlc, Rouble_2024}, but we will not cover the details. Transmission measurement is preferred to multiplex KIDs coupled to the same feed line. As $f$ scans across $f_r$, $S_{21}$ traces out a circle on the complex plane 
\begin{equation}
    S_{21} = 1-\frac{Q_r/Q_c}{1+2i Q_r(1-f/f_r)}
\end{equation}
where $Q_r$ and $Q_c$ are the total and coupling quality factors, respectively. Quality factors represent the ratio of the total stored energy to the energy lost to each port per period. The total loss is the sum of the internal loss and the power escaping through the coupling port, i.e., $Q_r^{-1} = Q_i^{-1}+Q_c^{-1}$, where $Q_i$ is the internal quality factor. 

The perturbation $\delta S_{21}$ is composed of changes from $f_r$ and $Q_i$ as the signal energy $\delta E$ changes the QP density $n_\mathrm{qp}$ in the inductor of the KID
\begin{equation}
    \delta S_\mathrm{21} = \left[\frac{\partial S_{21}}{\partial Q_i}\frac{d Q_i}{d n_\mathrm{qp}} + \frac{\partial S_{21}}{\partial f_r}\frac{d f_r}{d n_\mathrm{qp}}\right]\frac{d n_\mathrm{qp}}{dE} \delta E
\end{equation}
In the simple case when the probing tone $f$ is tuned to the quiescent $f_r$, the change through $Q_i$ is purely real, and the change through $f_r$ is purely imaginary. They are commonly referred to as dissipation and phase signals, respectively. They both scale with the change in QP density. 

The behavior of $d Q_i/d n_\mathrm{qp}$ and $d f_r/d n_\mathrm{qp}$ can be derived with the Mattis-Bardeen model \cite{Mattis:1958yid}, with the assumption that the QPs are thermally distributed regardless of thermal or external energy excitation. The derivation is detailed in \cite{Zmuidzinas_2012}, here we present the result:
\begin{equation}
    \delta S_\mathrm{21} = \frac{Q_i Q_c}{(Q_i+Q_c)^2}\frac{Q_i}{Q_\mathrm{qp}(f,T)n_\mathrm{qp}}[1-i\beta(f,T)]\frac{\delta N_\mathrm{qp}}{V_L}
    \label{eq:SC_S21response}
\end{equation}
where $Q_\mathrm{qp}$ represent the loss to the QP system, $\beta$ is the ratio between phase and dissipation signal, $\delta N_\mathrm{qp}$ is the number of QP excited by $\delta E$, and $V_L$ is the inductor volume of the KID. 

The response is proportional to $1/Q_\mathrm{qp}n_\mathrm{qp}$, which can be further break in to design parameters, material parameters, and general M-B theory dependence, writing as 
\begin{equation}
    \frac{1}{Q_\mathrm{qp}n_\mathrm{qp}} = \frac{\alpha\gamma}{2N_0\Delta_0}S_1(f,T)
    \label{eq:SC_Qqpnqp}
\end{equation}
where $\alpha$ is the kinetic inductance fraction, $\gamma$ is a material constant ranging from 1 to 1/3 depending on the type of the superconductor, $N_0$ is the single-spin electron density of state at the Fermi energy, $\Delta_0$ is the superconducting gap, and $S_1(f,T)$ is a general material-independent function dictated by the M-B theory \cite{Zmuidzinas_2012}. Both $\delta N_\mathrm{qp}$ and $\Delta_0^{-1}$ scales with $T_c^{-1}$, $S_1(f,T)$ has a dependence on $T_c^{1/2}$ given $hf$ and $k_\mathrm{B}T$ both are much less than $\Delta_0$. Thus, the dissipation response scales with $T_c^{-3/2}$. The phase response pick up an additional $T_c^{-1/2}$ from $\beta$ and it scales with $T_c^{-2}$. 

From the other factors in Eq.~\ref{eq:SC_S21response}, we can see that $\delta S_\mathrm{21}$ is maximized first with $Q_c$ matched to $Q_i$. Under this condition, $\delta S_\mathrm{21}\propto Q_i$, and $Q_i$ can be as high as $Q_\mathrm{qp}$, when the dielectric loss is negligible comparing to the loss to QP system. Since $Q_\mathrm{qp}n_\mathrm{qp}$ is a constant for a given KID design and an operation temperature, we see that $\delta S_\mathrm{21}\propto n_\mathrm{qp}^{-1}$. Various experiments\cite{Zmuidzinas_2012, Cardani:2020vvp, Mannila:2021dkm} observe the divergence of the QP density from the $e^{-\Delta/k_\mathrm{B} T}$ prediction by BCS theory at low temperatures. Lowering the operation temperature does not guarantee an exponentially growing signal, and understanding the origin of the residual QP density is critical. Finally, $\delta S_\mathrm{21}$ scales with $V_{L}^{-1}$, and same as TESs, reducing the inductor volume increases responsivity.

Now we consider the noises. The fundamental noise is caused by the QP number fluctuations due to generation and recombination, the GR noise, which directly change $\delta N_\mathrm{qp}$ in Eq.~\ref{eq:SC_S21response}. Again, following \cite{Zmuidzinas_2012}, the noise equivalent power spectrum depends on the excitation rate and the QP lifetime $\tau_\mathrm{qp}$
\begin{equation}
    S_{P}^\mathrm{GR}(\nu) = \frac{1}{1+(2\pi\nu\tau_\mathrm{qp})^2}\frac{4\Delta_0^2}{\eta_s^2}
    \left[ \frac{Q_i}{Q_\mathrm{qp}}\frac{\eta_aP_a}{\Delta_0} + \Gamma_\mathrm{th} +\Gamma_\mathrm{qp}\right]
\end{equation}
where $\nu$ is the signal frequency, $\eta_s$ is the efficiency of the signal energy to break Cooper pairs\footnote{Reference \cite{Zmuidzinas_2012} uses notation $\eta_o$ since the signal is optical photon in the context. Here the signal is generally defined, including phonons.}. The first term in the brackets correspond to excitation by readout power, where $P_a$ is the absorbed readout power and $\eta_a$ is its efficiency to break Cooper pairs. The second term $\Gamma_\mathrm{th} = \frac{1}{2}N_\mathrm{th}(\tau_\mathrm{max}^{-1}+\tau_\mathrm{th}^{-1})$ represents the thermal excitation rate of QP. $N_\mathrm{th}$ and $\tau_\mathrm{th}$ are the BCS-predicted QP number and lifetime, respectively, assuming no external excitations. $\tau_\mathrm{max}$ is the maximum QP lifetime at low temperatures when the residual QP density diverges from the BCS prediction. The last term $\Gamma_\mathrm{qp} = \frac{1}{2}N_\mathrm{qp}(\tau_\mathrm{max}^{-1}+\tau_\mathrm{qp}^{-1})$ represents the recombination rate. At low temperatures, the thermal excitation rate is negligible, GR noise is dominated by readout and other parasitic excitation powers. 

The second noise source is the amplifier noise. Traditionally, KIDs are readout by high electron mobility transistor (HEMT) amplifiers, which have a noise temperature around \SI{1}{\kelvin}. Superconducting parametric amplifiers can limit the noise to the standard quantum limit, i.e., adding only $1/2$ photon noise in the resonator bandwidth. Details of parametric amplifiers will be discussed in Sec.~\ref{sec:SC-Paramp}. In the case of using HEMT amplifiers with noise temperature $T_\mathrm{a}$, the noise equivalent power (NEP) spectrum of the dissipation signal is 
\begin{equation}
    \begin{split}
    S_{P}^\mathrm{a,diss} (\nu) =  & \frac{1}{1+(2\pi\nu\tau_\mathrm{qp})^2}\frac{1}{1+(2Q_r\nu/f_r)^2} \\
     &\frac{2\Delta_0^2}{\eta_s^2} \frac{n_\mathrm{qp}^2Q_\mathrm{qp}^2V_L^2}{Q_i^2} \frac{(Q_i+Q_c)^2}{Q_iQ_c} \frac{k_\mathrm{B}T_a}{P_a\tau_\mathrm{qp}^2}
     \end{split}
\end{equation}
and the phase signal has an additional suppression by $\beta$
\begin{equation}
    S_{P}^\mathrm{a,phase} (\nu) = S_{P}^\mathrm{a,diss} (\nu)/\beta^2
\end{equation}
Although it is clearly advantageous to use the phase readout to suppress the amplifier noise, another noise source, the two level system (TLS) noise, dominates the phase signal channel. Details will be discussed in Sec.~\ref{sec:SC_TLS}. 
The amplifier NEP is minimized when $Q_c$ matches $Q_i$, and it scales differently in the phase and the dissipation channels,  $S_{P}^\mathrm{a,diss}\propto\Delta_0^3Q_i^{-2}V_L^{2}P_a^{-1}\tau_\mathrm{qp}^{-2}$, and $S_{P}^\mathrm{a,phase}\propto\Delta_0^4Q_i^{-2}V_L^{2}P_a^{-1}\tau_\mathrm{qp}^{-2}$ \footnote{The additional factors of $\Delta_0$ couple in through $Q_\mathrm{qp}n_\mathrm{qp}$ and $\beta$, previously discussed after Eq.~\ref{eq:SC_Qqpnqp}}. In contrast to GR noise, a higher readout power is preferred to suppress amplifier noise.

Given a fixed amplifier noise temperature, the best energy resolution is achieved as the readout power increased to a point that the GR noise starts to dominate. It is possible that the KID becomes nonlinear as readout power increases, even before the GR noise dominates. There are two mechanisms of nonlinearity, one is due to the nonlinear kinetic inductance as the RF current approaching the critical current of the superconductor \cite{Zmuidzinas_2012, Swenson:2013jsa}, the other is due to $P_a$ exceeding the heat dissipation from QP to the substrate and rising the QP temperature \cite{de_Visser_2010, Wandui:2020xzv}. 

The first mechanism is non-dissipative, which also results in the parametric amplification. Carefully choosing detuning of the probing tone and power results in positive feedback that improves the signal to noise ratio \cite{Swenson:2013jsa}. The nonlinearity threshold of the internally stored power in the KID, $P_{\mathrm{th,int}}$, scales with $V_L\Delta_0^2Q_r^{-1}$ \cite{deRooij:2025lil}. Note that the stored power is $Q_i$ times the dissipation power, $P_\mathrm{th,int} = P_aQ_i$. In this scenario, the $S_{P}^\mathrm{a,diss}$ ($S_{P}^\mathrm{a,phase}$) scales as $V_L\Delta_0\tau_\mathrm{qp}^{-2}$ ($V_L\Delta_0^2\tau_\mathrm{qp}^{-2}$) before nonlinearity develops. Reducing the inductor volume will significantly improve resolution. QP density inversely relates to $\tau_\mathrm{qp}$, thus even when amplifier noise dominates over GR noise, it is still important to reduce excitation and increase QP lifetime. 

The second mechanism is dissipative, and the heat conduction is limited by the QP-phonon interaction for materials\cite{de_Visser_2010} with low-$T_c$ below about \SI{1}{\kelvin}. KID operated in this regime has thermal-electrical feedback as in TESs. But the elevated power usually significantly reduces $Q_i$ and resulting in worse resolution. The threshold power in this regime, $P_{\mathrm{th},a}$, scales as $P_{\mathrm{th},a} \propto V_L(T_c^5-T_b^5)$, same as TESs. 
%$P_{\mathrm{th},a}$ refers to the power absorbed in the resonator ($P_a$). 
If limited by dissipation, the $S_{P}^\mathrm{a,diss}$ ($S_{P}^\mathrm{a,phase}$)  scales as $V_L\Delta_0^{-2}Q_i^{-2}\tau_\mathrm{qp}^{-2}$ ($V_L\Delta_0^{-1}Q_i^{-2}\tau_\mathrm{qp}^{-2}$). Thus, for low-$T_c$ KIDs, using parametric amplifier to evade the amplifier noise is more important. 

In conclusion, a general rule to improve KID resolution is to reduce volume and QP number density. Although excellent resolution to photons has been demonstrated in sub-micrometer scale aluminum KIDs \cite{Day_2024}, it is hard to achieve high phonon resolution as the reduced inductor volume significantly reduces phonon collection efficiency. The RF circuitry adds complexity in the design comparing to TESs. Different approaches are considered, including isolation of the target crystal \cite{Cruciani:2022mbb}, and the QP-trapping assisted design as in TESs \cite{Golwala_2008}. Efficient QP trapping requires high QP diffusion distance in the phonon collector and a low-$T_c$ sensitive film \cite{Kaplan:1976zz}. With aluminum being the best choice of phonon collectors, KIDs fabricated from materials with $T_c$ well below \SI{1.2}{\kelvin}, around $O(100)\si{\milli\kelvin}$, \cite{Mazin_2022, Li:2025jil} are required. 

\subsubsection{Qubits as sensors}
\label{sec:SC-sensors-Qubit}
Superconducting Qubits are resonators that use Josephson junctions (JJs) to introduce nonlinearities, resulting in unevenly spaced energy levels. Due to this anharmonicity, the system can be effectively confined to the two lowest energy levels, typically denoted as $\ket{0}$ and $\ket{1}$, which form the basis for computation and quantum sensing. 

Couplings to the environment disrupt the delicate quantum states of Qubits. While such environmental interactions pose significant challenges for quantum computing, they simultaneously enable Qubits to function as high-sensitivity, low-threshold detectors. Understanding the sources of perturbations, the pathway of coupling, and their effects on Qubit states is essential for developing Qubit-based detectors. Depending on the signal channel, for example, phonon signals from light DM scattering and photon signals from axion haloscopes, Qubits can be engineered to preferentially couple to the specific channel and reject other perturbations. This knowledge is equally critical for isolating Qubits from environmental noise—a prerequisite for realizing robust quantum error correction and scalable quantum computing. More generally, same perturbation sources also couples to TESs and KIDs, and Sec.~\ref{sec:SC_BKG} will discuss the details.

Phonon or photon signals from dark matter can break Cooper pairs and excite QPs in a Qubit. Cooper pair box (CPB) and transmon Qubits are sensitive to charge distributions, as revealed in the Hamiltonian \cite{Koch:2007hay}. 
\begin{equation}
    \hat{H} = 4E_C(\hat{n}-n_g)^2-E_J\cos{\hat{\varphi}}
    \label{eq:H_CPB}
\end{equation}
The charge energy $E_C=e^2/2C_\Sigma$ is the energy required to shift one charge to the superconductor island on one side of the junction, and $C_\Sigma$ is the total capacitance of the island to the environment. The Josephson energy $E_J=h\Delta/8R_Ne^2$ is the energy required for one Cooper-pair to tunnel through the junction, and $R_N$ is the normal resistance of the junction. $\hat{n}$ measures the number of cooper-pairs on the island and $\hat{\varphi}$ measures the phase difference across the junction. Finally, $n_g$ is the number of induced charge pairs on the island and its change is detected as signal.
Tunneling of a single QP changes the parity and changes $n_g$ by $\pm1/2$, tunneling of a Cooper pair has no effect. 

The difference between CPBs and transmon Qubits is the ratio between $E_J$ and $E_C$. CPBs have $E_J/E_C<1$, where the charge energy dominates and the charge dispersion is as large as the energy level separations. To achieve low $C_\Sigma$, the design is limited to junctions with asymmetrical conductors on the two sides. The side that couples to the readout has to be small to reduce $C_\Sigma$, while the other side needs to be larger to absorb signal. transmon Qubits, more precisely, the offset charge sensitive (OCS) transmons \cite{Serniak:2019one, Ramanathan:2024hsf}, has $1\ll E_J/E_C\lessapprox50$, and the relaxed requirement of $C_\Sigma$ allows symmetrical designs to improve phonon collection efficiency. 

The parity can be measured in multiple ways. It can be measured as a change in the transition energy $E_{01}$, which is \cite{Fink_2024, Koch:2007hay}
\begin{equation}
    E_{01} \approx \hbar\omega_0+\hbar\chi_0\cos(2\pi n_g)
\end{equation}
where $\hbar\omega_0\approx\sqrt{8E_CE_J}-E_C$, and $\chi_0$ is the charge dispersion that also depends on $E_C$ and $E_J$.

The most straightforward way is to readout the transition frequency $E_{01}/h$ by coupling the Qubit directly to the feed line and prob at a frequency close to $\omega_0/2\pi$. One photon will be absorbed as the Qubit being excited from $\ket{0}$ to $\ket{1}$, which shows as a reduction in the transmission at the probing frequency, similar to reading out a KID with two alternating resonate frequencies \cite{Fink_2024}. The penalty of this method is that the coupling is limited to single-photon level, which adds difficulty to achieve high fidelity in resolving the parity states. Quantum-limited amplifiers are required to achieve \SI{99}{\percent} fidelity with \SI{100}{\kilo\hertz} readout bandwidth.

It can also be measured as a frequency shift in the coupled resonator in the dispersive limit. The additional readout resonator allows coupling to a stronger readout power to improve signal to noise ratio. In CPB type of detectors, the shift is calculated as an quantum capacitance shift \cite{Shaw_2009} of the Qubit island that loaded the readout resonator. For OCS type of detector, the frequency shift is protected by the opposite contributions from the $\ket{0}$ and $\ket{1}$ states, but amplified by the stronger coupling, thus giving a similar level of shift as CPBs \cite{Koch:2007hay}. In both types of detectors, direct readout of the charge parity with high fidelity is possible \cite{Ramanathan:2024hsf} and has been demonstrated \cite{Serniak:2019one}.

Finally, one can also use standard quantum computing techniques, such as the Ramsey sequence to map the even and odd states to $\ket{0}$ and $\ket{1}$ \cite{Riste:2012yem}. It offers high fidelity and speed. Microsecond level sampling speed has been demonstrated \cite{Li:2024dpf}, which allows the measurement of up to \SI{100}{\kilo\hertz} tunneling rates. Random Qubit de-excitations and excitations will also be recognized as tunneling events, and a reset pulse could be used to ensure the initial state is $\ket{0}$ \cite{Riste:2012yem}.

High fidelity is achievable with all the readout method mentioned above.
Thus, the detectors can be treated as a counting device of QP tunneling. The tunneling rates are proportional to the QP density around the junction. In CPB detectors, only one charge can tunnel onto the small island at a time and the asymmetry design causes different tunneling rates in the two directions. Practically, instead of resolving each fast tunneling incident, an average phase shift that is proportional to the tunneling rate can be measured with a slower readout \cite{Shaw_2009}. In OCS detectors, the rates in two directions are equal, which follows \cite{Ramanathan:2024hsf}
\begin{equation}
    \Gamma \approx \frac{16E_J\sqrt{k_\mathrm{B}T}}{\sqrt{2\pi}N_0\Delta^{3/2} h} n_\mathrm{qp} \equiv Kn_\mathrm{qp}
\end{equation}
where $N_0$ is the single-spin electron density of state at the Fermi energy, and $h$ is the Planck constant. 

Rate measurements are subject to Poisson fluctuations of the number of tunnelings in a given sampling time window $\Delta t$. This is the fundamental noise of the Qubit detectors. The excess QP density contributes to a background rate $2 K n_\mathrm{qp}$. To resolve a signal excitation, the measured tunneling events must increase by more than $5\sigma$ of the Poisson distributed background, $5\sqrt{2 K n_\mathrm{qp} \Delta t}$. Matching $\Delta t$ to the signal bandwidth helps to improve the SNR. Note that this is not degenerate from the GR noise, although both of them are related to excess QP backgrounds. GR noise is a result of fluctuations in $n_\mathrm{qp}$, and it adds on top of the Poisson fluctuations. 

Once again, we see the importance of reducing background QP density.
In theory, as the thermal QPs being suppressed at low temperatures, Qubit can sense a single Cooper pair breaking, which is much more sensitive than TESs and KIDs. A single phonon created from a \SI{10}{\kilo\electronvolt} light dark matter scattering can be detected. But in reality, the residue QP density contributes a background tunneling rate at least on the order of $O(10)\sim O(1)\si{\hertz}$\cite{Yelton:2025wsy}, which is the main background of a Qubit sensor. 

Other noise sources include TLS, amplifier noise, and charge jumps. TLS and amplifier noise not only affect the fidelity of the charge parity identification, they also affect Qubit coherence and more details will be discussed in Sec.~\ref{sec:SC_BKG}. Charge jumps are due to charge fluctuations of $n_g$ induced by external electric filed changes. It can be coupled through the DC biasing electrode that tunes $n_g$, or free charges in the substrate created by ionizing radiations.

Realizations of Qubit detectors are being actively pursued. CPB detectors have demonstrated sensitivity to single \SI{1.5}{\tera\hertz} photons \cite{Echternach:2017vtf}. Phonon sensing is more difficult, especially limited by the finite junction volume. QP trapping is necessary to achieve meaningful phonon collection efficiencies. In addition to increasing the collection area, QP trapping also confines the signal QPs in a small volume around the junction, increasing the probability of multiple tunnelings of a single QP, namely making a superconducting quasiparticle-amplifying transmon (SQUAT) \cite{Fink_2024}. A fabrication process for low-$T_c$ junction material for QP trapping is required. Hf junction can be a good candidate \cite{Morohashi_1992}. 
 
Many works originated from the quantum computing field has demonstrated sensitivity to radioactivity and cosmic rays in Qubit chips, but in order to do dark matter search, a proper energy calibration to calibrate the sensitivity of the chips needs to be demonstrated. Recent work exercised the energy scale modeling with phonon and QP simulations in a quantum computing chip \cite{Linehan:2025suv}. On the other hand, Qubits purposefully designed to be sensors are under development. 

\subsubsection{Other Josephson junction based devices}
Superconducting tunnel junctions (STJs) are large area superconductor-insulator-superconductor junctions that detect QP excitations through measuring the tunneling current. It is popular as X-ray spectrometers for the excellent energy resolution as low as \SI{12}{\electronvolt} full-width half maximum at \SI{6}{\kilo\electronvolt} \cite{Friedrich_2008, Kurakado_2000}.  

GHz photons can assist tunneling of sub-gap QPs to cross the junction. Ultra-sensitive single GHz photon counting has been demonstrated with this principle \cite{Pankratov:2024kdv}. 
The extra energy of the tunneling QP can emit multiple photons to achieve microwave photon multiplication \cite{Albert:2023wgy}. 
One can also use quantum interference effects to coherently convert $O(100)\si{\mega\hertz}$ photons to the standard \si{\giga\hertz} frequency range for quantum technologies. The devices are referred to as radio frequency quantum upconverters (RQUs) \cite{Chou:2023hcc}. 
 These techniques are specifically useful for axion detection in the $O(0.1)\sim O(1)$\si{\giga\hertz} range \cite{braggio2025quantum, Kuenstner:2022gyc} and approaching the GUT-scale \cite{DMRadio:2022jfv}.

In addition to photon sensing, single-phonon counting through coherent phonon-Qubit state swap is also proposed for sub-GeV light dark matter detection \cite{Linehan:2024btp}.

\subsubsection{Superconducting nanowire single-photon detectors}
Superconducting nanowire single-photon detectors (SNSPDs) detect individual photons that have energy above its threshold. SNSPDs operate with a DC current bias close to the critical current. When it absorbs an energetic photon, the broken Cooper pairs will create a local hot-spot. The supercurrent is repelled by the hot-spot as it expends and thermalizes, until the full nanowire section becomes normal. The current will be shunted to the readout circuit, creating a signal. Finally, soon after the current is removed, the normal section cools to superconducting states, and the current resets \cite{EsmaeilZadeh:2021gqi}. 

Unlike the sensors mentioned above, SNSPDs are digital counting devices and they can not resolve the energy of incident photons. The hot-spot formation requires localized energy deposition, so it is primarily used as photon detectors, for example, for terahertz axion detection~\cite{BREAD:2021tpx, Chiles:2021gxk}. As photon detectors, SNSPDs have achieved remarkable performance, including dark count rates below \SI{6.25e-3}{\per\second\per\square\centi\meter} \cite{Chiles:2021gxk}, thresholds below \SI{43}{\milli\electronvolt} \cite{Cai_2019}, picoseconds level timing jitter \cite{Korzh:2018oqv}, near-unity quantum efficiencies \cite{Reddy:2020usi} and high count rates \cite{Craiciu:2022akn}.
Recent works also demonstrated the possibility to look for direct scattering of sub-MeV dark matter detector in the nano-scale superconducting target, which takes advantage of the sub-eV threshold and low dark count rates \cite{Hochberg:2021yud}. The immunity to phonons from the substrate avoids the low-energy excess background that is present in athermal phonon detectors, but scaling the detectors to achieve sensitivities below the cosmological bounds is challenging. Multiplexing large arrays with kinetic inductance parametric up-converter (KPUP) is a promising solution \cite{Sypkens:2024oyx}.
The extremely low time jitter, high count rate, and the radiation hardness to be expected also make SNSPD a promising candidate for future collider experiments \cite{Chou:2023hcc, Lee:2023brm}. 

\subsection{Superconducting devices as quantum limited amplifiers and multiplexers}
\label{sec:SC-amps}
\subsubsection{SQUID}
\label{sec:SC-SQUID}
Superconducting quantum interference devices (SQUIDs) are Josephson junction-based magnetic field sensing devices. They utilize the flux quantization of magnetic field and the Josephson effect in superconductors to amplify small changes in magnetic field into voltage signals. 

There are two configurations. The first type of SQUID connects two JJs in parallel forming a superconducting loop \cite{Jaklevic:1964ysq}. The two JJs are biased with a constant current $I_b$, which is divided evenly between the two junctions when no external flux is present in the loop. As $I_b$ exceeds twice the critical current, $I_c$, of the JJ, a voltage difference across the junction appears. When an external magnetic field is applied, since the flux is quantized, a screening current is induced in the loop and cancels the external field. The screening current changes the distribution of $I_b$, leading to changes in the voltage across junctions. As the external field exceeds half of the flux quanta $\Phi_0=h/2e$, instead of canceling the field, the screening current adds to the field and increases the flux to $\Phi_0$. Thus, under a constant $I_b$, the current distribution hence the voltage drop oscillates with the flux of external magnetic field with a period of $\Phi_0$. SQUIDs in this configuration are usually designed to sense low-frequency signals below a few \si{\mega\hertz}, referred as dc-SQUIDs. 

The signal can be a changing magnetic field. A pick-up loop converts the field to an induced current, then feeds the signal to the vicinity of the SQUID loop. Applications include metallic magnetic calorimeter (MMC) readout \cite{Krantz:2023vpe} for light particle-like dark matter search \cite{vonKrosigk:2022vnf, Kim:2024xea}, nuclear magnetic resonance (NMR) readout for axion search \cite{Budker:2013hfa, Garcon:2017ixh}, and lumped LC resonator axion searches \cite{Sikivie:2013laa, Salemi:2021gck}.

The signal can also be a varying current. The current will run through the input coil which produces and threads magnetic field into the SQUID loop. It is particularly important for applications where a low input impedance preamplifier is needed. The most relevant use case is to readout TESs \cite{DeLucia:2024sxp} for light particle-like dark matter search \cite{CRESST:2022lqw, TESSERACT:2025tfw, SuperCDMS:2020aus}. 

Practically, a shunt resistor is added to each junction to remove the hysteresis in a dc-SQUID \cite{Clarke_1976}. The Johnson noise of the shunt resistors is the limiting noise source of dc-SQUIDs \cite{Tesche_1977}. Theoretically, it is possible to achieve quantum limited noise amplification by optimizing the SQUID parameters and operate at sufficiently low temperatures. The minimum noise is achieved with $2LI_c/\Phi_0\approx1$ \cite{Tesche_1977}, where $L$ is the inductance of the SQUID loop. Near-quantum limit operation has been demonstrated in a \SI{500}{\mega\hertz} high frequency dc-SQUID \cite{M_ck_2001}. Similar devices optimized for frequencies as high as \SI{7.4}{\giga\hertz} were also demonstrated \cite{M_ck_2003}. These devices are particularly interesting for axion searches with cavities \cite{Bradley:2003kg, Sikivie:2020zpn}. 

The second type consists of a single JJ in a superconducting loop \cite{Silver:1967zz}, which is inductively coupled to a readout resonating circuit at \si{\mega\hertz} to \si{\giga\hertz}, thus called rf-SQUID. When the flux induced by the RF current is less than a flux quanta, the rf-SQUID behaves like a parametric inductance that changes the resonant frequency of the readout resonator with the applied flux in the rf-SQUID loop \cite{Hansma_1973, Kleiner_2004}. It is especially useful for multiplexed readout of large sensor arrays. DC current signals from sensors can be coupled to different rf-SQUID loops, then couple to readout resonators at different frequencies and being read out by a single coax cable. Great progress has been made in the development of rf-SQUID as multiplexers \cite{Mates:2008bce, Hirayama_2013, Kempf_2017}. They are widely used to readout TESs \cite{Ullom_2015, Henderson:2018jlc} and MMCs \cite{Fleischmann:2009mey, Kempf_2017}.

\subsubsection{Parametric amplifiers}
\label{sec:SC-Paramp}
Driven by the demand of ultra low noise measurements of the weak microwave electromagnetic field in superconducting Qubit circuits for quantum computing, parametric amplifiers have been developed extensively and enables fast, precise measurements of quantum states with minimum perturbation. 

Parametric amplification is achieved by frequency mixing in a nonlinear media. A strong RF power, usually referred to as the pump, modulate the nonlinear media at frequency $f_p$. Signal at frequency $f_s$ will gain energy from the pump, either with one pump photon decay into a signal photon and an idler photon at frequency $f_i=f_p-f_s$, or with two pump photons convert to a signal photon and an idler photon at frequency $f_i=2f_p-f_s$. These nonlinear processes are called three-wave mixing and four-wave mixing, respectively. The gain grows exponentially with the interaction time of the signal wave in the nonlinear media, when the pump power is sufficiently large. The exponent is proportional to the pump power, and the gain is tunable by the pump, hence it is called parametric amplification.

In superconducting systems, in theory, the mixing processes are non-dissipative and the amplification introduces no additional thermal noise. It can be proved with quantum mechanics that a minimum of 1/2 photon noise will be added by any phase-insensitive amplifier, usually referred as the standard quantum limit (SQL), while even less added noise can be achieved in a phase-sensitive amplifier \cite{Caves_1985}. Here phase refers to the quadrature phase of the electromagnetic field, or say the conjugate amplitudes. In fact, superconducting parametric amplifiers have demonstrated near and beyond SQL performances \cite{Yurke:1989rli, Castellanos-Beltran:2008rap}. The quantum behavior originates from the simultaneous creation of the signal and the idler photons, which creates entanglements. This entanglement between $f_s$ and $f_i$ can be used in quantum non-demolition (QND) measurements, where the measurement is performed on the entangled quadrature different from the signal. In the special case of three-wave mixing when $f_p=2f_s$, signal and idler tones are degenerate, the entanglement creates phase-sensitive amplification that evades the SQL. It is often referred to as squeezing, where the vacuum noise in the same quadrature as the signal is deamplified and the other quadrature is amplified. In general, superconducting parametric amplifiers are nonlinear and dissipationless systems that provide a versatile playground for QND techniques \cite{Vijay:2011mof, Grimsmo_2021}. They are critical for both quantum computing and sensitive microwave photon sensing for axion and dark photon searches \cite{Malnou:2018dxn, DMRadio:2022jfv}.

The nonlinearity can be achieved by adding Josephson junctions (JJs) to the circuit, hence called Josephson parametric amplifier (JPA). The nonlinearity can also be achieved with kinetic inductance materials pumped close to their critical current, which is the principle of kinetic inductance parametric amplifiers (KIPAs). 
%Over the years, developments of kinetic inductance parametric amplifiers (KIPA) also gained momentum and matured as a complimentary technology in certain use cases. 
For both of the implementations, the core design consists of a resonator around $f_s$ containing the nonlinear component, which extends the interaction time of the signal photon while it oscillates in the resonator. The obvious downside of resonant amplifiers is the narrow bandwidth. They also require expensive cryogenic circulators and isolators to separate the input and output signals, as they are single-port and operate in reflection mode. It is possible to increase the bandwidth to $O(100)$MHz by directly connecting the resonator to the readout with an impedance-transformed transmission line \cite{Mutus:2014kef, Roy:2015vsi, Qing:2023qhs}.

The broadband design consists of a long nonlinear transmission line. The length needs to be on the order of $O(100)$ wavelengths to provide sufficient gain. This is called traveling wave parametric amplifier (TWPA). There are three basic design principles: increasing inductance and capacitance to reduce the wave speed, engineering the dispersion relation to phase-match $f_p$, $f_s$, and $f_i$, and creating stopping bands to suppress high-order nonlinear products \cite{Chaudhuri:2017bmn}. Both Josephson TWPA (JTWPA) and kinetic inductance TWPA (KITWPA) have been realized and both achieved near SQL wide band amplifications \cite{Macklin:2015xkk, Faramarzi:2024alp}. The reflections in the RF circuit causes gain ripples and depletion of pump power, limiting the overall bandwidth. One solution is to improve impedance matching by integrating the bias-tees and diplexers into a single compact packaging. Fundamentally, the ripples can be suppressed with the Floquet-mode amplification, which prevents the forward-backward wave coupling and sensitivity to out-of-band impedance mismatching \cite{Peng:2022gwy}. It is also possible to use frequency conversion to avoid amplification of backward waves and completely remove the bulky and magnetic circulators \cite{Malnou:2024hap}. 

JPA is most suitable for low-power and sensitive measurements, for example, measuring the states of superconducting Qubits. JJs have strong nonlinearity, hence a small pump power can produce sufficient gain. By substituting the single JJ in the resonator with a SQUID loop and adding flux to the loop, the JPA frequency can be tuned \cite{Yamamoto:2008lpo, Aumentado_2020}. The tunability is essential for resonant axion searches with tunable cavities. More advanced design with flux tuning is realized, such as the field-programmable Josephson amplifier (FPJA) \cite{Lecocq:2016vjp}. It is a three-port device which can be programmed as an amplifier, frequency converter, or a circulator. Other clever designs can also achieve frequency tuning without flux input \cite{Castellanos-Beltran:2008rap}. Currently, JPA has become commercially available and it is routinely used as the basic building blocks of high fidelity quantum measurement systems.

KIPA requires higher pump power, as the kinetic inductance nonlinearity is weaker than JJs'. But it offers several advantages. KIPA has larger power handling capability and higher gain. KIPA can handle power up to \SI{-60}{dBm} \cite{Parker_2022}, but JPA (JTWPA) is typically limited to \SI{-110}{dBm} (\SI{-100}{dBm}) by the critical current of JJs \cite{Yamamoto:2008lpo, Castellanos-Beltran:2008rap, OBrien:2014grw}. For this reason, KITWPA is ideal for multiplex readout of large resonator arrays. KIPA also has higher resistance to magnetic field \cite{Parker_2022, Frasca:2023nlk} comparing to JPA \cite{Janssen:2024xzp}. Since the common choice of kinetic inductance material, such as NbTiN, has higher $T_c$ than the commonly used Al in JPAs, KIPA can operate at higher temperature and in higher frequency bands. Recent work has demonstrated wide band parametric gain in the Ka-band ($27$ to \SI{40}{\giga\hertz}) and proposed KITWPAs up to $O(100)\si{\giga\hertz}$ \cite{Tan_2024}. KITWPA in the GHz to sub-GHz region has also been fabricated \cite{Faramarzi_2025}, which benefits the operation of spin qubits at sub-GHz \cite{Oakes:2022zsx}, far-IR single-photon counting KIDs \cite{Day_2024}, and ultra-light axion dark matter search in $0.8\sim2$ \si{\micro\electronvolt} \cite{Zhai:2025cwk}. 

In conclusion, superconducting parametric amplifiers are powerful and versatile tools to realize quantum measurement and operations. They are quickly shifting from stand-alone research topics to general applicants such as HEMTs in labs. They will enable QND measurements for quantum computing and axion and dark photon searches.

\subsection{Background and noises in superconducting quantum devices}
\label{sec:SC_BKG}

The development of particle-like light dark matter detectors primarily focuses on reducing backgrounds and noises. In previous sections, we have mentioned several sources of backgrounds and noises. Here we will discuss some general considerations and the current best understanding.

Traditionally, perturbations in the detectors that are discrete in time and generate identifiable event pulses are referred to as backgrounds. Other perturbations are continuous or happen too fast to be individually resolved, which are referred to as noises. We will discuss the common sources of backgrounds and noises in superconducting quantum devices, and we will see that the separation between background and noise is not definitive. Increase in background rate and worsening of energy resolution happens simultaneously. 

\subsubsection{Ionizing radiation}
The first background is ionizing radiation from the environment and cosmic rays. This is the primary focus of WIMP dark matter detectors. But for light dark matter, benefiting from the increased number density, at the current stage, athermal phonon detectors are on the scale of grams \cite{TESSERACT:2025tfw, SuperCDMS:2020aus, CRESST:2022lqw}, only have a small cross section for $\gamma$ and neutron radiation. And the signal energy of interest is below \SI{100}{\electronvolt}, lower than most ionizing backgrounds. Quantum computing community also identified ionizing radiation as the reason of catastrophic whole-chip failures \cite{McEwen:2021wdg}. 
Following studies \cite{Harrington:2024iqm, Li:2024dpf} showed that cosmic ray can account for about \SI{17}{\percent} to \SI{18}{\percent} of total QP burst events, while the rest can be accounted for by environmental $\gamma$-rays. 

Although the primary impact is much more energetic than the dark matter signals and easily distinguishable, they generate secondary effects that are more problematic. They can produce Cherenkov and scintillating photons in passive materials in the same enclosure of the detector. High energy $\gamma$ can coherently Thomson scatter off nucleons and deposit energies in the region of interest. They can release free-charges that change the field configuration in the substrate which changes operation points of the detectors, especially Qubits \cite{Bratrud:2024qnk}. They may also create defects in the substrate lattice that randomly release energy days after the interaction \cite{Heikinheimo:2021syx, Nordlund:2024xqi}. Standard mitigation procedures, such as underground lab \cite{Cardani:2020vvp}, shielding \cite{Vepsalainen:2020trd, Li:2024dpf}, and material screening, should be taken to minimize cosmic rays and radioactive contamination in the dilution refrigerator. 

\subsubsection{Low energy excess events}
The second background is the infamous Low energy excess events (LEE). As the athermal detector threshold lowered to \si{\electronvolt} scale, a type of non-ionizing background emerges \cite{EDELWEISS:2016nzl, CRESST:2022lqw, SuperCDMS:2020aus, Fuss:2022fxe, Du:2020ldo}, which does not necessarily scale with the detector volume, decays over the course of weeks to months after the detector is cooled to millikelvin temperatures, and can be re-excited with thermal cycles. They have a common power-law to exponential energy spectrum that quickly decays with energy in a variety of detector materials \cite{Baxter:2025odk}. 

A series of studies indicate that packaging induced external stress is a major source of the LEE \cite{Anthony-Petersen:2022ujw}. Following studies with crystals instrumented with two identical TES channels further demonstrated that there are two distinct populations of LEE \cite{Anthony-Petersen:2024vdh}. One type of LEE localizes around a single channel, creating uncorrelated phonon pulses; the other type of LEE has energy evenly distributed between the two channels causing correlated signals. The distinction hints at different origins, where the uncorrelated LEE is likely from the film stress in the superconducting sensors and the correlated LEE is likely from the substrate crystal. More recently, the same group demonstrated that the correlated LEE rate scales with the detector volume, suggesting bulk distributed sources, such as defects from radiation damages \cite{TESSERACT:2025odn}. 

More importantly, in the same work, it is clearly shown that the correlated noise power between the two channels decays with time exactly the same way as the above-threshold energy rates. This observation suggested that the LEE has a spectrum extended well below the threshold. The sub-threshold events have a rate high enough to manifest as a correlated phonon shot noise. By comparing the reduction of shot noise power and the increase in TES bias power, the average energy scale of the events is estimated to be close to the aluminum superconducting band gap. This is a strong indication that LEE is a major cause of the excess QPs in superconducting devices. Qubit experiments confirmed similar LEE rates \cite{Yelton:2025wsy}. Further studies are urgently needed to understand the microscopic origin of the bulk LEE events. At the same time, new fabrication recipes should be pursued to reduce the stress of superconductor films. 

\subsubsection{Continuous noise excitations}
As we look into excitations of lower energies, we smoothly transition from above-threshold background to below-threshold noise. The sources of excitations include blackbody radiation from warmer stages in the cryostat, mechanical vibrations induced frictional rubbing and flux cutting through magnetic fields \cite{Kono:2023jgd}, dissipation of the readout power, and back-propagating RF noise power from amplifiers. Most of the sources can be mitigated through shielding \cite{Barends:2011uku} and filtering \cite{Serniak:2019one, Jin:2015xlk}, while dissipation of the readout power is inevitable. 

In TESs, the readout power sustains the electrical-thermal feedback and generates the TFN. The energy resolution can be increased by reducing the TES volume and $T_c$, which lower the readout power and in turn reduces TFN, until external excitations become a significant fraction of the dissipation power. 

In KIDs and Qubits, excessive QPs cause additional GR noise and the background tunneling rate in Qubits, which are the fundamental noise source that limits the resolution of KIDs and Qubits. The origin of the excessive QP density has long been a mystery \cite{Zmuidzinas_2012, Cardani:2020vvp, Mannila:2021dkm}. Readout power and external black body photons are major contributions. Although the readout frequency is below the superconducting gap, they can still create QP heating and break Cooper pairs \cite{Catelani:2011pfh, deVisser:2014dcn}. Reducing readout power reduces the signal to noise ratio as the noise will be amplifier-limited. In such cases, quantum-limited parametric amplifiers can significantly improve the resolution \cite{Ramanathan:2024frr, Aumentado_2020}. 

In the best isolated setups, excess QPs still present. The high energy backgrounds mentioned above will cause persisting QP populations. Calibration using internal $^{64}$Cu sources \cite{Vepsalainen:2020trd} suggests that the impact from ionizing radiation contributes to a major fraction of power required to generate the excess QP populations, which in terms of the excited QP per Cooper pair $x_{qp}$ is on the order of $10^{-9}\sim10^{-8}$ \cite{Serniak:2019one, Mannila:2021dkm}. But the exponential reduction in QP density over time observed by \cite{Mannila:2021dkm} can not be explained by radiation. This suggests that LEE also plays a significant role, similar to the observations in TESs \cite{Anthony-Petersen:2022ujw, Yelton:2025wsy}. 
On the other hand, theory models suggest additional origins of excess QPs from the fluctuation of the superconducting gap due to disorders \cite{Bespalov_2016, de_Rooij_2025}. 

Nevertheless, the origins of excess QPs remain an intriguing and critical question for pair-breaking detectors. The signal power from dark matter interactions must dominate over the parasitic excitations to be resolvable. 

\subsubsection{Two level system noise}
\label{sec:SC_TLS}
KIDs and Qubits have another noise source, the two level systems (TLS). TLSs are disorders in amorphous materials surrounding the circuit, usually located in oxidized metal and substrate surfaces. Phenomenologically, the behavior of TLSs is well described by the standard tunneling model (STM) \cite{1998}, where the TLS has two local minima at similar levels in the energy potential that correspond to two configurations of the system. The barrier between the minima is higher than the thermal environment and tunneling dominates the transition between the two configurations. In real systems, TLS can be tunneling atoms, tunneling electrons, spins or magnetic impurities, or other interesting condensed matter states \cite{Muller:2019tnu}. TLSs usually possess electrical dipole moments that couple to external fields.

KIDs are mostly affected by weakly coupled TLS ensembles which have broad random distributions of energy and tunneling rates. This interaction manifests as low-frequency phase noise, i.e. fluctuations of resonant frequencies, through the fluctuations of the dielectric constant of the amorphous materials in the vicinity of the capacitors of KIDs \cite{Noroozian:2009gq}. The TLS noise is predominantly phase noise \cite{Zmuidzinas_2012, Muller:2019tnu}. Although no evidence of dissipation noise was observed even below the standard quantum limit \cite{Gao_2011} in certain setups, later results did demonstrate dissipative TLS noise \cite{Neill:2013aqs}, and it was tentatively explained by calculations showing TLSs are in squeezed states along the dissipation quadrature \cite{Takei_2012}.

The spectrum density of TLS noise $S$ scales as the inverse square root of the circulating power in the resonator, $S\sim P_\mathrm{int}^{-1/2}$ \cite{Kumar_2008, Neill:2013aqs}. The internal loss due to TLS follows the same scaling \cite{Goetz_2016, Pappas:2011dka}. Both effects can be explained by the saturation of TLS under strong field \cite{Muller:2019tnu, Gao:2008ryx}. The temperature dependence follows $S\sim T^{-1-\mu}$, where $\mu$ varies from 0.2 to 0.7 \cite{Kumar_2008, Muller:2019tnu, Zmuidzinas_2012}. This scaling can be explained as the spectral broadening of TLSs near the resonant frequency at elevated temperatures \cite{Burnett_2016, Faoro_2015, Burin_2015}. Finally, the frequency dependence ranges from $S\sim f^{-1/2}$ \cite{Gao:2008ryx, Kumar_2008, Zmuidzinas_2012} to $S\sim f^{-1}$ \cite{Neill:2013aqs, Burnett:2014lra}. The latter is observed in more recent measurements at lower frequencies around 10Hz and below \cite{Lindstrom:2011yx}, and it is motivated by the generalized STM \cite{Faoro_2015}, while the weaker frequency dependence might result from instrumentation limitations \cite{Muller:2019tnu}. Although the nature of TLS needs further investigation, these simple scaling laws serve as a general guideline for KID design optimization.

TLS is more detrimental to Qubits. In addition to the weak-coupled TLS ensemble scenario, which causes phase noise and internal losses, TLSs in the narrow dielectric barrier in the junction can strongly couple to Qubits, directly causing Qubit dephasing, dissipation and excitation \cite{Muller:2019tnu}. Understanding the microscopic origin of TLS and reducing them is a critical path to the realization of logical Qubits and large-scale quantum computing \cite{Klimov:2023bbz, Siddiqi:2021cbf, Wang_2022}.  

\subsubsection{Outlook}
TES, KID, and Qubits all share the same physical implementation of superconductors on crystal substrate in cryogenic environments. Many of the detector-level questions, for example the LEE, can be better tackled when comparing the three technologies. Another direction that the KID and Qubit can learn from TES is the idea of matching the signal time scale to the sensor response time to improve the signal-to-noise ratio through careful engineering of the energy dissipation path \cite{Pyle:2015pya, Billard:2022cqd, CPD:2020xvi}. Implementation of QP trapping in RF devices \cite{Fink_2024} is an important first step. While TES being the leading technology in athermal phonon sensing, KIDs and Qubits are being actively pursued and have the potential to build detectors with higher target mass and lower energy threshold. 

%% file: sections/conclusions.tex
\section{Summary}
\label{sec:summary}
In this article, we have summarized recent developments in particle physics driven by advances in quantum technology, highlighting both the scientific opportunities enabled by these technologies and a pedagogical framework for engaging with this interdisciplinary field from the perspectives of particle physics and quantum information science.

We first discussed quantum simulation as an emerging subfield within particle physics. In principle, first-principles calculations of real-time dynamics---such as parton evolution at colliders, early-universe processes, and out-of-equilibrium dynamics in first-order phase transitions relevant for baryogenesis---can be performed on quantum computers using Hamiltonian formulations. These problems are intrinsically challenging for classical computational methods. However, as we reviewed, the quantum algorithms required to realize such simulations remain in a very early stage. Key ingredients include digitization schemes for non-Abelian gauge theories in higher dimensions, quantum-circuit constructions for the corresponding Hamiltonians, methods for preparing nontrivial initial states relevant for hadron collisions and out-of-equilibrium physics, strategies for extracting physical observables, and approaches for taking the continuum spacetime limit. Substantial progress on these fronts will require co-development of quantum algorithms and quantum hardware over the long term.

We then turned to the use of quantum technologies as precision detectors capable of probing extremely subtle signals of new physics, including dark matter interactions beyond gravity and high-frequency gravitational waves. We reviewed three major categories of devices:
\begin{enumerate}
\item \textbf{Electromagnetic detectors}, including resonant cavities and LC circuits implemented on platforms similar to superconducting quantum-computing architectures, targeting ultralight dark matter (such as axions and dark photons with masses below ${\sim}\,\mu\text{eV}$) as well as high-frequency ($\gg$ kHz) gravitational waves that induce detectable electromagnetic signals;
    \item \textbf{Quantum magnetometers}, including superconducting quantum interference devices (SQUIDs) and spin-based magnetometers employing atomic ensembles, which are sensitive to ultralight dark-matter fields that couple to spins and new forces;
    \item \textbf{Pair-breaking superconducting sensors}, which exploit Cooper-pair--breaking thresholds around $0.1\,\mathrm{meV}$ including Transition edge sensors, Kinetic Inductance
Device and  to detect phonons or single photons generated through axion-like or dark-photon dark-matter conversion.
\end{enumerate}

These quantum-enabled detection technologies, as well as others have already begun to explore previously inaccessible regions of parameter space, for example, in axion dark-matter searches, and will continue to improve as both device performance and their applications to particle-physics questions advance.

Overall, the rapid development of quantum simulation and quantum sensing demonstrates how quantum technologies are beginning to reshape the landscape of particle physics. Continued progress in these areas promises not only deeper insights into the fundamental laws of nature but also a new generation of tools capable of probing phenomena far beyond the reach of conventional methods.

\section*{Acknowledgements}
\addcontentsline{toc}{section}{Acknowledgements}
We are grateful to Yuxin Liu and Dan Zhang for carefully reading the manuscript and providing valuable comments.
This work is partially supported by the National Natural Science Foundation of China under Grant Nos. 12025507, 12450006, and 12522509, and by the National Key R\&D Program of China under Contract No. 2025YFA1614200. The work of MC (partially) was supported by the Department
of Energy through the Fermilab QuantiSED program in
the area of “Intersections of QIS and Theoretical Particle
Physics.” This manuscript has been authored by the
Fermi Forward Discovery Group, LLC, under Contract
No. 89243024CSC000002 with the U.S. Department of
Energy, Office of Science, Office of High Energy Physics.
The research of MC at Perimeter Institute is supported in
part by the Government of Canada through the Department of Innovation, Science and Economic Development,
and by the Province of Ontario through the Ministry of
Colleges and Universities. The work of YC is supported by the Shanghai Qiguang Natural Science Development Foundation.